\documentclass[journal=mamobx,manuscript=article,layout=traditional]{achemso}
\setkeys{acs}{articletitle = true}

\usepackage{graphicx}
\graphicspath{{figures/}}

\usepackage{dcolumn}
\usepackage{bm}
\usepackage[utf8x]{inputenc}
\usepackage{lineno}
\usepackage{lipsum}
\usepackage{resizegather}
\usepackage{amssymb}
\usepackage{amsmath}
\usepackage{booktabs}
\usepackage{xcolor}
\usepackage{lineno}
\usepackage[symbol]{footmisc}
\title{Dynamics of nanoparticles in polydisperse polymer networks: From free diffusion to hopping}

\author{Valerio Sorichetti}
\affiliation{Laboratoire de Physique Théorique et Modèles Statistiques (LPTMS), CNRS, Université Paris-Saclay, F-91405 Orsay, France}
\alsoaffiliation{Laboratoire Charles Coulomb, Univ. Montpellier, CNRS, F-34095, Montpellier, France}
\alsoaffiliation{IATE, Univ. Montpellier, INRAE, Institut Agro,  F-34060, Montpellier, France}
\email{valerio.sorichetti@universite-paris-saclay.fr}

\author{Virginie Hugouvieux}
\affiliation{IATE, Univ. Montpellier, INRAE, Institut Agro,  F-34060, Montpellier, France}
\email{virginie.hugouvieux@inrae.fr}

\author{Walter Kob}
\affiliation{Laboratoire Charles Coulomb (L2C), Univ. Montpellier, CNRS, F-34095, Montpellier, France}
\email{walter.kob@umontpellier.fr}

\SectionNumbersOn

\begin{document}

%%%%%%%%%%%%%%%%%%%%%%%%%%%%%%%%%%%%%%%%%%%
\begin{abstract}

Using molecular dynamics simulations we study the static and dynamic properties of spherical nanoparticles (NPs) embedded in a disordered and polydisperse polymer network. Purely repulsive (RNP) as well as weakly attractive (ANP) polymer-NP interactions are considered. It is found that for both types of particles the NP dynamics at intermediate and at long times is controlled by the confinement parameter $C=\sigma_N/\lambda$, where $\sigma_N$ is the NP diameter and $\lambda$ is the dynamic localization length of the crosslinks. Three dynamical regimes are identified: i) For weak confinement ($C \lesssim 1$) the NPs can freely diffuse through the mesh; ii) For strong confinement ($C \gtrsim 1$) NPs  proceed by means of activated hopping; iii) For extreme confinement ($C \gtrsim 3$) the mean squared displacement shows on intermediate time scales a quasi-plateau since the NPs are trapped by the mesh for very long times. Escaping from this local cage is a process that depends strongly on the local environment, thus giving rise to an extremely heterogeneous relaxation dynamics. The simulation data are compared with the two main theories for the diffusion process of NPs in gels. Both theories give a very good description of the $C-$dependence of the NP diffusion constant, but fail to reproduce the heterogeneous dynamics at intermediate time scales. 
\end{abstract}

%%%%%%%%%%%%%%%%%%%%%%%%%%%%%%%%%%%%%%%%%%%%%%%%%                  
\maketitle
%\pagewiselinenumbers

%%%%%%%%%%%%%%%%%%%%%%%%%%%%%%%%%%%%%%%%%%%%%%%%%   
%%%%%%%%%%%%%%%%%%%%%%%%%%%%%%%%%%%%%%%%%%%%%%%%%                  
\section{Introduction}
%%%%%%%%%%%%%%%%%%%%%%%%%%%%%%%%%%%%%%%%%%%%%%%%%     
%%%%%%%%%%%%%%%%%%%%%%%%%%%%%%%%%%%%%%%%%%%%%%%%%       

When a nanoparticle (NP) is embedded in a polymer network, its dynamics can slow down dramatically \cite{dell2014theory,cai2015hopping}. Understanding what factors govern this slowing down is of primary importance in many fields, such as material science (e.g. with application to thin films \cite{huang2010long,flier2012heterogeneous,bhattacharya2013plasticization} and polymer-based sensors \cite{huang2010long,riedinger2011nanohybrids,zhai2013highly}), biophysics \cite{amblard1996subdiffusion,wong2004anomalous,fritsch2010anomalous,stylianopoulos2010diffusion,peulen2011diffusion,yu2018rapid,cherstvy2019non,burla2020particle,debets2020characterising}, and medicine, in particular for applications to drug delivery \cite{cho2008therapeutic,riedinger2011nanohybrids,ward2011thermoresponsive}. While in recent years the diffusion of NPs in polymer solutions and melts has been the subject of numerous theoretical \cite{cai2011mobility,egorov2011anomalous,yamamoto2011theory,yamamoto2014microscopic,dong2015diffusion} and simulation studies  \cite{bedrov2003matrix,liu2008molecular,kalathi2014nanoparticle,patti2014molecular,li2014dynamic,kalathi2015rouse,volgin2017molecular,karatrantos2017polymer,chen2017coupling,chen2017effect,yamamoto2018theory,du2019study}, only  few investigations have dealt with the problem of NP diffusion in permanently crosslinked networks, despite its importance in many applications~\cite{netz1997computer,sonnenburg1990molecular,licinio1997anomalous,zhou2009brownian,godec2014collective,kamerlin2016tracer,kumar2019transport,chen2020nanoparticle,cho2020tracer}.  
In some of the earliest simulation studies, this network was simply modeled as an array of fixed obstacles \cite{netz1997computer}, which is clearly a far cry from a physically realistic description. Other authors have included connectivity and flexibility in the network model, but most of them considered only regular structures, in which the crosslinks are placed on the vertices of a regular lattice and connected either by chain segments \cite{sonnenburg1990molecular,xu2021enhanced} or directly by springs \cite{licinio1997anomalous, zhou2009brownian,godec2014collective,kumar2019transport}. In the latter case, since there is no actual strand connecting the crosslinks, strand dynamics and entanglement effects are not accounted for. Moreover, real-life networks such as hydrogels \cite{peppas1985structure},  vulcanized rubbers \cite{gehman1969network}, or 
networks produced by electron irradiation \cite{falcao1993structure},
are often disordered and polydisperse, with a continuous distribution of strand lengths, properties that lead to an additional complexity in the dynamics of the NP. Recently, a small number of simulation studies adopting more realistic models for the network has been published \cite{kamerlin2016tracer,chen2020nanoparticle,cho2020tracer}, however none of these studies has, to the best of our knowledge, taken explicitly into account the effect of disorder and polydispersity.

Also analytical studies dealing with NP dynamics in permanently crosslinked networks are rather scarce \cite{dell2014theory,cai2015hopping}. In entangled polymer liquids, the relaxation dynamics of particles of size larger than the tube diameter\cite{rubinstein2003polymer} $d$ can proceed through the release of the entanglements (constraint release \cite{yamamoto2014microscopic}), which happens on time scales of the order of the disengagement (reptation) time, $\tau_d \propto N^{3.4}$ \cite{rubinstein2003polymer}, with $N$ the degree of polymerization of the chains. In polymer solids containing irreversible crosslinks, like dry networks (e.g. rubbers) and gels, the constraint release mechanism is completely turned off and the only available process for the motion of large NPs is the one of hopping, \textit{i.e.}, activated motion triggered by local fluctuations of the entanglement/crosslink mesh \cite{dell2014theory,cai2015hopping}. Dell and Schweizer \cite{dell2014theory} have developed a theory of hopping based on a combination of a nonlinear Langevin equation and PRISM\cite{schweizer1997integral} theory \cite{schweizer2003entropic}, showing that the quantity which controls the NP dynamics is the so-called \textit{confinement parameter}, \textit{i.e.} the ratio $C=\sigma_N/d$ between the NP diameter $\sigma_N$ and the effective tube diameter $d$, resulting from both crosslinks and entanglements. If the effect of the entanglements can be neglected one has $d\approx \xi$, where $\xi$ is the average mesh size of the network \cite{cai2015hopping}. The conclusion that $C$ is the parameter controlling the NP dynamics has also been reached by Cai \textit{et al.} using scaling theory \cite{cai2015hopping}. We note, however, that the two approaches predict qualitatively different behaviors for the relevant dynamical quantities, such as the NP diffusion coefficient as a function of $C$, see below for details. The importance of the confinement parameter for the description of the diffusion of NPs in polymer networks has also been confirmed in experiments \cite{parrish2017network,parrish2018temperature,anderson2019filament,cherstvy2019non,wang2020controlled,burla2020particle} and simulations \cite{kumar2019transport,chen2020nanoparticle,cho2020tracer}. However, even recent simulations \cite{chen2020nanoparticle,cho2020tracer} have not explored the strong confinement regime, $C \gtrsim 3$, due to the extremely slow dynamics that characterizes it and hence the dynamics of the NP in this range of parameters is at present not known.

In this work, we present a simulation study of NP diffusion in polymer networks which are both disordered and polydisperse. In particular, we probe for the first time in simulations the strong confinement regime, considering confinement parameters up to $C \simeq 4$, for which NP motion is dramatically slowed down. The remaining part of the paper is organized as follows: In Sec.~\ref{sec:methods}, the model and the simulation method are presented. In Sec.~\ref{sec:structure}, we analyze the structural properties of the network and of the NPs for different NP diameters at low NP concentration. In Sec.~\ref{sec:dynamics}, we present the NP dynamics and in particular their diffusion coefficient and van Hove function, showing that for $C > 1$, NP diffusion proceeds through hopping motion. We conclude with a summary in Sec.~\ref{sec:summary}.

%%%%%%%%%%%%%%%%%%%%%%%%%%%%%%%%%%%%%%%%%%%%%%%%%   
%%%%%%%%%%%%%%%%%%%%%%%%%%%%%%%%%%%%%%%%%%%%%%%%%                  
\section{Model and simulation method}\label{sec:methods}
%%%%%%%%%%%%%%%%%%%%%%%%%%%%%%%%%%%%%%%%%%%%%%%%%        
%%%%%%%%%%%%%%%%%%%%%%%%%%%%%%%%%%%%%%%%%%%%%%%%%      

We performed molecular dynamics (MD) simulations of spherical NPs embedded in a polydisperse, disordered and permanently crosslinked polymer network. The network is generated following the procedure described by Gnan \textit{et al.} \cite{gnan2017silico}, initially developed for the simulation of microgels \cite{rovigatti2019numerical,ninarello2019modeling}, which is based on the self-assembly of particles with limited valence (``patchy" particles): $N_m= 4 \times 10^{5}$ particles (monomers) are placed in a cubic box of volume $V$ with periodic boundary conditions, and thus the monomer density is $\rho_m=N_m/V$. Of these particles, $N_\text{cl}=cN_m$, with $c=0.1$, play the role of crosslinks, in that they can form three bonds, whereas the $N_m-N_c$ others can only form two bonds (bivalent particles). These bivalent particles have patches on the opposite site of the particles, while the crosslinks have three patches forming 120 degrees with each other. Two crosslinks are forbidden to bind to each other, whereas a bivalent particle can bind to any other particle. A $NVT$ MD simulation is started and it is stopped when at least $99 \%$ of all the possible bonds are formed and a percolating network is generated. At this point, all the monomers which are not part of the percolating network (at most $0.4\%$ of all the monomers in the systems considered here) are removed. (Although  for the sake of computer time we chose to stop the reaction before reaching the fully-bonded ground state of the system, reaching this state is in principle possible by making a greater computational effort.) The system obtained from this procedure contains only a small number of dangling ends, \textit{i.e.}, it is an almost perfect network. We find that the number of monomers forming dangling ends is less than $2.5\%$ of the total. However, since a dangling end of length $n$ has a relaxation time which growth exponentially with $n$ \cite{curro1983theoretical,duering1994structure}, we remove recursively all the dangling ends in the system, so that at the end a fully-bonded network is obtained. We note that in this procedure the value of $N_m$ slightly decreases as well as the fraction $c$ of crosslinks since a crosslink becomes a bivalent particle whenever a dangling end attached to this crosslink is removed; however, given the small number of dangling ends, these changes can be considered to be negligible. 

Previous studies have found that the networks generated using this procedure have a chain length distribution $p(n)$ which decays exponentially in $n$, Ref.~\cite{sorichetti2021effect}, and is given by the Flory-Stockmayer expression \cite{flory1953principles,stockmayer1943theory},

\begin{equation}
p(n) = \frac{N_m-N_c}{\langle n\rangle^2} \left(1-\frac 1 {\langle n \rangle}\right)^{n-1},
\label{eq:pn_fs}
\end{equation}

\noindent where $n$ is the chain length, \textit{i.e.}, the number of beads between two crosslinks, see  Supplementary Material, Sec.~S2.1. We note that $p(n)$ is independent of the monomer density $\rho_m$ in the $\rho_m$-range here considered, as it is to be expected given the equilibrium nature of the assembly procedure \cite{gnan2017silico,sorichetti2021effect}. The mean chain length $N_x$ is obtained as 
$ N_x \equiv \langle n\rangle = {2 (c^{-1}-1)/3}$~\cite{rovigatti2017internal}, which 
for $c=0.1$ gives $N_x=6$ (the actual value is slightly larger, since the removal of the dangling ends makes that $c$ decreases slightly, as discussed above).

Once the dangling ends are removed, the topology is frozen and the interaction potential between the particles is changed from that of Gnan \textit{et al.}\cite{gnan2017silico} to the Kremer-Grest potential \cite{kremer1990dynamics}, in which all monomers interact via a Weeks-Chandler-Andersen (WCA) potential \cite{weeks1971role}, 

\begin{equation}
U_{mm}(r) = 
\begin{cases}
4 \epsilon \left[ \left(\frac \sigma r\right)^{12} -\left(\frac \sigma r\right)^6+ \frac 1 4 \right] & r \leq 2^{1/6} \sigma\\
0 & \text{otherwise}.\\
\end{cases}
\label{lj}
\end{equation}

\noindent In addition, bonded monomers interact via a finite extensible nonlinear elastic (FENE) potential,

\begin{equation}
U_{\text{bond}}(r)= -\frac {k r_0^2} 2 \ln \left[1-\left(\frac{r}{r_0}\right)^2\right],
\end{equation}

\noindent where $k=30 \epsilon/\sigma^2$ and $r_0 = 1.5 \sigma$. With this choice of parameters the minimum of the potential is at $r_b = 0.961$. The combined effect of the FENE and the WCA potentials prevents the chains from crossing each other at the thermodynamic conditions considered here \cite{kremer1990dynamics}. In the following, all quantities are given in Lennard-Jones (LJ) reduced units. The units of energy, length and mass are thus, respectively, $\epsilon$, $\sigma$ and $m$, where $\epsilon$, and $\sigma$ are defined by Eq.~\eqref{lj} and $m$ is the mass of a monomer. The units of temperature and time are, respectively, $T^*=\epsilon/k_B$ and $\tau^*=\sqrt{m \sigma^2/\epsilon}$, where $k_B$ is Boltzmann's constant, which we set equal to $1$.

\begin{figure}
\centering
\includegraphics[width=0.6 \textwidth]{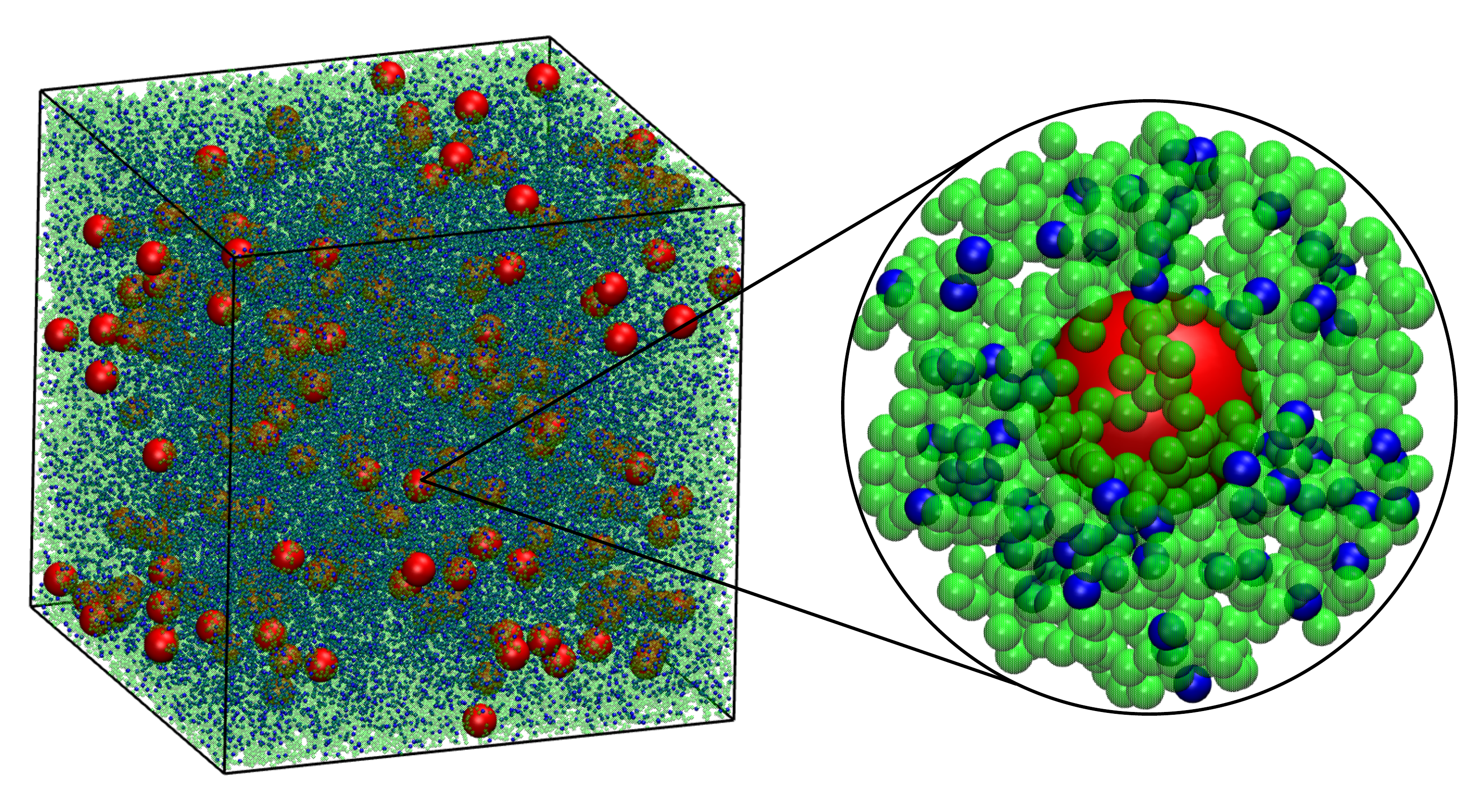}
\caption{\textit{Left}: Snapshot of the system for $\rho_{m0}=0.290$, $\sigma_N=6.0$ for the repulsive nanoparticles. NPs, crosslinks and bivalent monomers are represented, respectively, by red, blue and green spheres. For clarity, bivalent monomers are shown as transparent. \textit{Right}: Close-up of a NP with the surrounding polymer mesh. Shown are all monomers within a distance $3\sigma_N/2$ from the NP center.}
\label{fig:snapshot}
\end{figure}

Once the network is generated, we embed in it spherical nanoparticles and perform MD simulations to anneal the system. We consider NPs of diameter $\sigma_N$ ranging from $\sigma_N=1$ to $\sigma_N=10$ (see Supplementary Material, Sec.~S1). For a given set of parameters we use a single realization of each NP-enriched network, since the system is large enough to be self-averaging. Note that the network is the same for each value of $\sigma_N$. The interaction between monomers and NPs and between two NPs is given  by an expanded Lennard-Jones potential \cite{chen2017effect,sorichetti2018structure,chen2018coupling,chen2020nanoparticle}:

\begin{gather}
\mathcal U_{N\alpha}(r) = 
\begin{cases}
4 \epsilon\left[ \left(\frac \sigma {r-\Delta_{N\alpha}}\right)^{12} -\left(\frac \sigma  {r-\Delta_{N\alpha}}\right)^6 \right] +E_{N\alpha} &  r \leq \Delta_{N\alpha} + r^c_{N\alpha}\\
0 & \text{otherwise},\\
\end{cases}
\label{eq:expanded_lj}
\end{gather}

\noindent with $\alpha \in \{N,m\}$ and where for the NP-monomer interaction $\Delta_{Nm} =(\sigma_N+\sigma)/2-\sigma=(\sigma_N - \sigma)/2$ and for the NP-NP interaction $\Delta_{NN} = \sigma_N - \sigma$. The addition of $E_{N\alpha}$ ensures that at the cutoff distance $\Delta_{N\alpha}+r^c_{N\alpha}$ the potential $\mathcal U(\Delta_{N\alpha}+ r^c_{N\alpha})$ is continuous. The NP-NP interaction is purely repulsive ($r^c_{NN}=2^{1/6}$), whereas for the NP-monomer interaction we consider both repulsive ($r^c_{Nm}=2^{1/6}$) and attractive ($r^c_{Nm}=2.5$) interactions.  In the following, we will refer to attractive NPs as ANP and to repulsive NPs as RNP. We assume that the NPs have the same mass density as the monomers, $\rho_\text{mass} =6 m/\pi \sigma^3$, and therefore the mass of the NPs is $m_N = m (\sigma_N/\sigma)^3$. In Fig.~\ref{fig:snapshot} we show a snapshot of the system for $\sigma_N=6$ and density $\rho_{m0}=0.290$ (defined below) for the case of RNPs.

All the simulations are carried out using the LAMMPS software \cite{lammps,plimpton1995fast}. The simulation box is cubic and periodic boundary conditions are applied in all directions. Initially, the centers of the NPs are inserted at random positions in the network. The possible overlaps are then removed using the ``fast push-off'' method \cite{auhl2003equilibration}, that increases the NP diameter from $0$ to $1$; subsequently the NP size is gradually increased until the diameter reaches the value $\sigma_N$. After the NPs have reached the desired size, we perform an $NPT$ run with Nos\'e-Hoover chains\cite{tuckerman2010statistical} and allow the system to reach pressure $P=0$ at temperature $T=1.0$. During this $NPT$ run, the box sides are coupled so that they fluctuate together, i.e.~$L_x=L_y=L_z$.  Once the system has adjusted to $P=0$, we perform another short $NPT$ run, during which the mean volume $V$ is measured. Subsequently we switch to the $NVT$ ensemble by fixing the system's volume to $V$ and perform an equilibration run before starting production at constant volume, i.e.~the dynamics of the particles is not perturbed by unphysical volume fluctuations due to a barostat. Since the Nos\'e-Hoover thermostat does not produce a realistic dynamics \cite{tuckerman2010statistical}, we switch to a Langevin thermostat, so that the force experienced by particle $i$ (monomer or NP) is given by \cite{schneider1978molecular}

\begin{equation}
m_i \ddot {\mathbf r}_i = - \mathbf \nabla_i U (\{ \mathbf r_k \}) - m_i \gamma_i \dot {\mathbf r}_i + \ \boldsymbol{\zeta}(t).
\label{eq:langevin}
\end{equation}

\noindent 
Here $\mathbf r_i$ is the position vector, $m_i$ the mass, and $U (\{ \mathbf r_k \})$ is the total interaction potential acting on the particle, with $\{ \mathbf r_k \}$ representing the set of coordinates of all the particles in the system. The second term on the right-hand side of Eq.~\eqref{eq:langevin} represents viscous friction, with $\gamma_i$ the friction coefficient. The term $\boldsymbol{\zeta}$ is a stochastic force which represents the collisions with solvent molecules, and satisfies $\langle \boldsymbol{\zeta}(t) \rangle = 0$ and $\langle \zeta_\alpha (t) \zeta_\beta (t') \rangle =  {2m_i\gamma_i k_B T \delta_{\alpha,\beta} \delta(t-t')}$, with $\zeta_{\alpha}$ its spatial components. The monomer friction coefficient is $\gamma_m=0.1$, whereas the NP friction coefficient is $\gamma_N = \gamma_m (m \sigma_N/m_N \sigma) = \gamma_m (\sigma/\sigma_N)^2$, so that an isolated monomer and an isolated NP experience the same solvent viscosity \cite{sorichetti2018structure}. Although with this thermostat hydrodynamic interactions are neglected, we expect that due to the slow dynamics these interactions are not relevant. The velocity Verlet algorithm is employed to integrate the equations of motion, and the integration time step is $\delta t = 0.006$. The duration of the equilibration run is between $3 \times 10^4$ and $1.5 \times 10^6$ time units, depending on the system considered. We note that the relaxation time of the NPs is expected to increase at least exponentially in $\sigma_N/d$, where $d$ is an effective tube diameter resulting from both crosslinks and entanglements \cite{dell2014theory,cai2015hopping}. For this reason, not all systems presented here have reached equilibrium, as it will be discussed below. Nevertheless, we decided to include also the results for these systems since they are affected only weakly by aging effects and hence are still instructive. The duration of the production runs is between $6 \times 10^4$ and $3 \times 10^6$, depending on the NP diameter and on the network density.

In order to avoid that the addition of the NPs influences significantly the structural and dynamical properties of the network and to study single-NP dynamics, the number $N_N$ of NPs embedded in the network is kept small. The value of $N_N$ is chosen in such a way that the total NP volume, $V_N\equiv \pi N_N \sigma_N^3/6$, is $2\%$ of the mean volume of the neat system $V_0$, \textit{i.e.}, the NP volume fraction of the unrelaxed system is

\begin{equation}
\phi_{N0} \equiv \frac {\pi \sigma_N^3 N_N}{6 V_0} = 0.02\quad .
\end{equation}

\noindent
To make sure that we are indeed probing the dilute limit, we have simulated also some systems with $\phi_{N0}=0.005$, but, unless explicitly specified, the shown results are for $\phi_{N0}=0.02$. No significant difference in the dynamical properties is found between these two values, confirming that the dilute limit is already reached at $\phi_{N0}=0.02$. We consider three different starting networks, with monomer densities $\rho_{m0}=N_m/V_0 = 0.190$, $0.290$ and $0.375$ in the neat state. We note that in general $V_0$ is different from the final volume $V$ (and therefore $\rho_0 \neq \rho$ and $\phi_{N0} \neq \phi_N$), since the addition of NPs can cause swelling or shrinkage (depending on $N_N$ and $\sigma_N$) of the network, as also observed in non-crosslinked nanocomposites \cite{sorichetti2018structure}. This point is discussed in more detail in the Supplementary Material, Sec.~S2.2, where one can also find the details of the simulated systems (Sec.~S1).

%%%%%%%%%%%%%%%%%%%%%%%%%%%%%%%%%%%%%%%%%%%%%%%%%   
%%%%%%%%%%%%%%%%%%%%%%%%%%%%%%%%%%%%%%%%%%%%%%%%%                  
\section{Static properties}\label{sec:structure}
%%%%%%%%%%%%%%%%%%%%%%%%%%%%%%%%%%%%%%%%%%%%%%%%%   
%%%%%%%%%%%%%%%%%%%%%%%%%%%%%%%%%%%%%%%%%%%%%%%%%   
In this section we discuss the structural properties of the system, notably the radial distribution function, the static structure factor, and the pore size distribution.

%%%%%%%%%%%%%%%%%%%%%%%%%%%%%%%%%%%%%%%%%%%%%%%%%                  
\subsection{Radial distribution function and static structure factor}\label{sec:rdf_sq}
%%%%%%%%%%%%%%%%%%%%%%%%%%%%%%%%%%%%%%%%%%%%%%%%%  

\begin{figure}
\centering
\includegraphics[width=0.48 \textwidth]{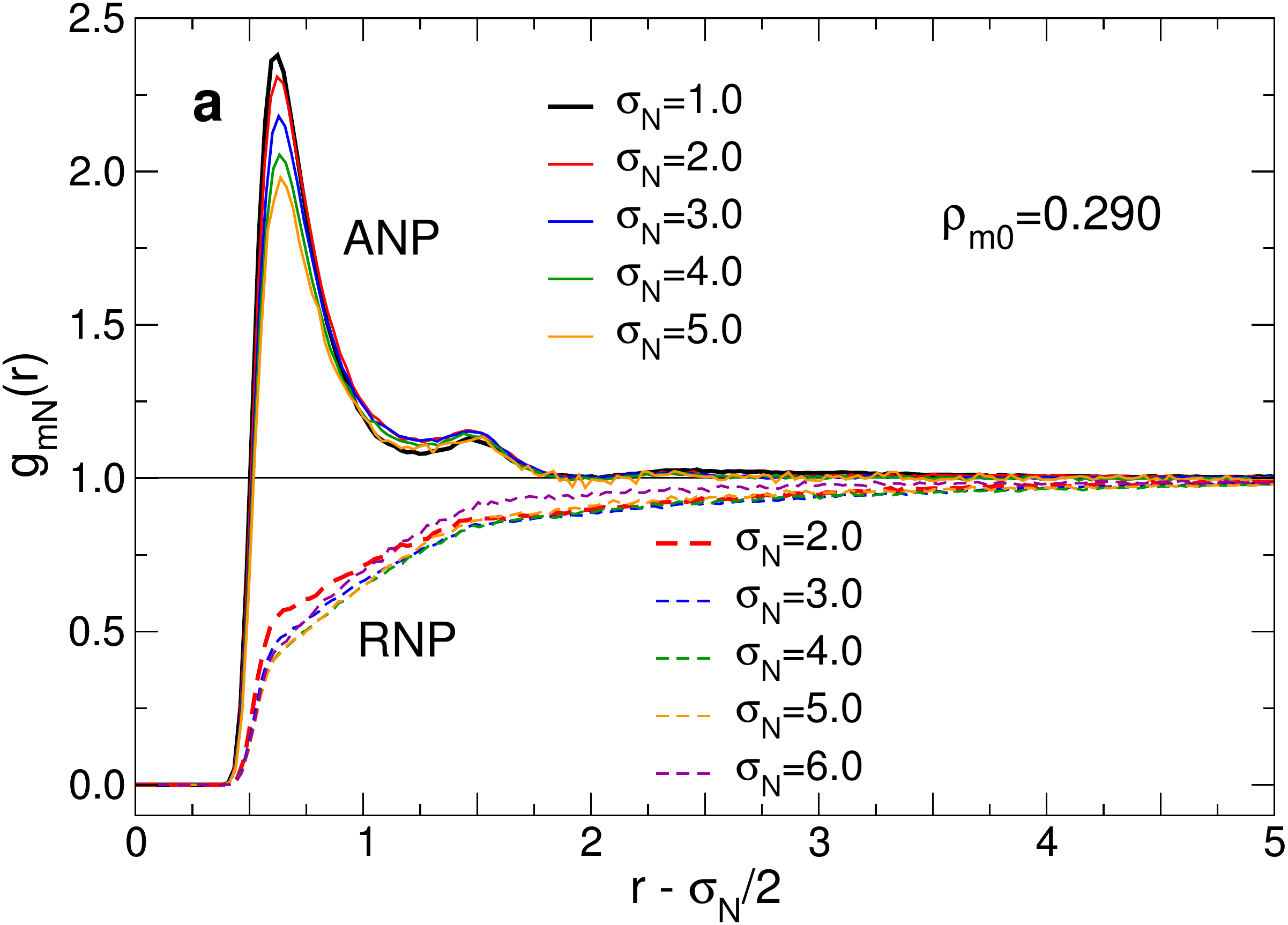}
\includegraphics[width=0.48 \textwidth]{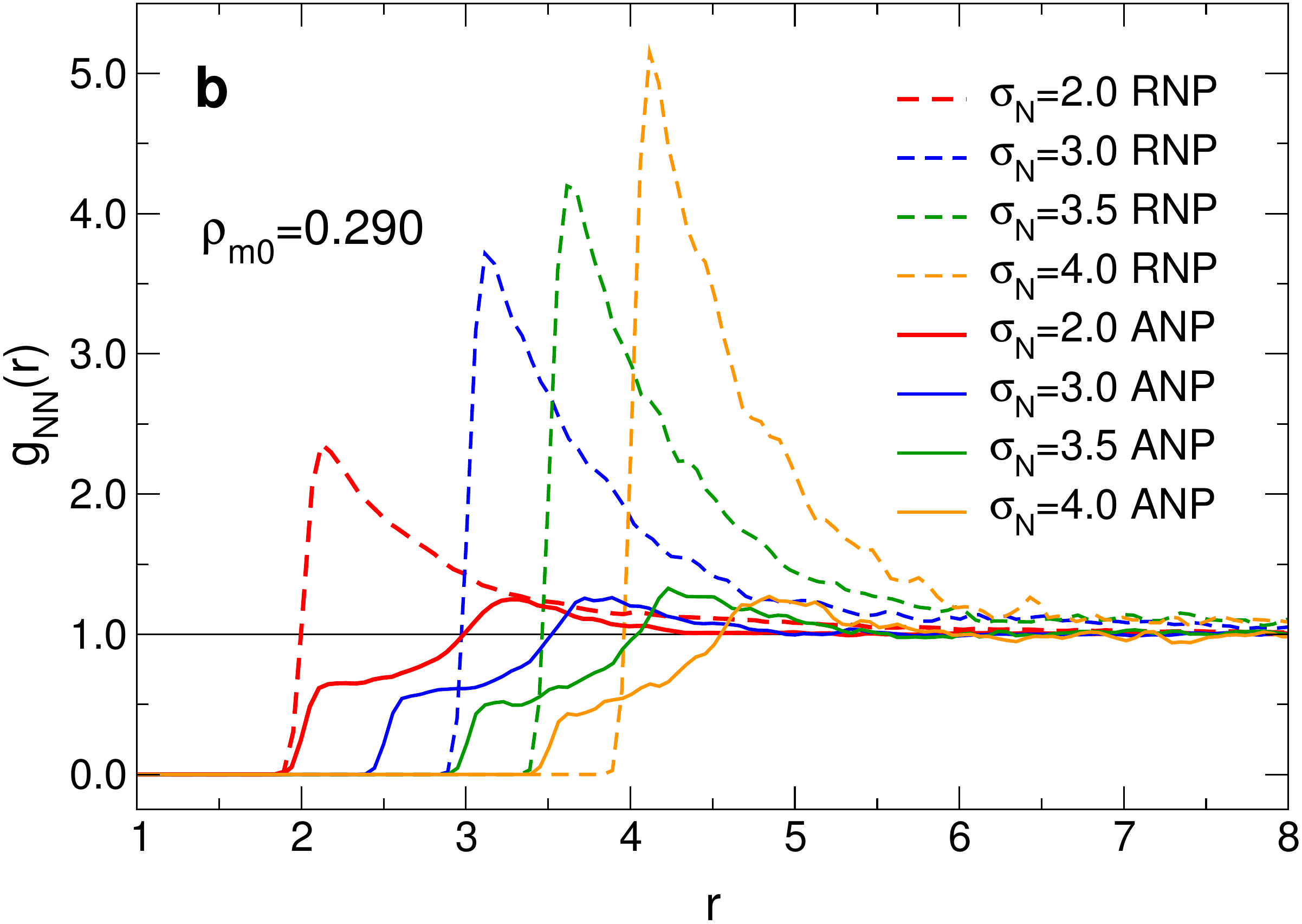}
\caption{Monomer-NP (\textbf{a}) and NP-NP (\textbf{b}) radial distribution functions for ANPs and RNPs embedded in the $\rho_{m0}=0.290$ network.}
\label{fig:rdf_mono_np_rho02}
\end{figure}

The simplest way to characterize the structure of the system is by means of the radial distribution function (RDF).
In Fig.~\ref{fig:rdf_mono_np_rho02}a we show for the systems with $\rho_{m0}=0.290$ the monomer-NP radial distribution function, defined as \cite{binder2011glassy}

\begin{equation} 
g_{mN}(r) \equiv \frac{N_m + N_N}{4 \pi N_m N_N (\rho_m+\rho_N) r^2} \sum_{k=1}^{N_m} \sum_{j=1}^{N_N}  \langle \delta(|\mathbf r + \mathbf r_k- \mathbf r_j|)\rangle,
\label{eq:rdf_mn}
\end{equation}

\noindent
where $\rho_m$ and $\rho_N$ are, respectively, the monomer and NP number density. Note that on the horizontal axis we plot $r-\sigma_N/2$ in order to remove the trivial $\sigma_N$ dependence of the curves. One recognizes that for repulsive NPs (RNPs) as well as for attractive NPs (ANPs), the data collapse approximately onto a master curve. For the ANPs, $g_{mN}(r)$ displays a main peak at $r\simeq(\sigma_N+1)/2=r_1$, which corresponds to the first shell of monomers touching the NP, followed by a smaller peak at $r\simeq(\sigma_N+3)/2=r_2$, corresponding to the second shell. Since this is not a dense system, the presence of these peaks demonstrates that, as expected, monomer-NP contacts are favored. One can see that the height of the main peak increases slightly if $\sigma_N$ is decreased: This can be explained by the fact that the total surface of the NP increases with decreasing $\sigma_N$, making that more monomers can touch this (energetically favorable) surface, thus increasing the height of the nearest neighbor peak. Furthermore this attraction has the  effect that neighboring strands are pulled closer to each other, making that on overall the system shrinks (see Supplementary Material, Sec.~S2.2).

For repulsive NPs, $g_{mN}(r)$ is significantly smaller than $1$ for $r_1 < r < r_2$, in agreement with the fact that monomer-NP contacts are unfavorable. A consequence of this correlation hole is that the effective radius of repulsive NPs is larger than that of ANPs with the same $\sigma_N$, as we will discuss below, and hence the mobility of the ANP is higher than the one of the RNP, see next section. For the other two networks ($\rho_{m0}=0.190$ and $0.375$) we find qualitatively similar results (see Supplementary Material, Sec. S2.3).

In Fig.~\ref{fig:rdf_mono_np_rho02}b we show for the same systems the NP-NP radial distribution function, which is obtained by setting in Eq.~\eqref{eq:rdf_mn} $m=N$. For the RNPs, the RDF displays a pronounced single peak at the contact distance  $r_c \equiv \sigma_N + (2^{1/6}-1) = \sigma_N + 0.122$  \cite{sorichetti2018structure}. From the rather large value of $g_{NN}(r_c)$ and the absence of further peaks one can deduce that the RNPs tend to form clusters, likely by filling the largest holes in the mesh. The height of this peak increases with increasing $\sigma_N$, which is simply due to the fact that the NP density is lower, at constant NP volume fraction, for larger NPs (see also Supplementary Material, Sec.~S2.3).  For the ANPs, one finds at $r =\sigma_N + 0.122$   a weak shoulder and a small peak at $r= \sigma_N + (2^{7/6}-1) =\sigma_N+1.245$, which corresponds to a configuration in which two neighboring NPs are separated by a polymer strand \cite{sorichetti2018structure}. This is a consequence of the fact that the NPs are well dispersed, and that each NP is surrounded by a layer of polymers. Such local structures are expected, since it is known that weak attractive polymer-NP interactions lead to a good NP dispersion in nanocomposites \cite{hooper2006theory,liu2011nanoparticle,meng2013simulating,karatrantos2015polymer,chen2020nanoparticle}. For a more detailed discussion of $g_{NN}(r)$, we refer to the Supplementary Material, Sec.~S2.3.

\begin{figure}
\centering
\includegraphics[width=0.48 \textwidth]{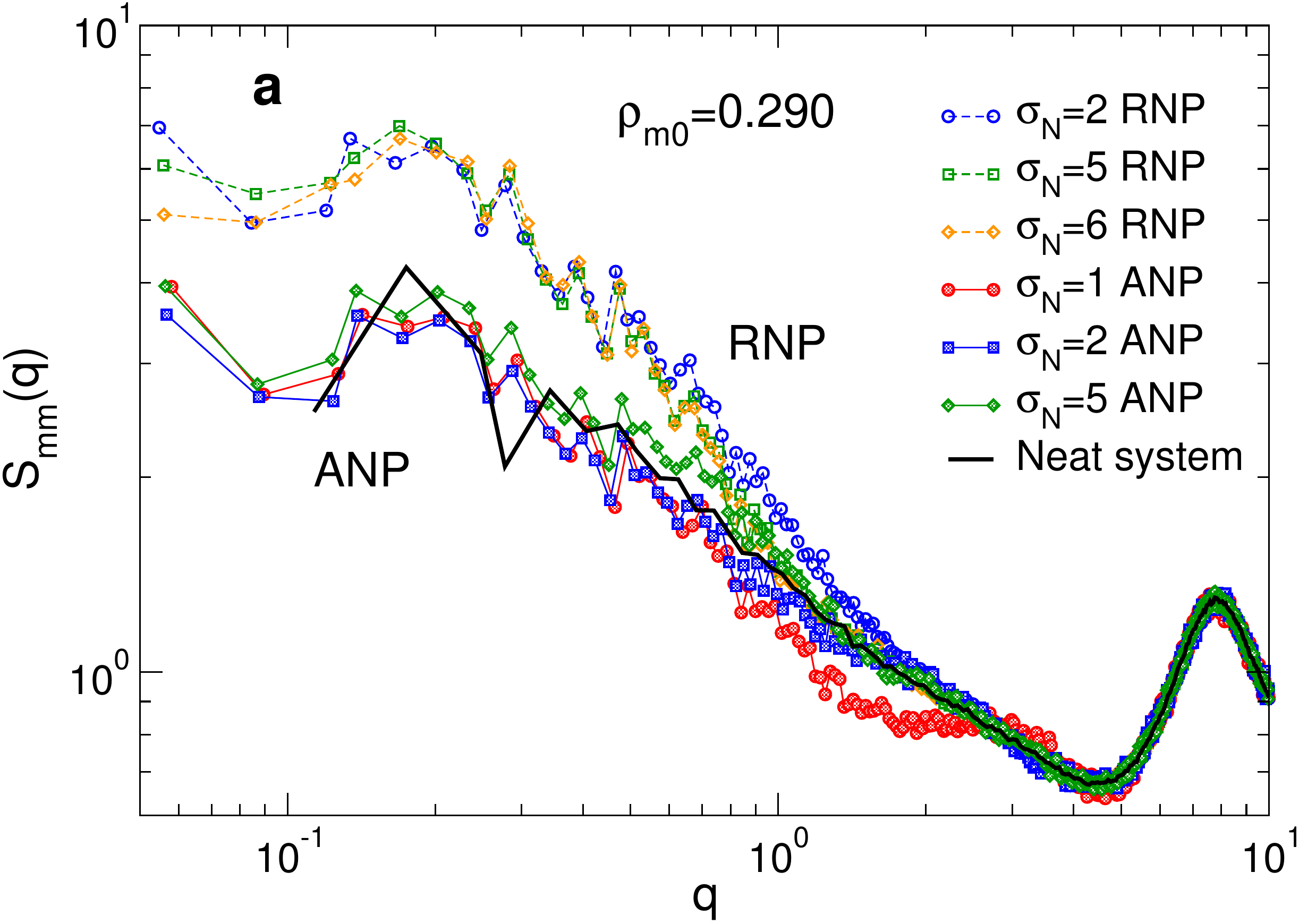}
\includegraphics[width=0.48 \textwidth]{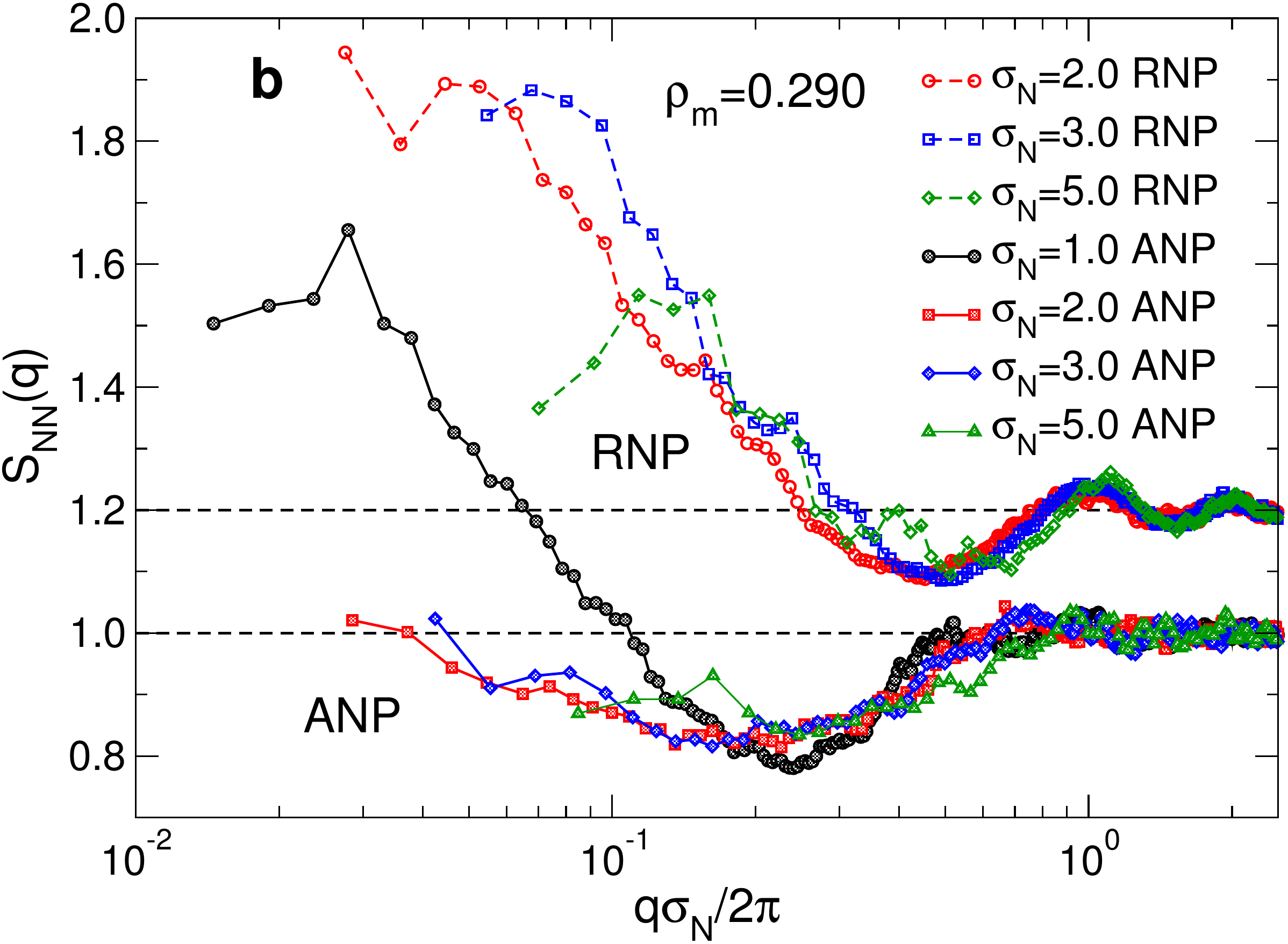}
\caption{(\textbf{a}) Monomer-monomer structure factor for $\rho_{m0}=0.290$ and different values of $\sigma_N$. (\textbf{b}) NP-NP structure factor as a function of the rescaled wavevector $q \sigma_N /2 \pi$, for $\rho_{m0}=0.290$ and different values of $\sigma_N$. The curves for RNPs are shifted up by $0.2$ for clarity.}
\label{fig:sq_rho02}
\end{figure}

Although the volume fraction of NPs added to the network is less than $2\%$ in all the systems we considered, the presence of the NPs can still modify the structure of the network. In order to quantify this effect, we study two quantities: The monomer-monomer structure factor $S_{mm}(q)$ \cite{hansen1990theory} and the pore size distribution $P(r)$ \cite{gelb1999pore,bhattacharya2006fast,sorichetti2020determining}.

The structure factor is defined as \cite{hansen1990theory}

\begin{equation} 
S(\mathbf q) = \frac 1 N\sum_{k,j=1}^{N} \langle \exp [-i \mathbf q \cdot (\mathbf r_k - \mathbf r_j)] \rangle,
\label{eq:sq}
\end{equation}

\noindent where $\mathbf q$ is the wavevector. Since our configurations are isotropic,  we will consider the spherically averaged structure factor $S(q)$, with $q=|\mathbf q|$.

In Fig.~\ref{fig:sq_rho02}a we show $S_{mm}(q)$ for $\rho_{m0}=0.290$, for both ANPs and RNPs. The addition of ANPs leaves $S_{mm}(q)$ basically unchanged with respect to the neat system (thick black line\footnote{The structure factor for the neat system has been computed for a system of $5 \times 10^4$ instead of $4 \times 10^5$ particles.}), except for $\sigma_N=1$, for which we observe a shoulder at around $q \simeq 3$, \textit{i.e.}, $r \simeq 2 \pi / q \simeq 2$, which results from configurations with two monomers separated by a NP. We note that for small $q$, the structure factor shows a power-law, $S_{mm}(q) \propto q^{-\alpha}$, indicating that the network has a fractal nature  \cite{roldan2017connectivity}. For the case of RNPs, the presence of the NPs influences the structure factor much more strongly in that $S_{mm}(q)$ increases noticeably for $q \lesssim 1$, \textit{i.e.}, $r \gtrsim 6$. This is a consequence of the fact that the addition of RNPs causes a swelling of the network on intermediate length scales, as we will also discuss below when analyzing the pore size distribution (see also Supplementary Material, Sec.~S2.2). This interpretation is also in agreement with the fact that the modification of $S_{mm}(q)$ is independent of NP size. Note that we do not consider RNPs with $\sigma_N=1$, since they induce an excessive swelling of the network, as discussed in the Supplementary Material, Sec.~S2.2. For additional details on $S_{mm}(q)$, see also Sec.~S2.4 in the Supplementary Material.

In Fig.~\ref{fig:sq_rho02}b we show the NP-NP structure factor $S_{NN}(q)$ for $\rho_{m0}=0.290$ as a function of the rescaled wavevector $q\sigma_N/2 \pi$. (The corresponding data for the other densities are shown in the Supplementary Material~Sec.~\ref{sec:sq_si}.) Studying this quantity is useful to determine whether or not the NPs are distributed homogeneously in the system. For the ANPs, with the exception of $\sigma_N=1$, the data fall on a master curve. The curve  is basically flat for $q\sigma_N/2 \pi > 1/2$, $S_{NN}(q) \simeq 1$, like that of a gas, signaling that the NPs are homogeneously dispersed and aggregation is essentially absent. The only exception is $\sigma_N=1$, for which a peak of modest height appears at $q  \to 0$, signaling a weak NP aggregation in which several NPs (occasionally) fill the holes in the mesh. For the RNPs, on the other hand, the presence of weak aggregation is clear for all the systems, in that all the curves fall on a master curve with a main peak at $q \to 0$ and a smaller one at $q\sigma_N/2 \pi =1$, which results from direct contacts between NPs. The main peak at small $q$ is likely due to the fact that these NPs fill the largest cavities of the network. However, for $\sigma_N$=5 we find that the main peak is no longer at the smallest accessible wave-vector, likely because there are not sufficiently large cavities to accommodate a substantial number of large NPs. These trends are consistent with what we have concluded from the radial distribution function $g_{NN}(r)$ (Fig.~\ref{fig:rdf_mono_np_rho02}b). For additional details on $S_{NN}(q)$, see also Sec.~S2.5 in the Supplementary Material.

%%%%%%%%%%%%%%%%%%%%%%%%%%%%%%%%%%%%%%%%%%%%%%%%%                  
\subsection{Pore size distribution}\label{sec:psd}
%%%%%%%%%%%%%%%%%%%%%%%%%%%%%%%%%%%%%%%%%%%%%%%%%  

\begin{figure}
\centering
\includegraphics[width=0.48 \textwidth]{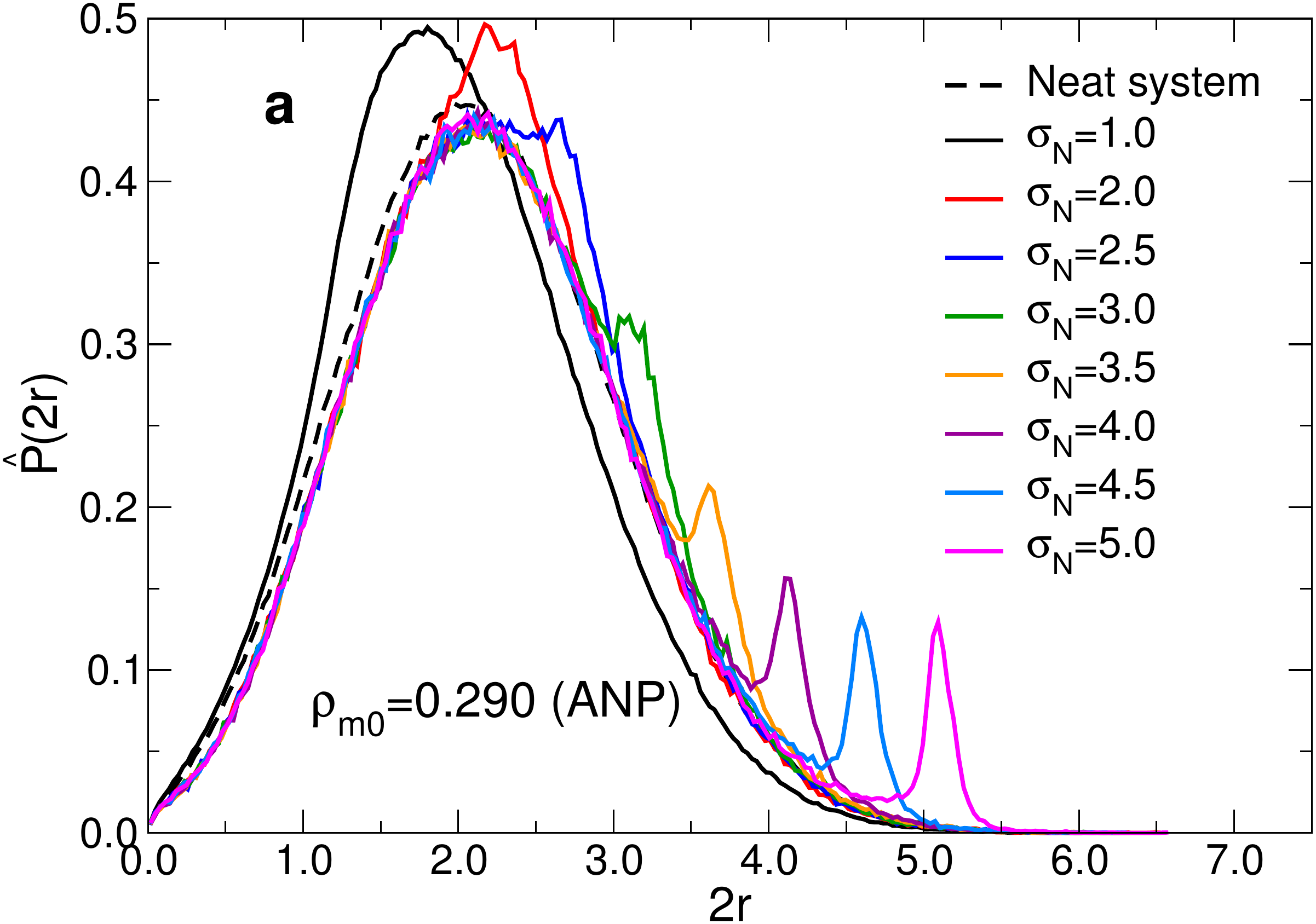}
\includegraphics[width=0.48 \textwidth]{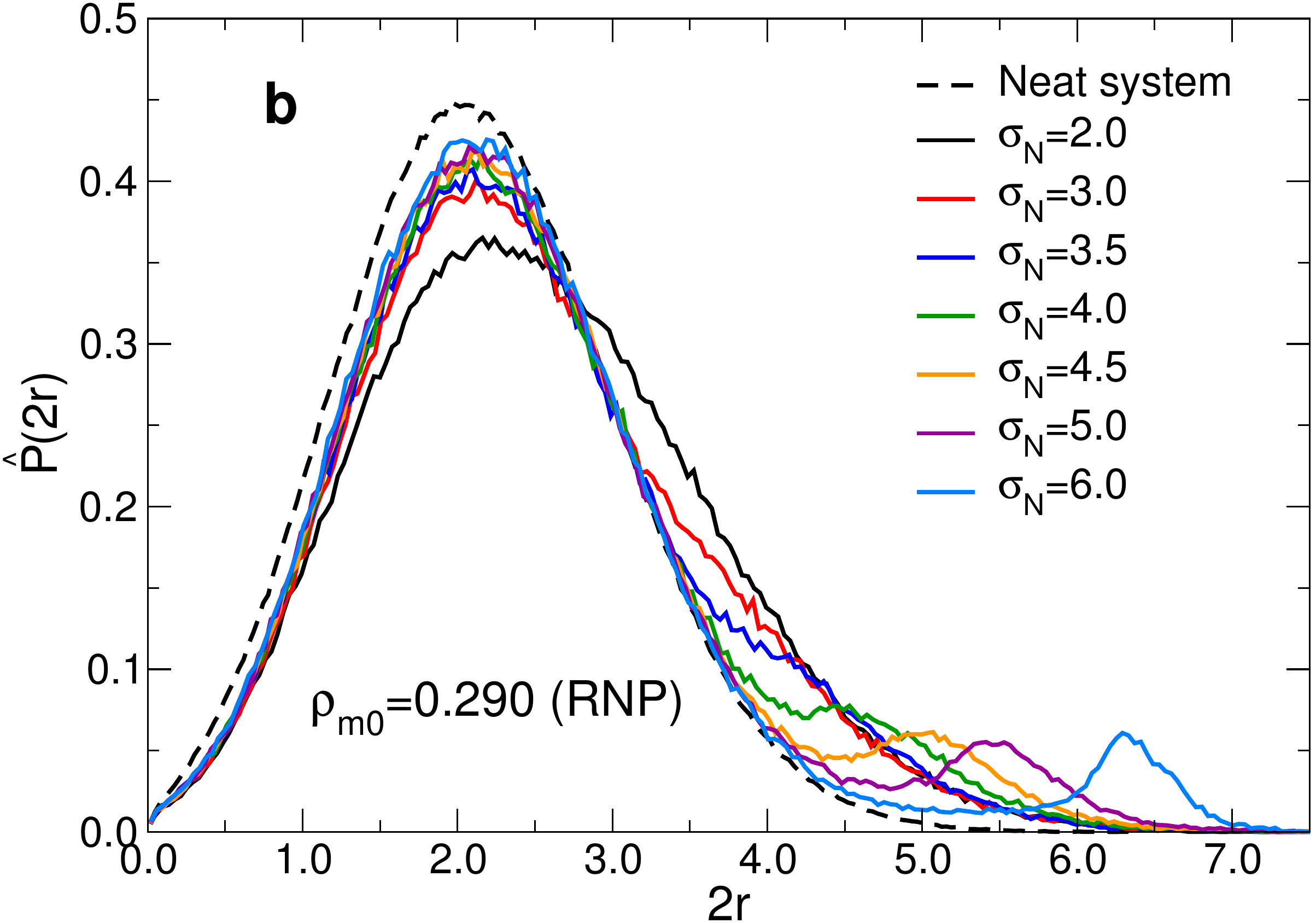}
\caption{Distribution of pore diameters, $\hat P(2r)$, for $\rho_{m0}=0.290$ and for different NP diameters.
Dashed black line: $\hat P_0(2r)$ (neat system distribution). (\textbf{a}): ANP (\textbf{b}): RNP. }
\label{fig:psd_rho02}
\end{figure}

A further quantity that is useful for the characterization of the network structure is the pore size distribution (PSD) $P(r)$~\cite{sorichetti2020determining}. The PSD is calculated by randomly sampling points in the void space and finding the radius $r$ of the largest sphere that can be inserted without touching a polymer strand~\cite{gelb1999pore,bhattacharya2006fast,sorichetti2020determining}. In Fig.~\ref{fig:psd_rho02} we show $P(r)$ for $\rho_{m0}=0.290$, while the data for the other densities are presented in the Supplementary Material, Sec.~\ref{sec:psd_si}. (To facilitate the comparison with the NP diameter $\sigma_N$, we actually show the distribution of the pore diameters, $\hat P(2r)$, a distribution that is trivially related to $P(r)$.) For the sake of comparison we have also included in the graph the PSD of the neat system, $P_0(r)$ (dashed black line). For the ANPs (Fig.~\ref{fig:psd_rho02}a) and $\sigma_N>1$, $P(r)$ is basically identical to $P_0(r)$, with the exception of a sharp peak appearing at $2r = \sigma_N$, which results from the NPs creating ``cavities'' in the network. For $\sigma_N=1$, $P(r)$ is slightly shifted to smaller $r$ with respect to $P_0(r)$, with the main peak shifting to smaller $r$: The network contracts as a consequence of the attractive monomer-NP interaction, as already observed in non-crosslinked nanocomposites \cite{sorichetti2018structure}. For the RNPs (Fig.~\ref{fig:psd_rho02}b), $P(r)$ shifts to the right at small $r$, rationalizing the swelling of the network, and a broad peak appears centered at $2r \simeq \sigma_N+\delta$ \textit{i.e.}, shifted with respect to the peak observed for ANPs. For $\rho_{m0}=0.290$, one finds $\delta \simeq 0.5$, but this value decreases with increasing network density (see Sec.~S2.6 in the Supplementary Material). This shifting of the peak is related to the fact that the RNPs have a larger effective diameter than the ANPs, in agreement what discussed for $g_{mN}(r)$ (Fig.~\ref{fig:rdf_mono_np_rho02}a). The peak is significantly broader than the one seen for the ANPs for the same value of $\sigma_N$: This is likely due to the fact that for ANPs, the network strands are attracted to the surface of the NPs, giving thus rise to a well defined distance, whereas the RNPs repel the strands and locally deform the network. Since the amplitude of this deformation will depend on the local properties of the mesh, the resulting distribution of the pore size is broad. 

The position of the main peak of $P(r)$, $r_\text{max}$ (most probable pore radius), can be taken as a measure of the mean mesh size $\xi$ of the system, \textit{i.e.}, $\xi \equiv r_\text{max}$. In a previous study \cite{sorichetti2020determining}, it was found that for purely polymeric systems $r_\text{max}$ as well as the mean $\langle r \rangle$ can be taken as reliable estimates of $\xi$. However, for the networks studied in the present work, $r_\text{max}$ provides a more meaningful value for the mean mesh size, since $P(r)$ can have a double peak structure because of the presence of the NP. We note that $\xi$ has a weak dependence on the NP diameter $\sigma_N$ and on the character of the monomer-NP interaction. However, this dependence can be neglected for all but the smallest value of $\sigma_N$, and hence $\xi \simeq \xi_0$, where $\xi_0$ is the mean mesh size of the neat system.

%%%%%%%%%%%%%%%%%%%%%%%%%%%%%%%%%%%%%%%%%%%%%%%%%   
%%%%%%%%%%%%%%%%%%%%%%%%%%%%%%%%%%%%%%%%%%%%%%%%%                  
\section{Dynamic properties}\label{sec:dynamics}
%%%%%%%%%%%%%%%%%%%%%%%%%%%%%%%%%%%%%%%%%%%%%%%%%   
%%%%%%%%%%%%%%%%%%%%%%%%%%%%%%%%%%%%%%%%%%%%%%%%%   

%%%%%%%%%%%%%%%%%%%%%%%%%%%%%%%%%%%%%%%%%%%%%%%%%                  
\subsection{Theoretical background}\label{sec:theory}
%%%%%%%%%%%%%%%%%%%%%%%%%%%%%%%%%%%%%%%%%%%%%%%%%   

The dynamics of NPs in permanently crosslinked networks has been studied among others by Dell and Schweizer \cite{dell2014theory} using the nonlinear Langevin equation (NLE) theory \cite{schweizer2003entropic}, and independently by Cai, Panyukov and Rubinstein \cite{cai2015hopping} using scaling theory. Dell and Schweizer considered a polymer network characterized by an effective tube diameter $d$ resulting from both the crosslinks and the entanglements, with $d \approx b (N_e^{\text{eff}})^{1/2}$ \footnote{Here and in the following, we will use $\approx$ to signify equivalence apart from a dimensionless constant of order $1$, and $\simeq$ to signify a numerical approximation.} where $b$ is the Kuhn length and $N_e^{\text{eff}}$ is the effective entanglement length \footnote{The effective entanglement length, $N_e^{\text{eff}}$, can be obtained experimentally from the plateau modulus $G_0$, since $G_0 \approx \rho k_B T/N_e^{\text{eff}}$}. Making some simplifying assumptions regarding the structure of the polymeric matrix \cite{dell2014theory}, and introducing the confinement parameter $C\equiv \sigma_N/d$, one finds that the onset of localization happens at $C_c$ slightly larger than $1$, as intuition suggests (the exact value of $C_c$ depends on the density and compressibility of the matrix). For $C$ slightly above $C_c$, the mean hopping time, \textit{i.e.}, the mean time between two hopping events, increases approximately exponentially with $C$, whereas for stronger confinements this increase becomes stronger than exponential, making that for large $C$ the NPs can be considered as indefinitely trapped on experimental time scales. The authors concluded therefore that hopping can be observed experimentally only in the weak confinement regime.

Cai \textit{et al.} \cite{cai2015hopping} considered a generic network with both crosslinks and entanglements, containing non-sticky NPs. Depending on the ratio between the tube diameter resulting from topological entanglements, $d_e \approx b N_e^{1/2}$ ($N_e=$ entanglement length), and the one resulting from crosslinks, $d_x \approx b N_x^{1/2}$ ($d_x$ being equivalent to the geometrical mesh size $\xi$, $N_x=$ mean strand length), they identified two dynamical regimes: the entangled regime ($N_e<N_x$, or equivalently $d_e<d_x$), where entanglements dominate, and the unentangled regime ($N_e>N_x$), where crosslinks dominate. Since all our systems are unentangled, see Sec. \ref{sec:msd}, we will only consider the latter regime for which one has $d \simeq d_x$. In such unentangled networks, NPs of diameter $\sigma_N<d$ can freely diffuse through the mesh, whereas NPs of diameter $\sigma_N>d$ can only diffuse via hopping processes. The resulting diffusion coefficient is predicted to have the following $C$ dependence:

\begin{equation}
D_N \propto \frac{b^2}{\tau_x} \frac{\exp\left(-C^2\right)}{C},
\label{eq:dhop_x}
\end{equation}

\noindent
where $\tau_x \propto N_x^2$ is the Rouse relaxation time of a network strand and $C \equiv \sigma_N / d$. 

 %%%%%%%%%%%%%%%%%%%%%%%%%%%%%%%%%%%%%%%%%%%%%%%%%                  
\subsection{Mean-squared displacement and diffusion coefficient}\label{sec:msd}
%%%%%%%%%%%%%%%%%%%%%%%%%%%%%%%%%%%%%%%%%%%%%%%%% 

\begin{figure}
\centering
\includegraphics[width=0.472 \textwidth]{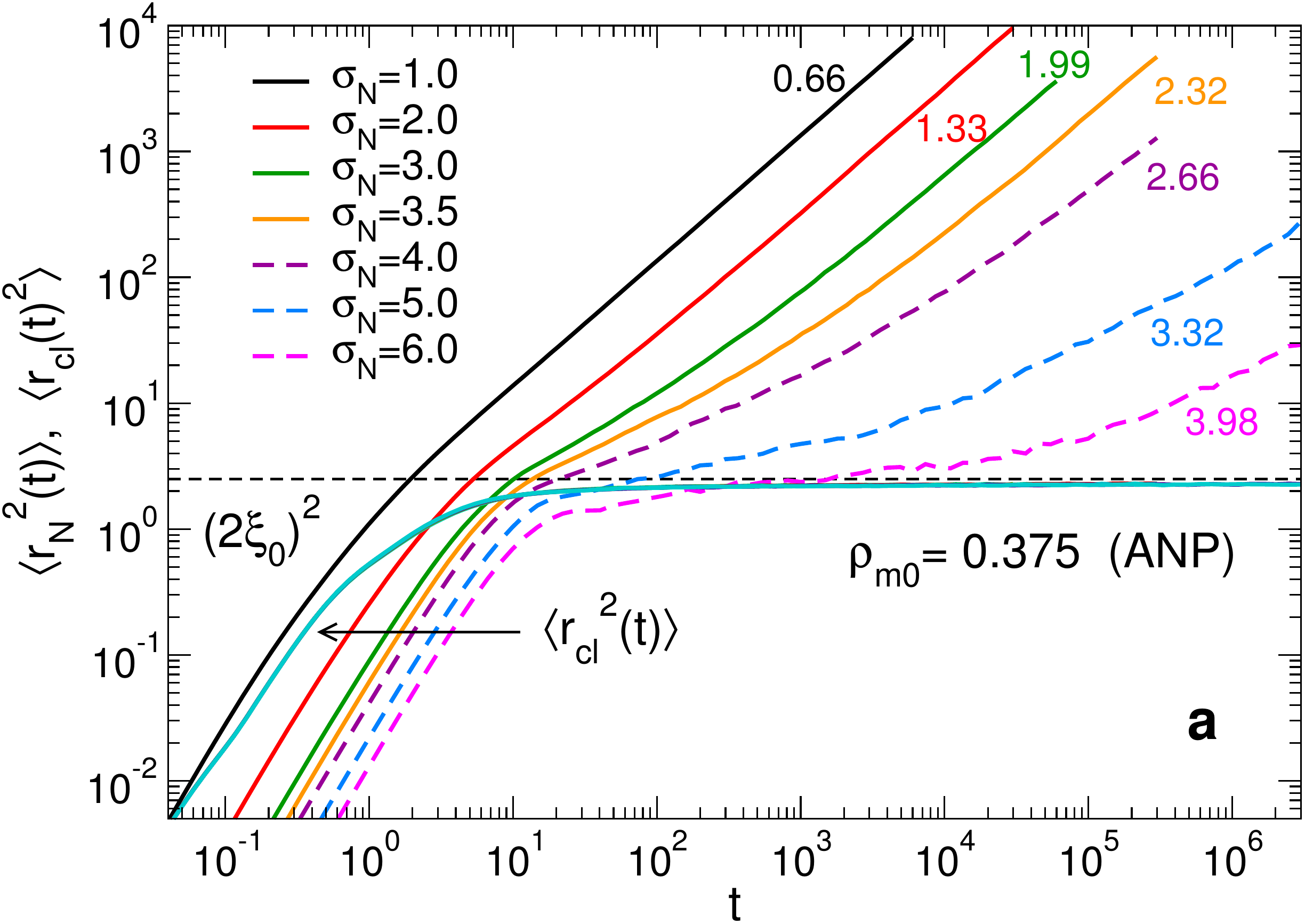}
\includegraphics[width=0.48 \textwidth]{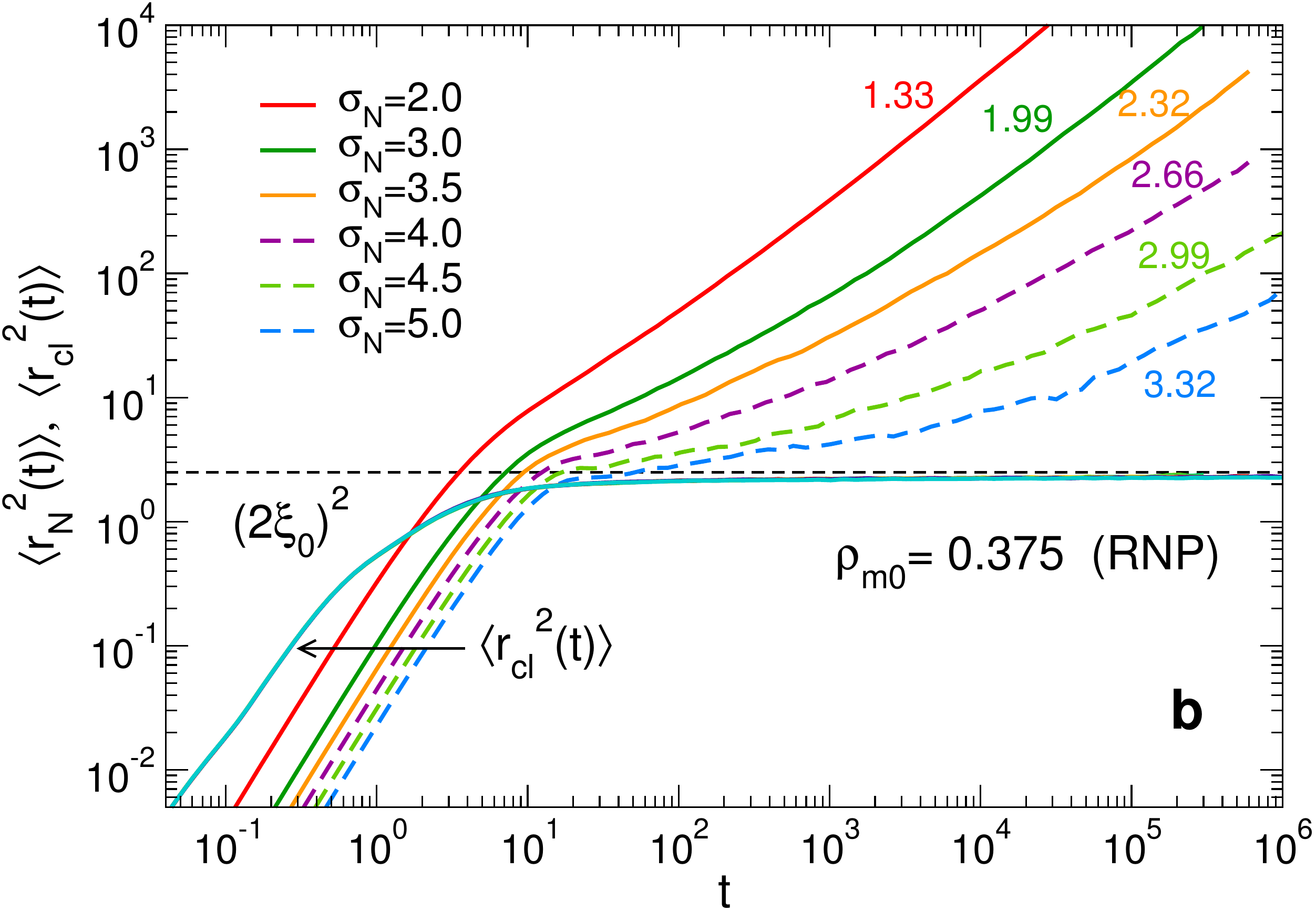}
\caption{MSD, $\langle r_N^2(t) \rangle$, of the ANPs (\textbf{a}) and the RNPs (\textbf{b})  and of the crosslinks, $\langle r_\text{cl}^2(t) \rangle$, for $\rho_{m0}=0.375$ and for different NP diameters. Dashed curves represent systems which have not reached the diffusive regime (see discussion in the text). Dashed horizontal line: $(2 \xi_0)^2$ (see Tab.~\ref{tab:lambda_xi}) Labels give the values of the confinement parameter $C$, Eq.~\eqref{eq:confinement}.}
\label{fig:msd_np_cl_rho03}
\end{figure}

In order to test the validity of the theoretical predictions summarized in Sec.~\ref{sec:theory}, we have determined the NP diffusion coefficient from the mean-squared displacement (MSD). In Fig.~\ref{fig:msd_np_cl_rho03}, we show the time dependence of the MSD, 

\begin{equation}
\langle r_N^2(t) \rangle \equiv {\langle |\mathbf r_{N}(t) - \mathbf r_{N} (0)|^2 \rangle},
\end{equation}

\noindent
for ANPs (a) and RNPs (b), for the densest network ($\rho_{m0}=0.375$), and compare it with the MSD of the crosslinks, $\langle r_\text{cl}^2(t) \rangle$. (The corresponding curves for the other densities are in the Supplementary Material, Sec.~\ref{sec:np_msd_si}.) Dashed lines denote systems for which the NPs have not reached the diffusive regime within the duration of the simulation, \textit{i.e.}, $\langle r_N^2(t) \rangle = 6 D_N t$, where $D_N$ is the diffusion coefficient of the NPs. We stress that the MSD of the crosslinks is shown for all values of $\sigma_N$ but since the curves superimpose very well, only one curve is visible. We can therefore conclude that this quantity is basically independent of $\sigma_N$, confirming that the NP volume fraction we consider is sufficiently low to represent the dilute-NP limit \cite{chen2020nanoparticle}. Due to the fixed network topology, the MSD of the crosslinks quickly reaches a plateau, $\langle r_\text{cl}^2(t) \rangle = \lambda^2 = \text{const.}$, the height of which defines the (squared) \textit{localization length} \cite{zaccarelli2005model}, \textit{i.e.}~the mean amplitude of the fluctuations of the crosslinks around their equilibrium position: 

\begin{equation}
\lambda \equiv \left[ \lim_{t \to \infty} \langle r_\text{cl}^2(t) \rangle \right]^{1/2}.
\label{eq:lambda}
\end{equation}

\noindent
For $\rho_{m0}=0.375$, we find $\lambda=1.51$, a value that is close to $2 \xi_0$ (see horizontal dashed line in the graph), the most probable pore diameter $r_\text{max}$ in the neat state (see Fig.~\ref{fig:psd_rho02}). Since for the other values of $\rho_{m0}$ we find the same relation, see Tab.~\ref{tab:lambda_xi}, we can conclude that the amplitude of the oscillations of the crosslinks around their equilibrium positions are basically equal to the mean mesh size of the network, a result that is certainly very reasonable. We also note that $\lambda$ is proportional to the effective tube diameter $d$ of the system, since $d \approx \lambda (N_e/N_x)^{1/4}$ \cite{hsu2016static,chen2020nanoparticle}. Assuming $N_e = 85$, \cite{hoy2009topological} and since $N_x=6$, we obtain $d \approx 1.94 \times \lambda$. We also remark in passing that $N_e \gg N_x$, \textit{i.e.}, all our systems are in the unentangled regime.

\begin{table}[t]
\centering
\caption{Properties of the neat polymer networks. $\rho_{m0}$: monomer density. $\lambda$: crosslink localization length, Eq.~\eqref{eq:lambda}. $\xi_0$: mean mesh size in the neat state ($2\xi_0$ = most probable pore diameter). $d$: Estimate for the effective tube diameter, $d \approx \lambda (N_e/N_x)^{1/4} = 1.94 \times \lambda$ ($N_e=85,N_x=6$). }
$\begin{array}{@{\hspace{1.5 em}} c @{\hspace{1.5 em}} c @{\hspace{1.5 em}} c @{\hspace{1.5 em}}  c @{\hspace{1.5 em}}}
\toprule
\rho_{m0} & \lambda & 2 \xi_0 & d\\
\midrule
0.190 & 3.12 & 3.06 & 6.05\\
0.290 & 2.01 & 2.06 & 3.90\\
0.375 & 1.51 & 1.58 & 2.93\\
\bottomrule
\end{array}$
\label{tab:lambda_xi}
\end{table}

Since previous studies have shown that it is useful to discuss the dynamics as a function of the so-called \textit{confinement parameter} \cite{dell2014theory,cai2015hopping,parrish2017network} we define 

\begin{equation}
C \equiv \frac{\sigma_N}{\lambda}.
\label{eq:confinement}
\end{equation}

\noindent
Note that this definition is slightly different from the one of previous studies in that we use here $\lambda$ as the length scale for the normalization instead of the tube diameter $d$. This choice is motivated by the fact that $\lambda$ can be measured with high precision from the simulations while $d$ is a length scale that rests on a theoretical concept that allows to define $d$ only up to a numerical prefactors. As discussed above, however, for all simulated systems we have $\lambda \approx d/1.94$, a relation which allows to express our results also in terms of $d$. We also note that $\lambda \simeq 2 \xi_0$, where $2 \xi_0$ is the mesh size of the neat system (Tab.~\ref{tab:lambda_xi}).

The values of $C$ are reported in the labels of the curves in Figs.~\ref{fig:msd_np_cl_rho03}a and b. One can see that for $C \lesssim 1$, the MSD becomes diffusive, \textit{i.e.}, $\langle r_N^2(t) \rangle \propto t$, immediately after the ballistic regime, irrespective of the type of NP. More interesting is the dynamics of the NPs with $C \gtrsim 1$, in that their MSD shows between the ballistic regime and the diffusive behavior at long times, a subdiffusive regime, $\langle r_N^2(t) \rangle \propto t^{\beta}$, with an exponent $\beta<1$ that depends on $C$. This subdiffusive regime begins when $\langle r_N^2(t) \rangle \approx d^2 \sigma/ \sigma_N \approx 3.76 \times \lambda^2 \sigma/\sigma_N$ \cite{dell2014theory,cai2015hopping} (see Supplementary Material, Sec.~S3.1 and discussion below). Note that for the largest value of $C$, the systems do not reach the diffusive regime neither during equilibration nor during production, since the simulation time required would be prohibitively long (dashed lines). Although strictly speaking these systems have not fully equilibrated, we have not observed any noticeable time dependence in any of the structural and thermodynamic quantities (\textit{i.e.}, no aging is observed), and have therefore decided to include them in the analysis as well. Moreover, as discussed in Sec.~\ref{sec:theory}, for strong confinements we expect the typical hopping time of the NPs to diverge at least exponentially in $C$. This implies that, even in experiments, a slight increase of $C$ will prevent the system from reaching equilibrium. Therefore, we expect that our data for the extreme confinement regime is still useful for the comparison with experimental data.

The MSD from the simulations can be compared with the theoretical predictions by Cai \textit{et al.} \cite{cai2015hopping}, who for the case of large particles in an unentangled network predict four regimes in the dynamics: After the initial ballistic regime, one finds a subdiffusive regime $\langle r_N^2(t) \rangle \propto t^{\beta}$ with $\beta=1/2$, which lasts up to the Rouse relaxation time of a network strand, $\tau_x \approx \tau_0 N_x^2$, where $\tau_0\approx m\gamma_m \sigma^2 / k_BT$ is the monomer relaxation time \footnote{We recall that $m$ is the monomer mass, $\sigma$ the diameter, and $\gamma_m$ the friction coefficient.}. For $t>\tau_x$, the NP is trapped by the mesh, and the MSD shows a plateau $\langle r_N^2(t) \rangle \approx d^2 \sigma / \sigma_N$. For times larger than the hopping time, \textit{i.e.}~$t>\tau_x N_x^{1/2} \exp(-\sigma_N^2/d^2)$, the NP finally escapes from the local trap formed by the mesh and starts to diffuse via hopping motion. Comparing these predictions with our simulation data shows that we do not observe the intermediate subdiffusive $t^{1/2}$ regime, which is likely
 due to the fact that the average length of the network strands, $N_x=6$, is too small to be in the scaling regime. For additional details on the MSD of the NPs, we refer to Sec.~S3.2 of the Supplementary Material.

\begin{figure}
\centering
\includegraphics[width=0.48 \textwidth]{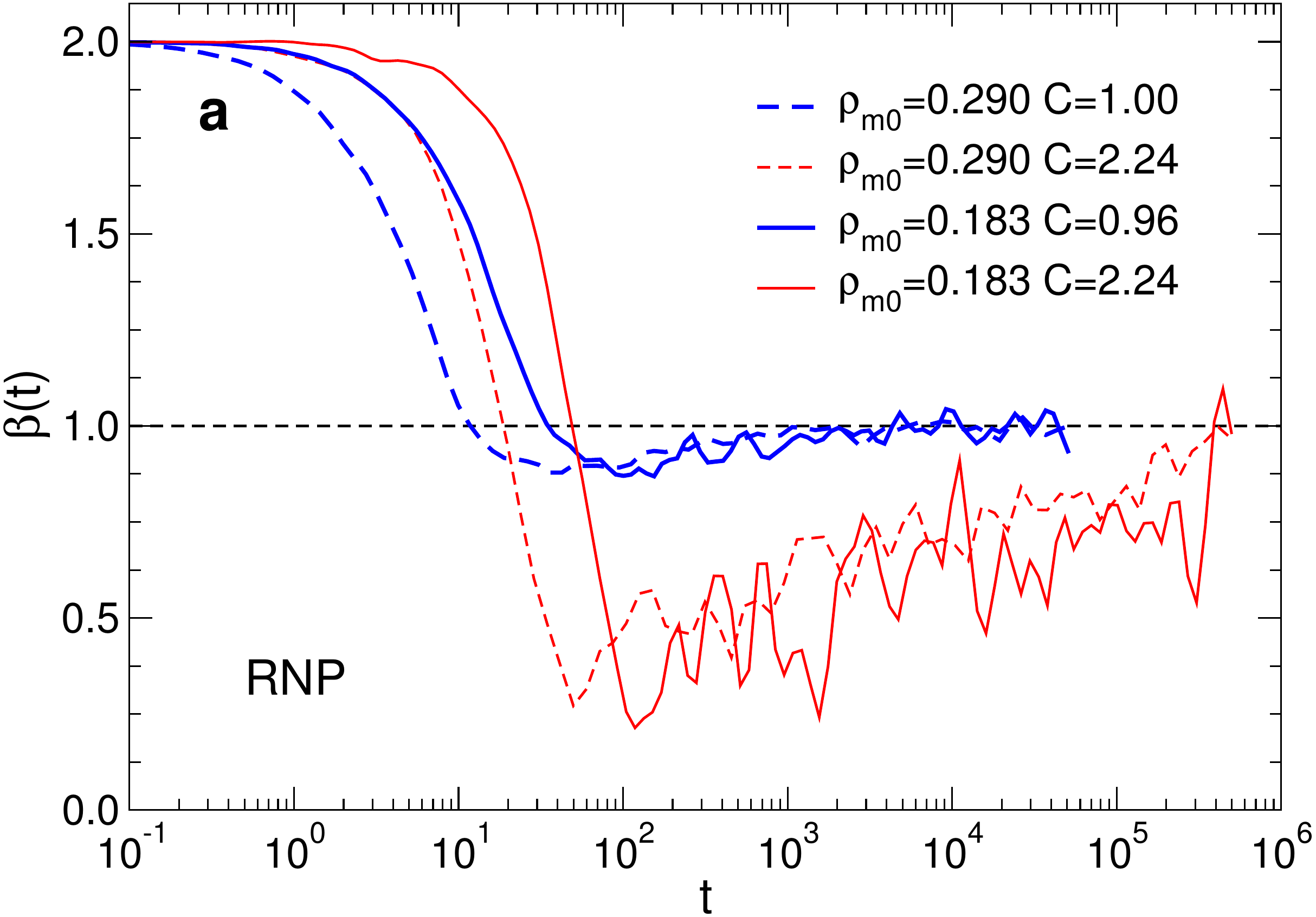}
\includegraphics[width=0.48 \textwidth]{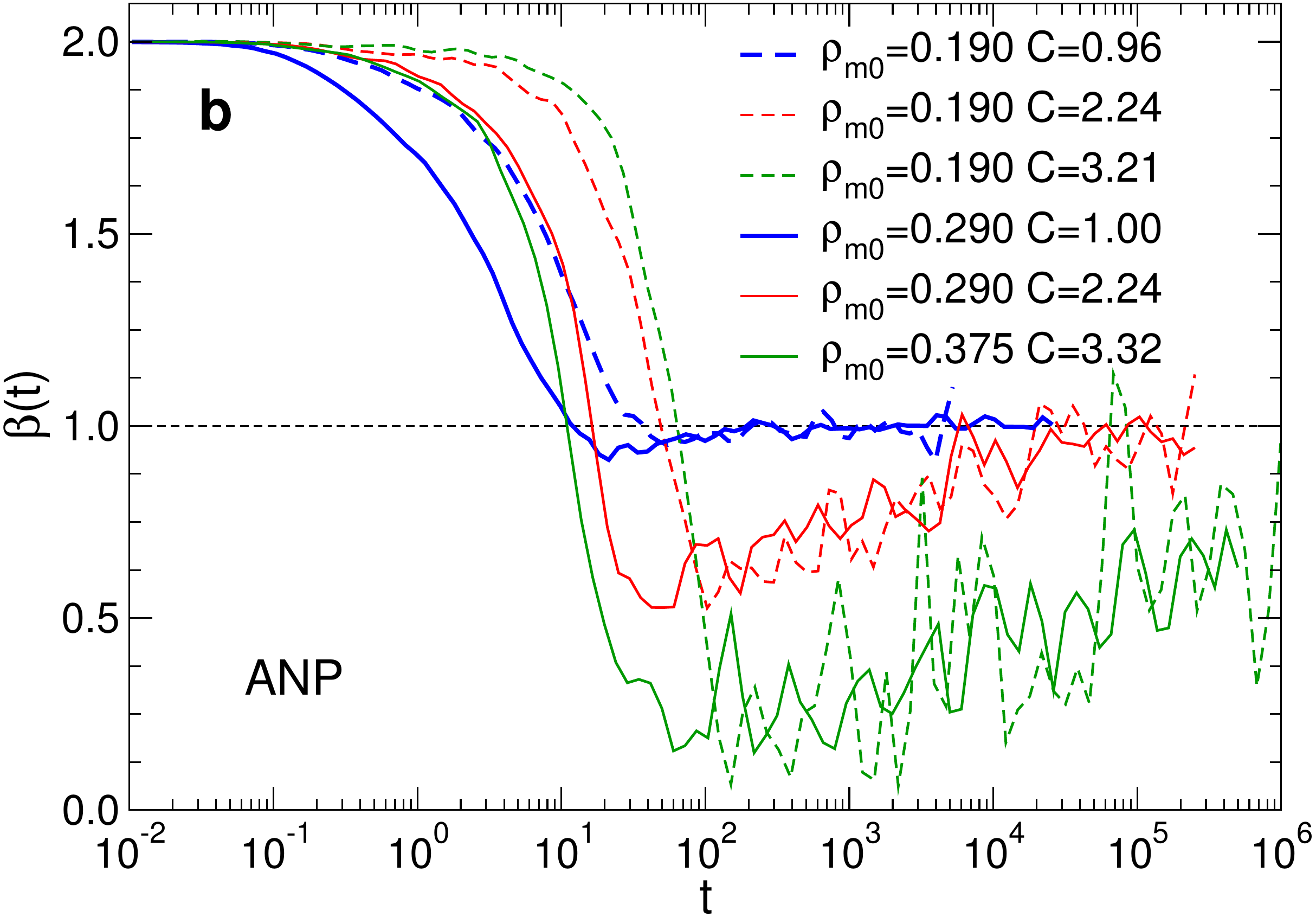}
\caption{Apparent subdiffusive exponent, Eq.~\eqref{eq:beta}, for systems of RNPs (\textbf{a}) and ANPs (\textbf{b}) for different $\rho_{m0}$ and similar values of $C$.}
\label{fig:beta}
\end{figure}

In order to study the dependence of the subdiffusive dynamics on $\sigma_N$, we consider the apparent subdiffusive exponent $\beta(t)$, defined as

\begin{equation}
\beta(t) \equiv \frac{\text{d}\ln [\langle r^2_N(t) \rangle]}{{\text{d} \ln(t)}}.
\label{eq:beta}
\end{equation}

\noindent
In Fig.~\ref{fig:beta}, we show $\beta(t)$ for systems with different $\rho_{m0}$ and similar values of $C$, for the RNPs (a) and the ANPs (b). The short-time ballistic regime ($\beta(t)=2$) is followed by a sharp decrease of $\beta(t)$ when the MSD of the NPs leaves the ballistic regime. For $C\leq 1$ the exponent becomes unity, thus indicating the diffusive motion. If the confinement parameter is larger than 1.0 the exponent $\beta$ drops to a value below unity and subsequently increases slowly with time. (Note that $\beta(t)$ is independent of the density $\rho_{m0}$ but decreases if $C$ is increased, which is further evidence that $C$ is the relevant parameter for the dynamics.) In the subdiffusive regime the time dependence of $\beta$ is compatible with a logarithmic dependence, a result that indicates that in this time window the dynamics of the NP is very heterogeneous. This result can be rationalized by the fact that the time at which a given particle starts to leave the cage formed by the mesh depends on the details of this cage (number of strands, their length,...) giving rise to a very broad distribution of local relaxation times. Note that if this type of disorder is absent it can be expected that the escape times from this cage is not broadly distributed, making that for a long time the NP will stay confined and the MSD shows a plateau. Hence we conclude that the plateau that is predicted in \citet{cai2015hopping} is washed out due to the intrinsic disorder of the gel. For more details on the apparent subdiffusive exponent, we refer to the Supplementary Material, Sec.~S3.1.

\begin{figure}
\centering
\includegraphics[width=0.48 \textwidth]{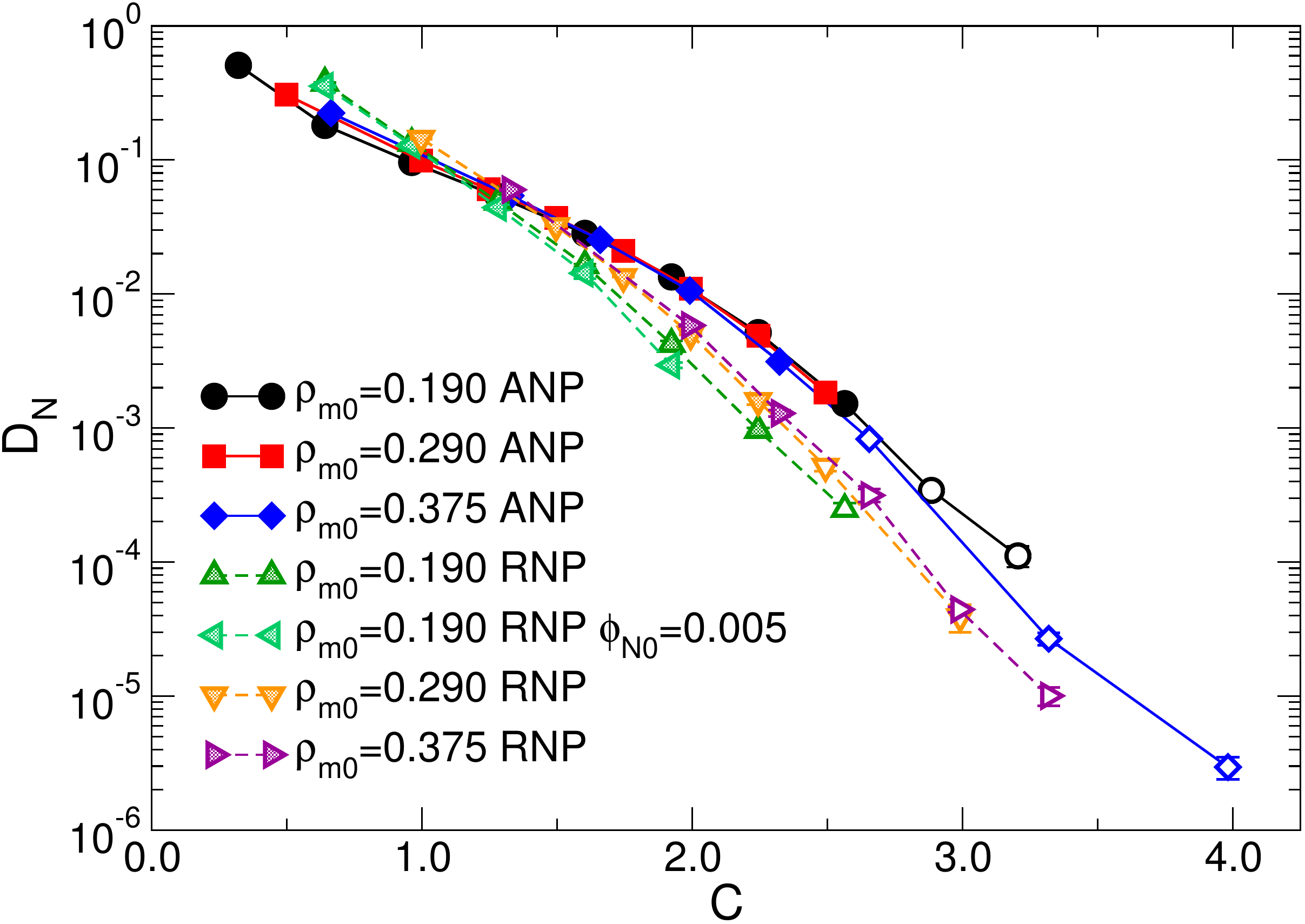}
\caption{NP diffusion coefficient as a function of the confinement parameter $C$, Eq.~\eqref{eq:confinement}. Unless specified, $\phi_{N0} = 0.02$. Solid lines: ANPs. Dashed lines: RNPs. Open symbols: Upper bounds for $D_N$.
}
\label{fig:dnp}
\end{figure}

As discussed above, the NPs are eventually able to escape from the mesh and start to diffuse. The dynamics in the diffusive regime can be characterized by the diffusion coefficient of the NPs, which can be obtained from the MSD via the Einstein relation \cite{hansen1990theory}:

\begin{equation}
D_N = \lim_{t \to \infty} \frac 1 {6t} \langle r_N^2(t)\rangle.
\end{equation}

\noindent
In Fig.~\ref{fig:dnp}, we show the NP diffusion coefficient $D_N$ as a function of the confinement parameter $C$ for all the systems studied here. For the systems which have not reached the diffusive regime, we report an upper bound for $D_N$, represented by open symbols. One sees that for the ANPs as well as the RNPs the data falls nicely on a master curve when plotted as a function of $C$, which indicates that the confinement parameter is the relevant quantity determining also the long-time diffusive dynamics. We also note that a relatively small increase of $C$ causes a dramatic decrease of $D_N$. For example, for the ANPs in the $\rho_{m0}=0.375$ system, $D_N$ decreases from $2.23\times 10^{-1}$ to less than $2.95 \times 10^{-6}$ (upper bound) when going from $C=0.664$ to $C=3.98$, \textit{i.e.}, a reduction of five orders of magnitude.

It is interesting to observe that for $C<1$ (weak confinement), the diffusivity of ANPs is smaller than that of RNPs, while for $C>1$ the reverse is true. This can be rationalized as follows: In the weak confinement regime the NPs can freely diffuse through the mesh, and do not have to wait for the relaxation of the local mesh. Thus in this regime, an attractive monomer-NP interaction will causes an increase of the effective friction felt by the NP, and hence the $D_N$ is smaller for ANPs. For $C>1$ (strong and extreme confinement), these short-range effects lose their importance since they are dominated by the confinement effects induced by the mesh on the NP. In this regime it is thus the effective diameter of the NP that is the relevant quantity and, as we will discuss below, this diameter is smaller for the ANPs than the one for the RNPs, thus explaining why at large $C$ the diffusion of the RNPs is slower than the one of the ANPs.

In order to verify that the NP volume fraction ($\phi_N \simeq \phi_{N0} = 0.02$) is sufficiently low to be in the single-NP (dilute) limit, we also include in Fig.~\ref{fig:dnp} data for $\rho_{m0}=0.190$ and NP volume fraction $\phi_N \simeq \phi_{N0} = 0.005 $. One can see that the difference between the higher and lower volume fraction is negligible, confirming the validity of our choice.

\begin{figure}
\centering
\includegraphics[width=0.48 \textwidth]{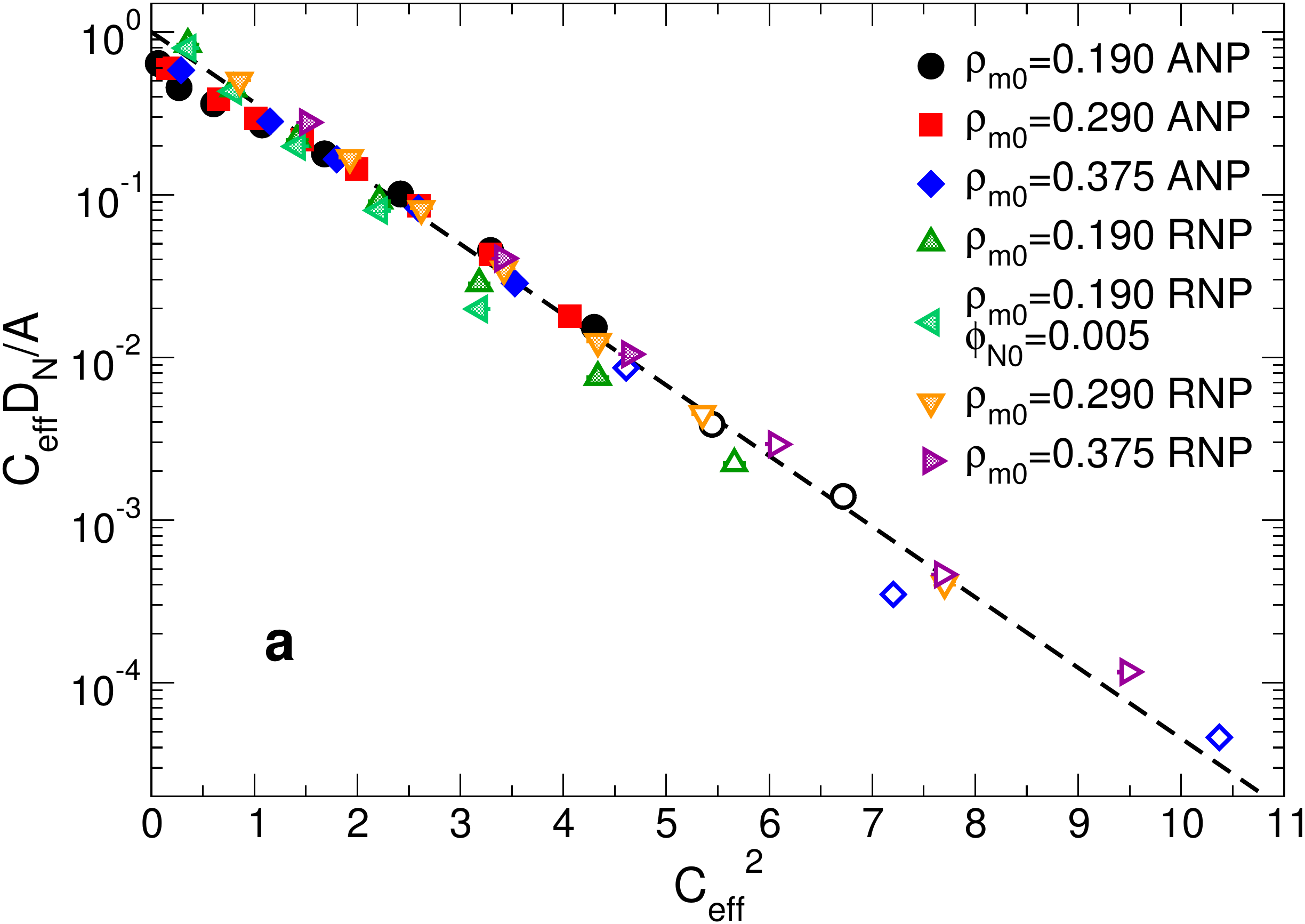}
\includegraphics[width=0.48 \textwidth]{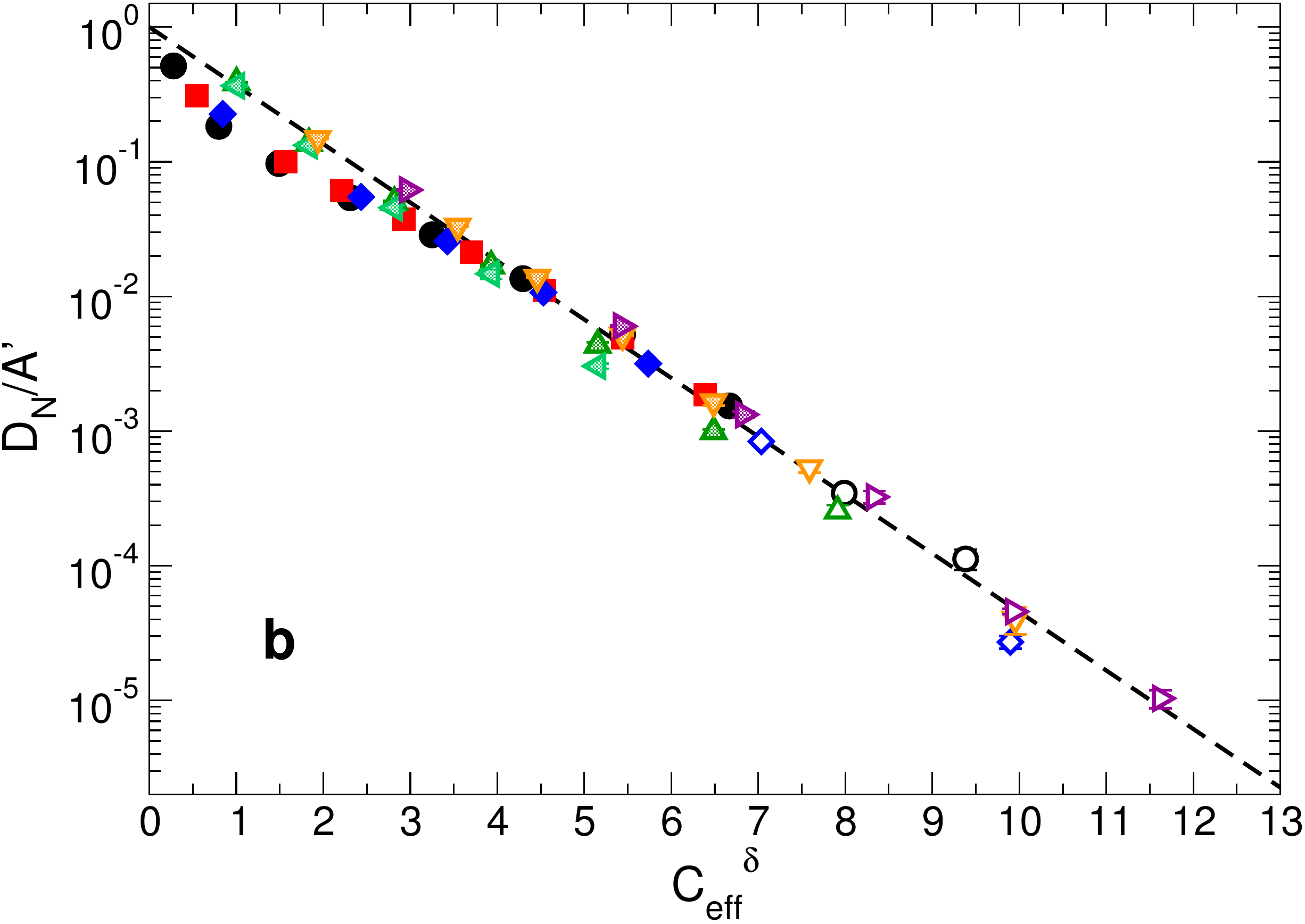}
\caption{Same data as in Fig.~\ref{fig:dnp}, rescaled according to Eq.~\eqref{eq:dn_cpr} (\textbf{a}) and Eq.~\eqref{eq:dn_ds} (\textbf{b}). $C_\text{eff}\equiv BC$ (\textbf{a}) and $B'C$ (\textbf{b}) is the effective confinement parameter. (\textbf{a}): RNPs: $A=0.265, \ B=0.928$. ANPs: $A=0.206, \ B=0.809$. (\textbf{b}): RNPs: $A'=0.970, \ B'=1.56, \ \delta=1.49$. ANPs: $A'=0.989, \ B'=1.35, \ \delta=1.53$. Dashed lines: exponential with slope $-1$.}
\label{fig:dnp_rescaled}
\end{figure}

As discussed above, all our systems are in the unentangled regime, \textit{i.e.}~$N_x<N_e$: According to the theory of Cai \textit{et al.} \cite{cai2015hopping} we thus expect $D_N$ to be given by (see Eq.~\eqref{eq:dhop_x}):

\begin{equation}
D_N = A \frac{\exp[-(BC)^2]}{BC} \equiv A \frac{\exp(-C_\text{eff}^2)}{C_\text{eff}}.
\label{eq:dn_cpr}
\end{equation}

\noindent
where $A,B$ are positive constants and $C_\text{eff}\equiv BC$ is the effective confinement parameter. Fitting the data in Fig.~\ref{fig:dnp} with the expression \eqref{eq:dn_cpr}, we find for the RNPs $A=0.265$ and $B=0.928$, so that $C_\text{eff} \simeq C$, confirming that for the RNP $\sigma_N/\lambda$ represents indeed the relevant confinement parameter (Eqs.~\eqref{eq:dhop_x} and \eqref{eq:confinement}). 
For the ANPs, we restrict the fit to $C>2$, in order to probe the intermediate-strong confinement regime, for which hopping is expected to take place. We find $A=0.206$ and $B=0.809$, so that once again $C_\text{eff} \simeq C$. We note, however, that the ANPs behave as if they experience an effective confinement parameter $C_\text{eff} = B C = B \sigma_N /\lambda$, \textit{i.e.}, their effective size is smaller, as discussed for $g_{Nm}(r)$ (Sec.~\ref{sec:rdf_sq}). Note that we find for both types of particles a value of $B$ which is close to unity, indicating that $C=\sigma_N/\lambda$ is the relevant parameter for a quantitative prediction of $D_N$.

In Fig.~\ref{fig:dnp_rescaled}a, we show the same data as in Fig.~\ref{fig:dnp}, rescaled according to Eq.~\eqref{eq:dn_cpr} and using the fit parameters mentioned above for the RNPs and the ANPs. One can see that the data fall very well on a master curve, indicating that the expression by Cai \textit{et al.} for the diffusion constant does give a good description of the data. The only exceptions to this are the ANPs at small values of $C$ in that one can note some small but systematic deviation from the theoretical curve. This disagreement is likely due to the fact that for the ANPs the mechanism of diffusion at small $C$ is dominated by the small-scale friction felt by the NPs, as discussed above.

Finally, we compare our data with the theoretical prediction by Dell and Schweizer \cite{dell2014theory}. For $C\gtrsim C_c$ , diffusion proceeds \textit{via} hopping (see Sec.~\ref{sec:vanhove} below), and the hopping diffusion coefficient can be estimated as $D_N \approx \Delta_h^2 / \tau_h$, where $\Delta_h^2$ is the mean jump length and $\tau_h$ the mean hopping time\cite{dell2014theory}. Since the dependence of $\Delta_h$ on $C$ is weak for small values of $C-C_c \simeq C-1$,
to a first approximation one has $D_N \propto \tau_h^{-1} \propto \exp(-F_B/k_B T)$, where $F_B$ is the free-energy barrier associated with the hopping process. This free energy barrier is found to increase with $C$ as a power law for small values of $C-C_c$, so that $D_N$ is predicted to be given by

\begin{equation}
D_N = A' \exp[-(B'C)^\delta] = A' \exp[- C_\text{eff}^\delta] ,
\label{eq:dn_ds}
\end{equation}

\noindent
with $A',B',$ and $\delta$ positive constants. Fitting the data in Fig.~\ref{fig:dnp} with Eq.~\eqref{eq:dn_ds}, we find for the RNPs $A'=0.970$, $B'=1.56$, and $\delta=1.49$, whereas for the ANPs (restricting also in this case the fit to $C>2$, see discussion above) we find $A'=0.989$, $B'=1.35$, and $\delta=1.53$.\ Thus the value of the exponent $\delta$ we find is independent of the type of particles considered and close to the one from the theory, \textit{i.e.}, $1.7\pm0.1$, valid for value of $C$ that are not too large \cite{dell2014theory}. Furthermore we note that when using Eq.~\eqref{eq:dn_ds}, one finds (again) that the effective diameter of the ANPs is slightly smaller than that of the RNPs.

In Fig.~\ref{fig:dnp_rescaled}b, we report the data from Fig.~\ref{fig:dnp}, but now rescaled according to Eq.~\eqref{eq:dn_ds}. We observe an excellent agreement with the theoretical prediction, in that the data collapse on the same master curve. Also in this case, there are small deviations from the master curve at small $C$ for the ANPs, for the reasons discussed above. We conclude this analysis by observing that, since both Eq.~\eqref{eq:dn_cpr} and Eq.~\eqref{eq:dn_ds} are in very good agreement with our data, we cannot, at present, conclude that one of the two theoretical descriptions is significantly better than the other. In order to draw any further conclusion, longer simulations are needed, so that more precise estimates of $D_N$ in the $C \gtrsim 2.5$ regime (strong/extreme confinement) can be obtained.

Before concluding this discussion, we also mention that the confinement parameter is not only able to describe the dynamics of the relaxation dynamics of the NP but also, e.g., the amount of shrinking/expansion of the network due to the NP (see Sec.~S2.2 of the Supplementary Material for more details). We thus conclude that $C$ is not only a relevant parameter for the dynamics, but also for certain static properties of the NP-gel system.

 %%%%%%%%%%%%%%%%%%%%%%%%%%%%%%%%%%%%%%%%%%%%%%%%%                  
\subsection{Analysis of the hopping dynamics: van Hove function}\label{sec:vanhove}
%%%%%%%%%%%%%%%%%%%%%%%%%%%%%%%%%%%%%%%%%%%%%%%%% 

\begin{figure}
\centering
\includegraphics[width=0.75 \textwidth]{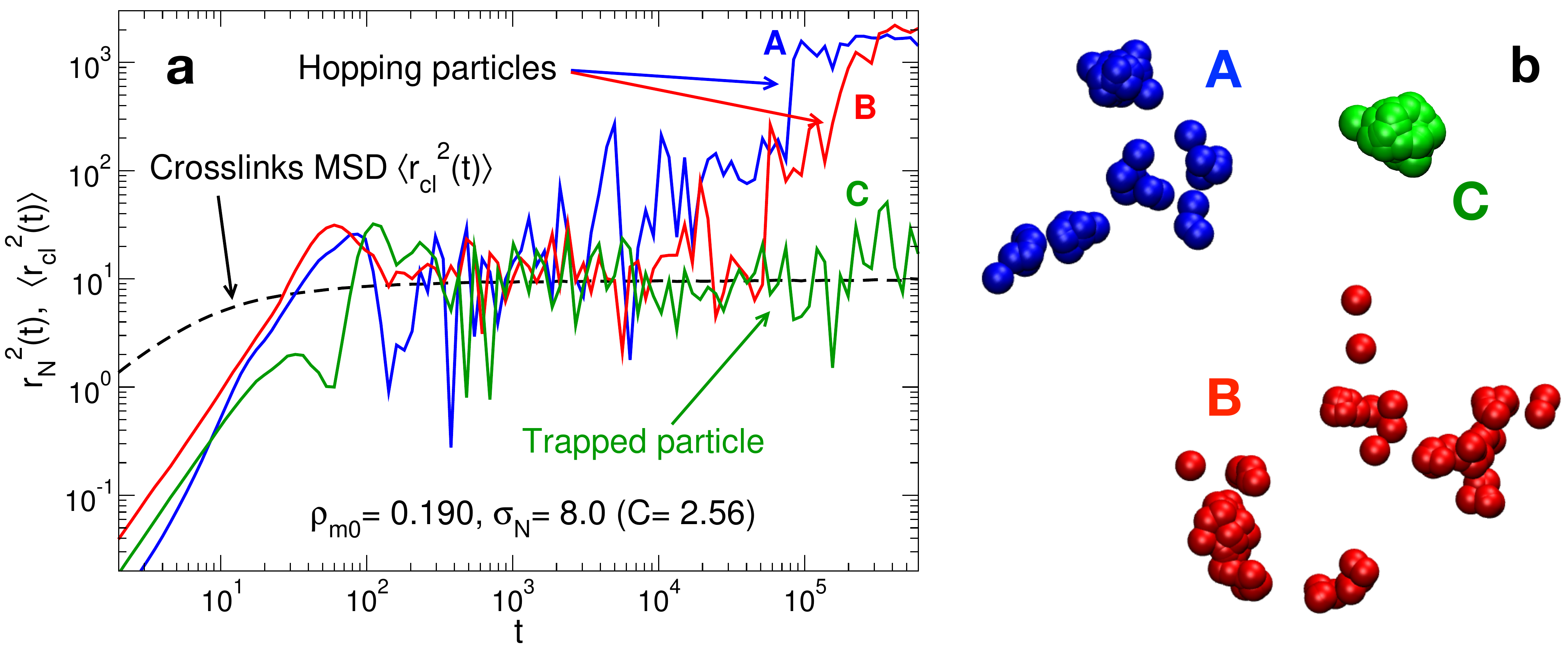}
\caption{(\textbf{a}): Squared displacement of three RNPs denoted A,B, and C (solid lines) and MSD of the crosslinks (dashed line) for $\rho_{m0}=0.190$ and $\sigma_N=8.0$ ($C=2.56$). (\textbf{b}): Trajectories of the NPs A,B and C, created by taking snapshots of the NP position at regular time intervals of length $\Delta t = 6 \times 10^3$ from the time $t=0$ to $t=6 \times 10^5$. To facilitate visualization, the NPs are represented as having half their real diameter.}
\label{fig:msd_np_single}
\end{figure}

To provide a better understanding of the hopping dynamics, we compare in Fig.~\ref{fig:msd_np_single}a the squared displacement (SD) $r_N^2(t)$ of three RNPs (labeled A, B and C) with the MSD of the crosslinks. One can see that particle C is trapped by the mesh, in that it displays, after the brief ballistic regime, a completely flat SD, with a localization length approximately equal to the localization length of the crosslinks, $r_\text{cl}^2(t) \simeq \lambda$. Also the A and B particles are initially trapped by the mesh, but eventually manage to escape. The SD of the NPs consists in a series of abrupt jumps, each of which is followed by a flat region. This type of interrupted progression is one of the typical signatures of a hopping process. In order to clarify the nature of the motion, we show in Fig.~\ref{fig:msd_np_single}b snapshots of the particles taken at regular time intervals of duration $\Delta t = 6 \times 10^3$, from the time $t=0$ to $t=6 \times 10^5$. These 100 snapshots clearly reveal the existence of hopping motion, in that the positions of the mobile particles A and B at different times form well-separated clusters. The same qualitative behavior is observed, under similar conditions, for the ANPs.

\begin{figure}
\centering
\includegraphics[width=0.48 \textwidth]{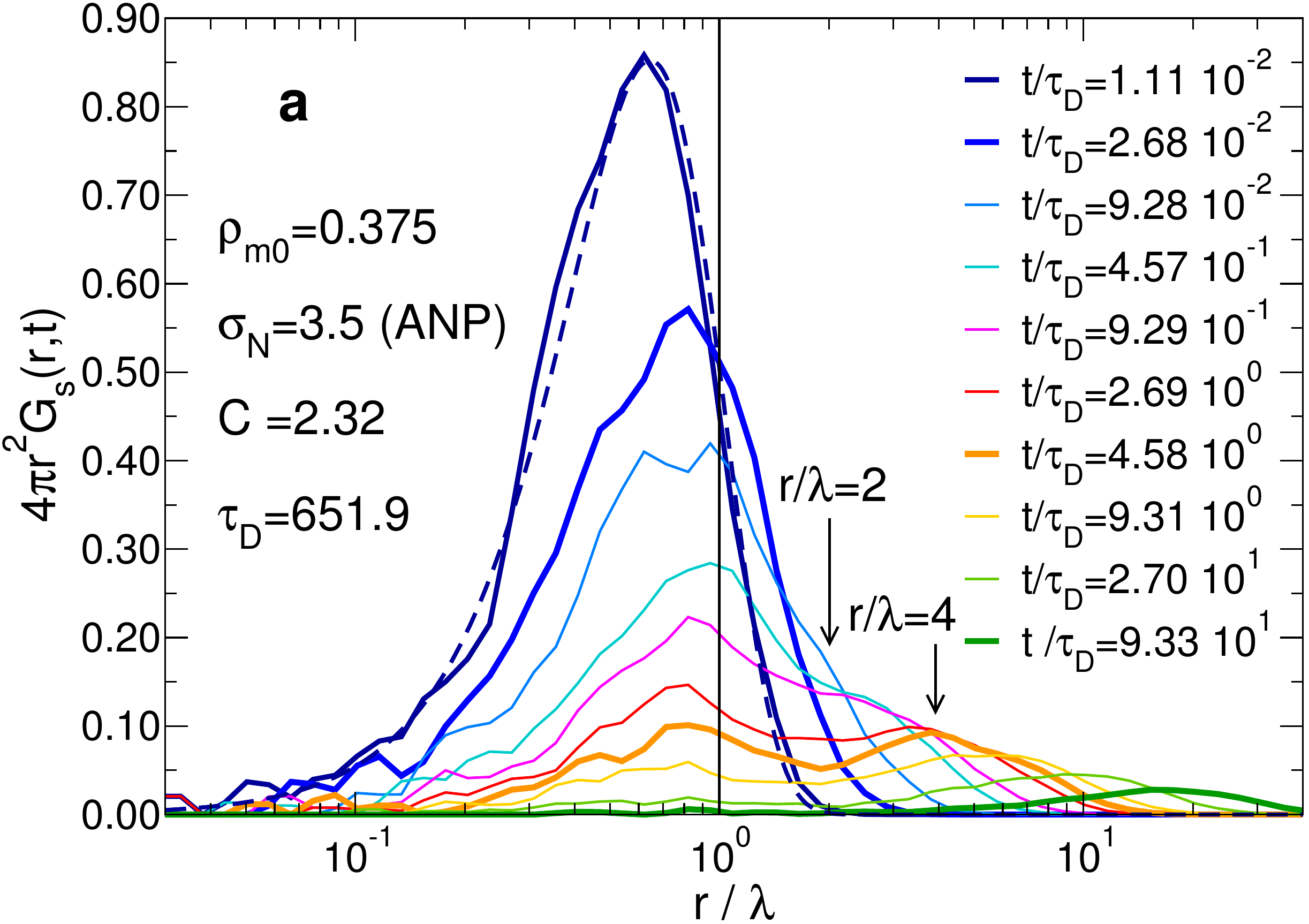}
\includegraphics[width=0.48 \textwidth]{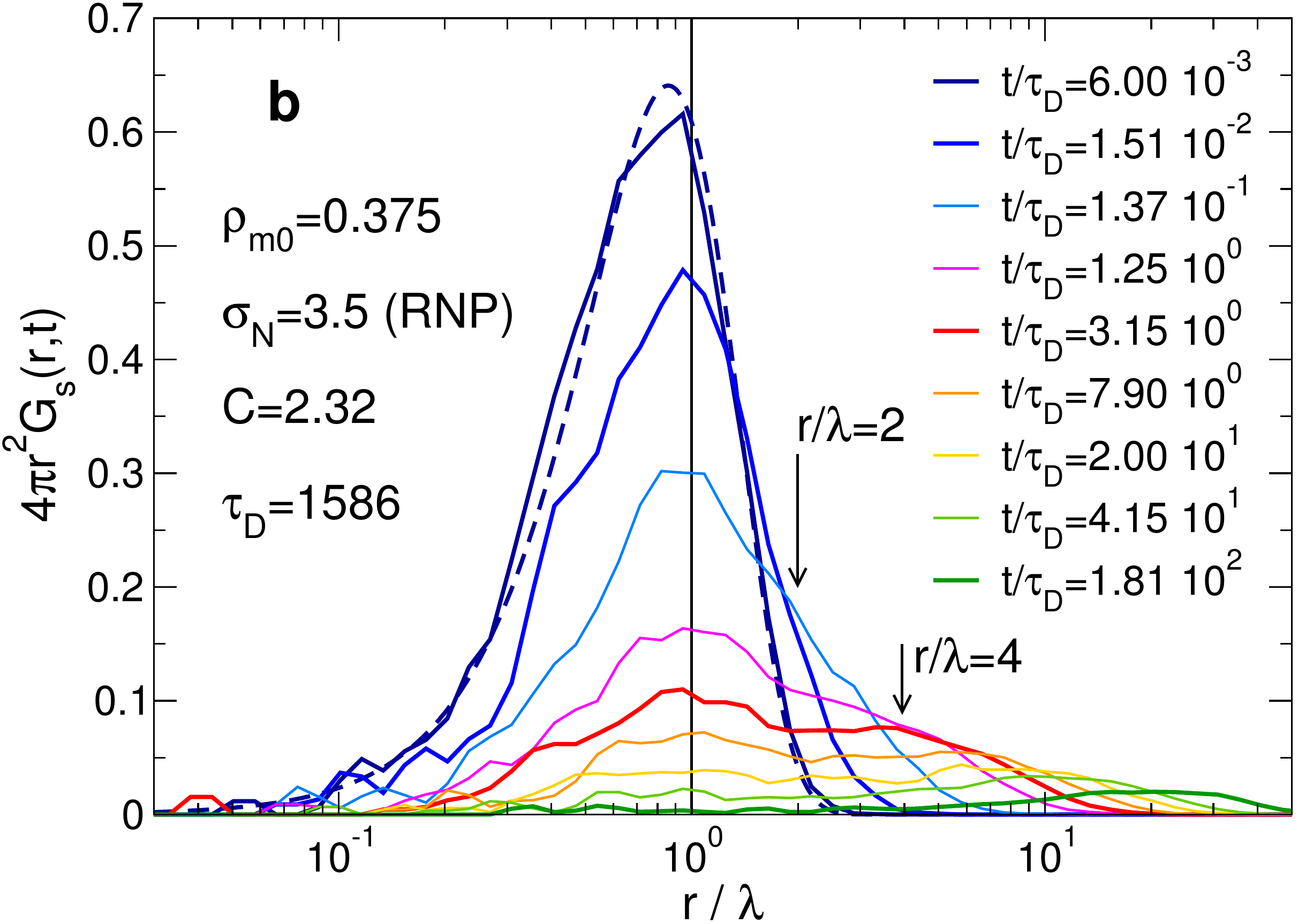}
\caption{Self part of the van Hove function, Eq.~\eqref{eq:g_self}, for $\rho_{m0}=0.375$, $\sigma_N=3.5$ for the ANPs (\textbf{a}) and the RNPs (\textbf{b}). $\tau_D \equiv \sigma_N^2/6D_N$ is the NP diffusion time. Dashed lines: Gaussian approximation, Eq.~\eqref{eq:g_self_Gaussian}.}
\label{fig:np_vhs}
\end{figure}

In order to characterize the hopping dynamics in a quantitative manner, we probe the self part of the van Hove function, which is defined as \cite{binder2011glassy}:

\begin{equation}
G_{s,N} (\mathbf r, t) \equiv \langle \delta [\mathbf r - (\mathbf r_{N,i} (t) - \mathbf r_{N,i} (0)) ] \rangle \quad .
\label{eq:g_self}
\end{equation}

Since our configurations are isotropic, we consider the spherically-averaged van Hove function, $G_{s,N}(r,t)$. From $G_{s,N}$ one can define the distribution of the displacements, $\Delta_N(r,t) dr = 4 \pi r^2 G_{s,N}(r,t) dr$, which represents the probability to find a particle at time $t$ at distance between $r$ and $r+dr$ from its original position. At short times the motion is ballistic and hence $G_{s,N}(r,t)$ is a Gaussian \cite{hansen1990theory} with variance $2 \langle r_N^2(t)\rangle/3$:

\begin{equation}
G_{s,N}(r,t) = \left( \frac 3 {2 \pi \langle r_N^2(t) \rangle} \right)^{3/2} \exp \left( - \frac {3 r^2}{2 \langle r_N^2(t)\rangle}\right).
\label{eq:g_self_Gaussian}
\end{equation}

At long times the NP motion becomes uncorrelated and Gaussian behavior is recovered again~\cite{hansen1990theory}. In Fig.~\ref{fig:np_vhs}, we show $\Delta_N(r,t)$ for $\rho_{m0}=0.375$, $\sigma_N=3.5$ for the ANPs (a) and the RNPs (b). Let us introduce the diffusion time, 

\begin{equation}
\tau_D \equiv \frac{\sigma_N^2}{6 D_N},
\label{eq:taud}
\end{equation}

\noindent
which represents the time it takes a NP with diffusion coefficient $D_N$ to move over a distance of the order of its diameter. At short times ($t \ll \tau_D$), both for ANPs and RNPs, $G_{s,N}(r,t)$ is well described by Eq.~\eqref{eq:g_self_Gaussian} (dashed curves in Fig.~\ref{fig:np_vhs}), and $\Delta_N(r,t)$ displays a single peak at  $r= \left[2 \langle r_N^2(t)\rangle/3\right]^{1/2}$, which progressively moves towards larger $r$ values. When the position of the peak approaches the crosslink localization length $\lambda$, the NP interacts strongly with the mesh and the dynamics is sub-diffusive, see Fig.~\ref{fig:msd_np_cl_rho03}. For longer times $\Delta_N(r,t)$ maintains the peak at $r \simeq \lambda$, while at the same time developing a weak shoulder at $r \simeq 2 \lambda$. This double peaked distribution is the clear signal of heterogeneous hopping dynamics \cite{kob1995testing}: While some particles are still trapped in their initial cage (primary peak at $r\simeq \lambda$), other particles have escaped from the cage and started a diffusive motion. The presence of the second peak indicates that the escaping NPs perform ``jumps'' with a typical length comparable to the mesh size, as predicted by Cai \textit{et al.} \cite{cai2015hopping}. In simulations of NP-charged networks with a regular structure, multiple peaks can be observed at $r$-values which are multiples of the mesh size \cite{cho2020tracer}. In contrast to this, we do not find in our system such multiple peaks, likely because our network is disordered. For $t > \tau_D$, the primary peak eventually disappears completely as all the particles have escaped from their initial cage.

\begin{figure}
\centering
\includegraphics[width=0.48 \textwidth]{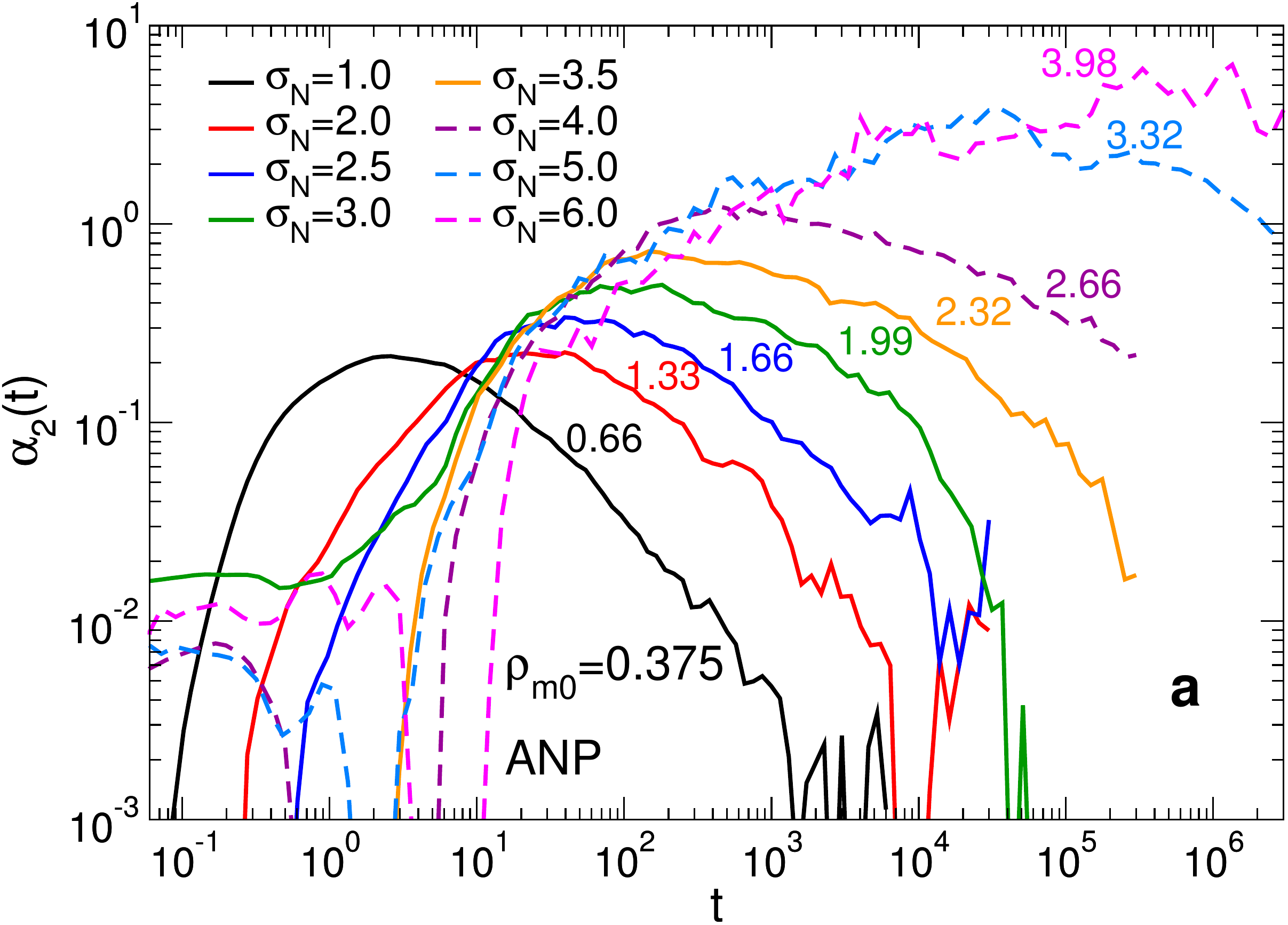}
\includegraphics[width=0.48 \textwidth]{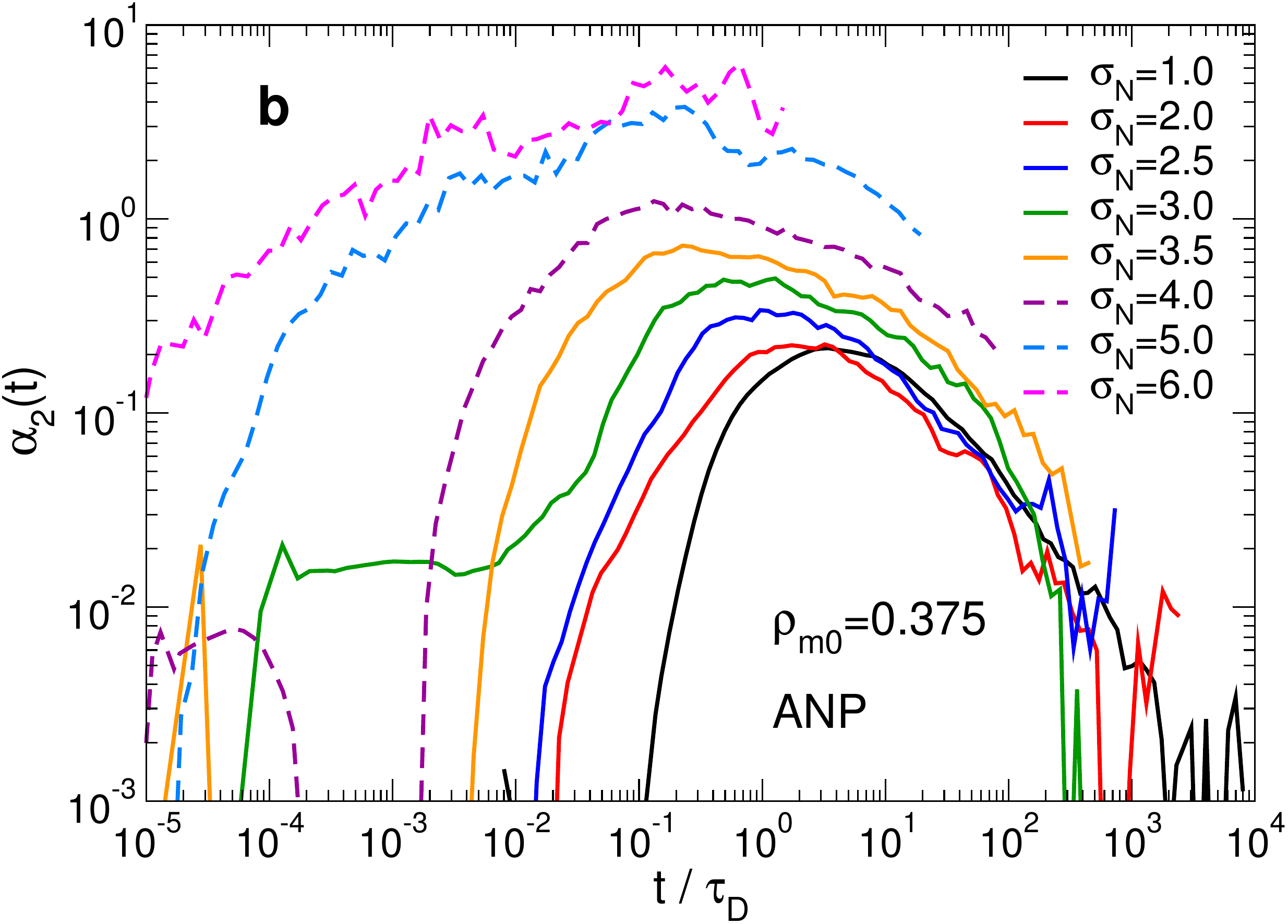}
\caption{(\textbf{a}): Non-Gaussian parameter for the ANPs with different values of $\sigma_N$ at $\rho_{m0}=0.375$, shown as a function of time (\textbf{a}) and of the rescaled time $t/\tau_D$, with $\tau_D \equiv \sigma_N^2/6D_N$ (\textbf{b}). Labels in (\textbf{a}) give the value of the confinement parameter $C$. Dashed curves represent systems which have not reached the diffusive regime.}
\label{fig:nongaussian}
\end{figure}

The heterogeneity of the relaxation dynamics can be characterized by probing the deviations of $G_s(r,t)$ from a Gaussian, which can be quantified using the non-Gaussian parameter \cite{kob1995testing}, 

\begin{equation}
\alpha_2 (t) \equiv \frac {3 \langle r^4(t) \rangle}{5 \langle r^2(t) \rangle^2} -1.
\end{equation}

\noindent
Thus if $\alpha_2= 0$, the dynamics is Gaussian, whereas a value which is appreciably different from $0$ signals strong heterogeneities in the dynamics. In Fig.~\ref{fig:nongaussian} we report $\alpha_2(t)$ for $\rho_{m0}=0.375$ for the ANPs (dashed lines denote systems where the diffusive regime was not reached). At short times $\alpha_2$ is very small, \textit{i.e.}, the dynamics of the NPs is Gaussian because of the ballistic motion. This regime is followed by an increase of $\alpha_2$, which corresponds to the onset of the subdiffusive regime in the MSD. For $C \lesssim 2$, this increase is very modest, and $\alpha_2$ remains smaller than $1$. This result is in agreement with the fact that for small $C$ the MSD crosses over from the ballistic regime directly to the diffusive regime, see Fig.~\ref{fig:msd_np_cl_rho03}. However, for strong confinement, $C \gtrsim 3$, $\alpha_2$ reaches values significantly larger than $1$. These values are comparable to those obtained from experimental studies of polymer networks containing NPs \cite{parrish2017network} or in simulations of deeply supercooled liquids \cite{kob1995testing}, \textit{i.e.}, systems which show pronounced dynamical heterogeneities. In our system, as in experimental ones \cite{parrish2017network}, this pronounced non-Gaussian dynamics is directly related to the structural heterogeneity of the network on the length scale of the NP size: Some NPs diffuse freely through the gaps in the network, whereas other NPs end up trapped in the mesh for a long time. The time at which $\alpha_2(t)$ reaches its maximum also increases strongly with $C$, \textit{i.e.}, the time needed to make one hopping movement depends as expected strongly on the confinement. In Fig.~\ref{fig:nongaussian}b we plot $\alpha_2$ as a function of the rescaled time $t/\tau_D$, where $\tau_D$ is defined in Eq.~(\ref{eq:taud}). We observe that for $C \lesssim 3$ the descending parts of all the curves fall on the same master curve, confirming that it is indeed the time scale $\tau_D$ which controls the long-time dynamics. We also note that for small $C$-values, $\alpha_2(t)$ reaches its maximum for $t\simeq \tau_D$, as also observed in polymer solutions containing NPs \cite{sorichetti2018structure}. For extreme confinement, one cannot conclude whether or not the Gaussian dynamics is recovered at long times, since we are unable to reach the diffusive regime. From this graph one also recognizes that the width of the peak in $\alpha_2(t)$ increases quickly with increasing $C$, a further indication that the dynamical heterogeneity of the system is increasing. We also mention that the time dependence of $\alpha_2$ for the RNPs display the same qualitative behavior as the one for the ANPs, and the same is true when different values of $\rho_{m0}$ are considered (see Supplementary Material, Sec. S3.3).

\begin{figure}
\centering
\includegraphics[width=0.49 \textwidth]{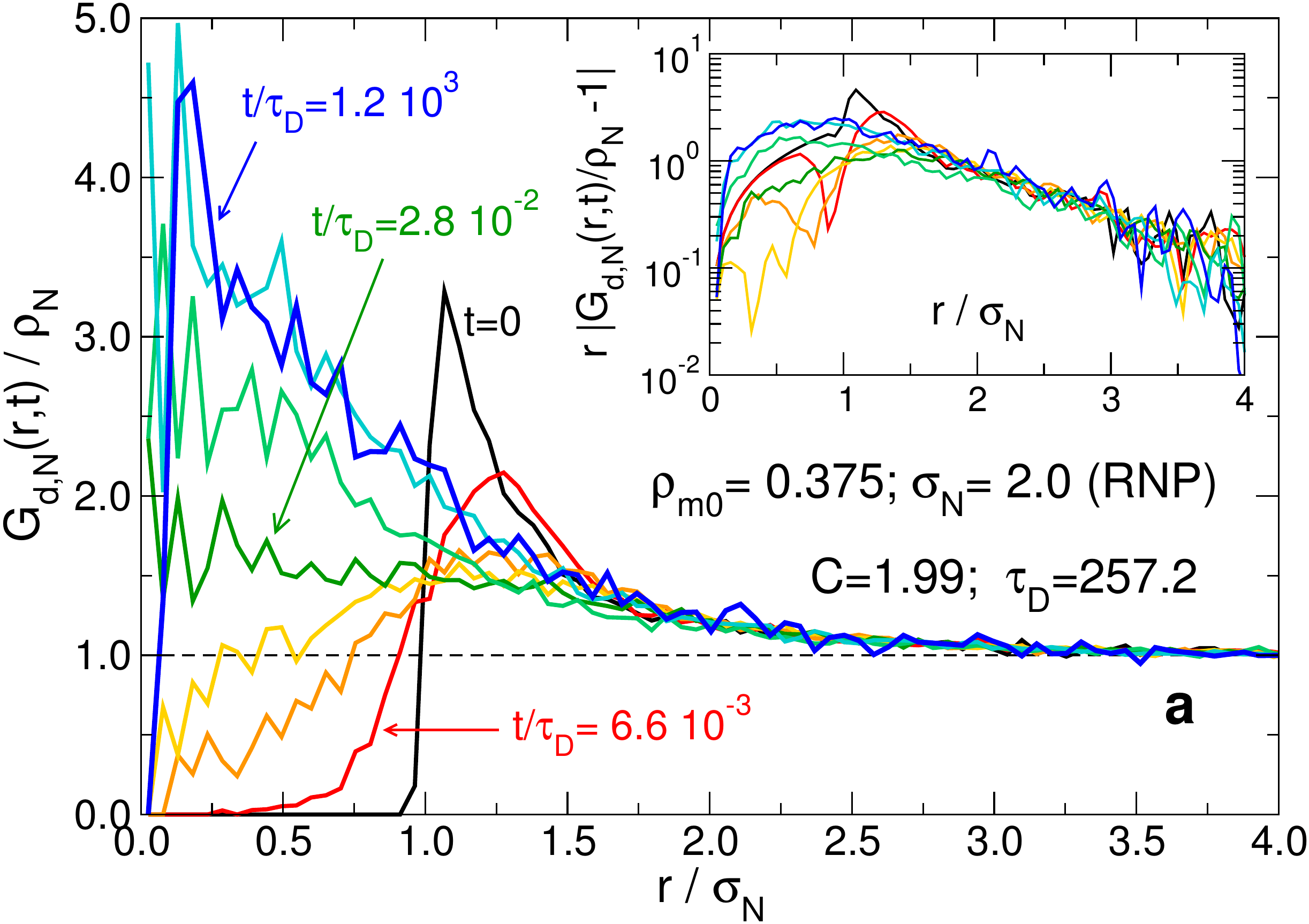}
\includegraphics[width=0.49 \textwidth]{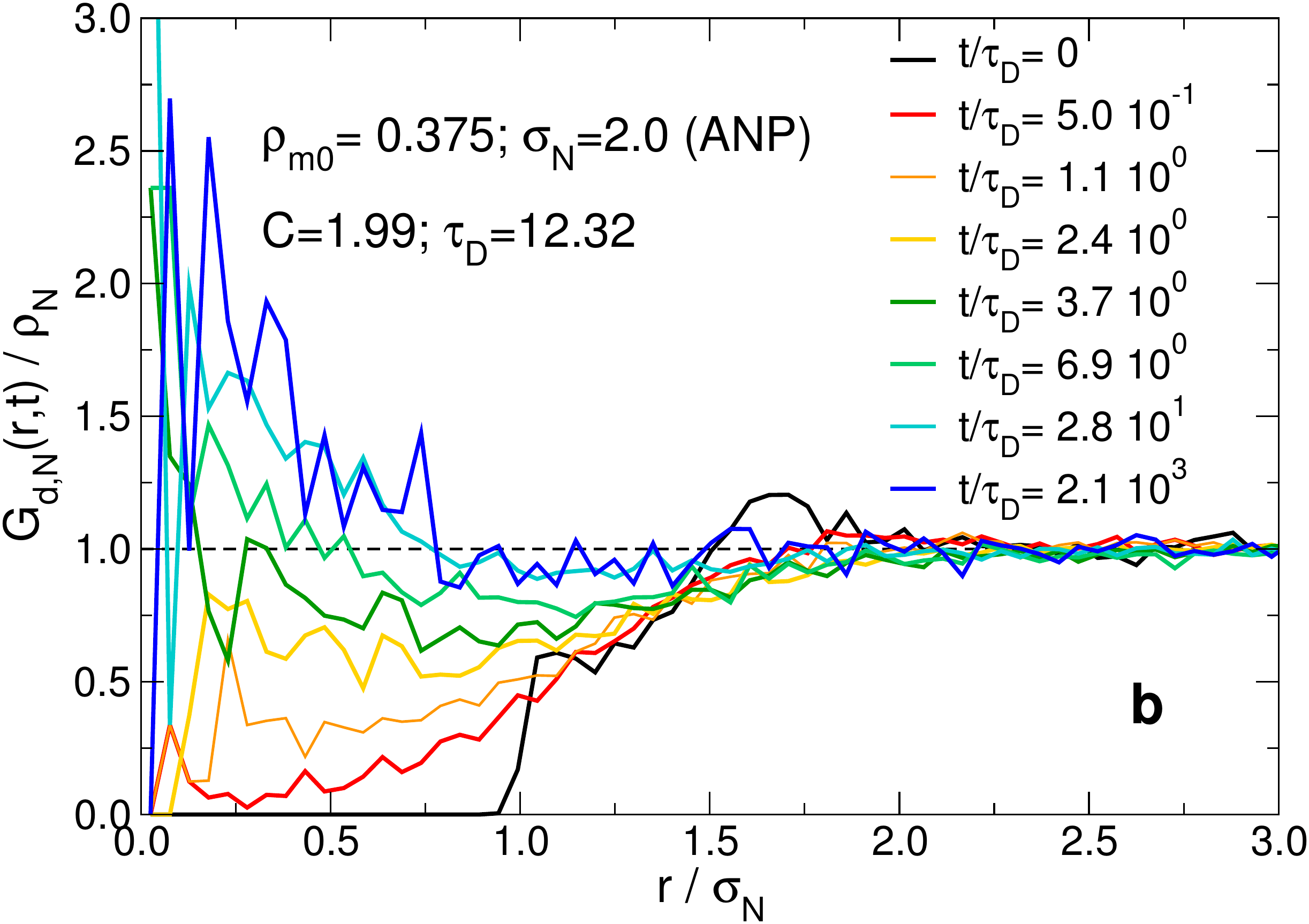}
\includegraphics[width=0.493 \textwidth]{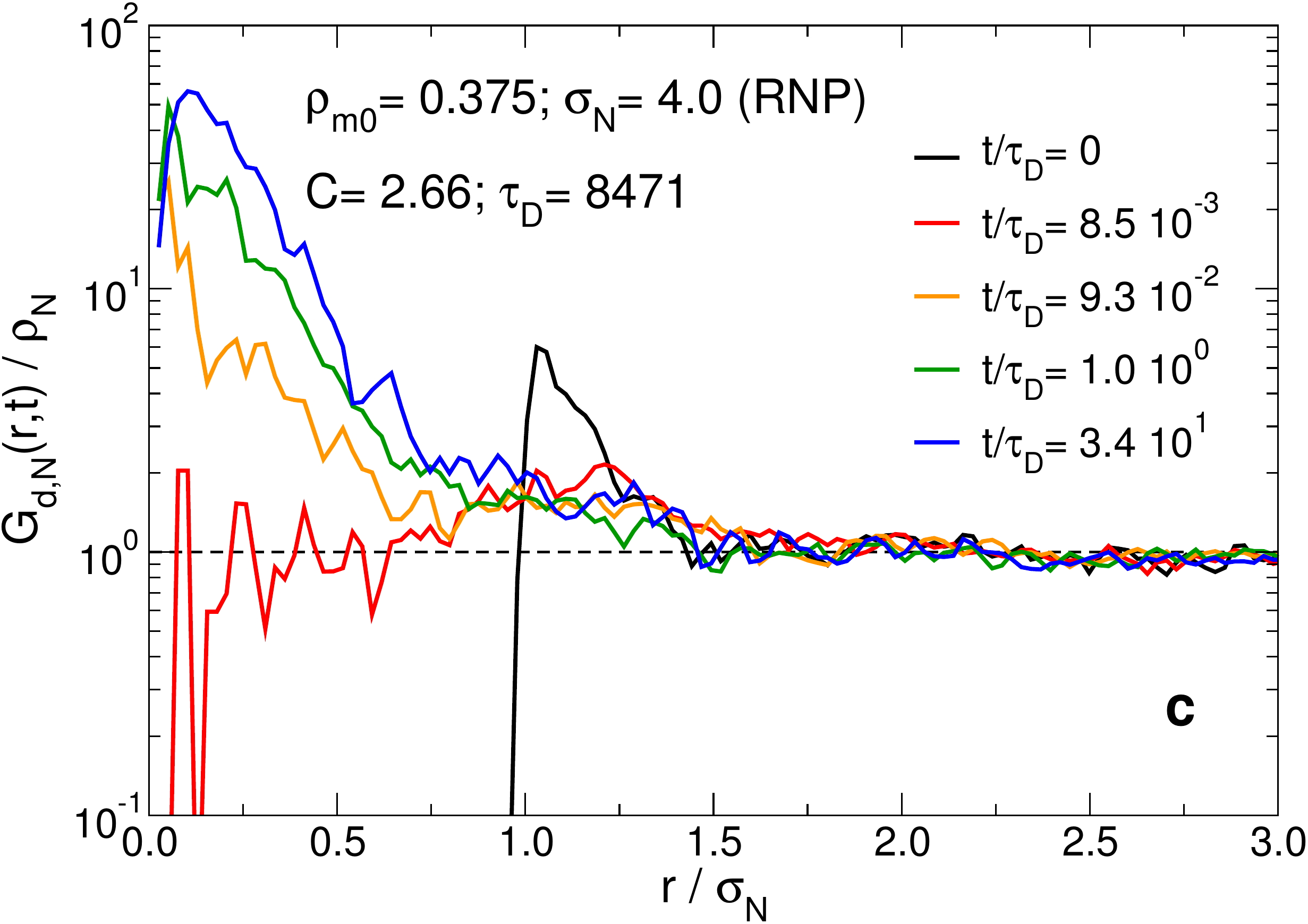}
\includegraphics[width=0.493 \textwidth]{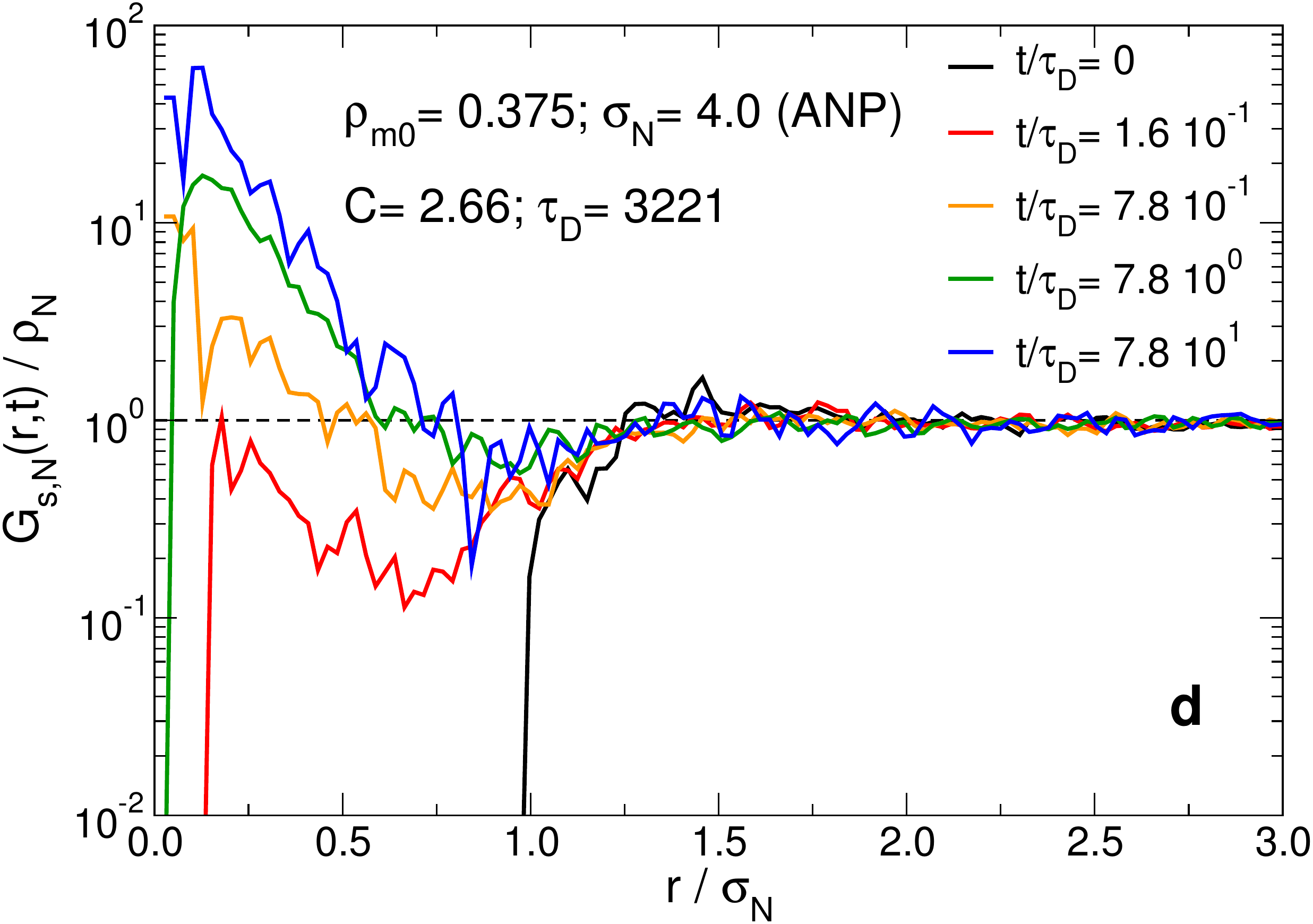}
\caption{Distinct part of the van Hove function $G_{d,N}(r,t)$ for $\rho_{m0}=0.375$, $\sigma_N=2.0$ (\textbf{a, b}) and $\rho_{m0}=0.375$, $\sigma_N=4.0$ (\textbf{c, d}). (\textbf{a, c}) show the RNP and (\textbf{b, d}) show the ANP. $\tau_D \equiv \sigma_N^2/6D_N$ is the NP diffusion time. \textit{Inset of (a)}: $r|G_d(r,t)-1|$ on a logarithmic scale.}
\label{fig:vhd_np}
\end{figure}

To probe the collective dynamics of the NPs we investigate the distinct part of the van Hove function which is given by \cite{hansen1990theory}

\begin{equation}
G_{d,N} (\mathbf r, t)  \equiv \frac 1 {N_N} \sum_{\substack{k=1\\ j\neq k}}^{N_N} \langle \delta [\mathbf r - (\mathbf r_{N,j} (t) - \mathbf r_{N,k} (0)) ] \rangle.
\label{eq:g_distinct}
\end{equation}

\noindent
For our isotropic system the quantity $4 \pi r^2 \rho_N G_{d,N}(r,t)dr$ is thus proportional to the probability to find at time $t$ a particle at a distance between $r$ and $r+dr$ from the position occupied by another particle at time $0$. For $t=0$ one recovers the NP-NP radial distribution function $g_{NN}(r)$. In Fig.~\ref{fig:vhd_np}a we show $G_{d,N}(r,t)/\rho_N$ for a system with RNP with $\rho_{m0}=0.375$ and $\sigma_N=2$ ($C=1.99$). One observes that with increasing $t/\tau_D$ the correlation hole at $r<\sigma_N$ is slowly filled and the contact peak at $r/\sigma_N \simeq 1$ is washed out. If the NPs were able to completely explore the system's volume at $t \gg \tau_D$, we would expect $G_{d,N}(r,t)/\rho_N\simeq 1$, however this is not what we find, as shown in the inset of Fig.~\ref{fig:vhd_np}a. The  slope on the right hand side of the contact peak becomes independent of time for all values of $t/\tau_D$, even though the NPs reach the diffusive regime (Fig.~\ref{fig:msd_np_cl_rho03}b) and the dynamics is basically Gaussian (see Supplementary Material). At the same time we find at long times a pronounced peak at $r\approx 0$, \textit{i.e.}, a probability that is enhanced with respect to the one for an ideal gas.
This behavior is due to the permanent nature of the network: A given region of space that at $t=0$ is occupied by a NP is likely to have a NP also at infinitely long times, since the local network structure makes that this spot has a higher probability than average or, put otherwise, the frozen in disorder of the network makes that the stationary distribution of the NP is non-uniform in space. 
Qualitatively the same behavior is observed for the ANP, Fig.~\ref{fig:vhd_np}b. Although in this case there is no pronounced nearest neighbor peak, see discussion in the context of Fig.~\ref{fig:rdf_mono_np_rho02}, we find a marked peak at small distances even at long times, \textit{i.e.}, the signature that certain regions in the network are highly preferential sites for the NP. This is thus evidence that the NP spend a substantial time in these sites and when they leave them they move quickly to another favorable site, \textit{i.e.}, that there is hopping dynamics~\cite{kob1995testing,roldan2017connectivity}.

For stronger confinement, hopping motion becomes even more relevant, and the peak at $r=0$ is dramatically more pronounced, as shown in Figs.~\ref{fig:vhd_np}c-d, where we report $G_d(r,t)/\rho_N$ for $\rho_{m0}=0.375$ and $\sigma_N=4$ ($C=2.66$) for RNPs (c) and ANPs (d) (note the logarithmic scale for the ordinate). Although the  statistics is worse than for the case of smaller values of $C$ and hence does not allow us to study accurately the long-range behavior of  $G_d(r,t)$, we clearly observe the appearance of a peak at $r\simeq0$. At long times, this peak grows significantly, signaling that NP motion is dominated by hopping dynamics.

 %%%%%%%%%%%%%%%%%%%%%%%%%%%%%%%%%%%%%%%%%%%%%%%%%   
  %%%%%%%%%%%%%%%%%%%%%%%%%%%%%%%%%%%%%%%%%%%%%%%%%                                 
\section{Summary and conclusions} \label{sec:summary}
 %%%%%%%%%%%%%%%%%%%%%%%%%%%%%%%%%%%%%%%%%%%%%%%%%                  
%%%%%%%%%%%%%%%%%%%%%%%%%%%%%%%%%%%%%%%%%%%%%%%%% 

We have carried out molecular dynamics simulations of a disordered and polydisperse polymer network in which we have embedded nanoparticles at a fixed volume fraction $\phi_N \simeq 0.02$. Three different networks were studied, all with trivalent crosslinks, with densities in the neat state $\rho_{m0}=0.190,0.290$, and $0.370$.
The size of the NP ranged between 1 and 10, allowing to probe weak as well as extreme confinement by the network mesh, and we considered NP-polymer interactions that were either attractive or repulsive.

The static structure factor indicates that the ANPs are well dispersed in the network, whereas RNPs show a weak tendency to cluster. From the analysis of the pore size distribution of the gel, $P(r)$, one can conclude that the NP locally deform the mesh and cause the appearance of a secondary peak in $P(r)$, the location of which is directly related to the size of the NPs. 

The analysis the MSD of the NP shows that the confinement parameter $C=\sigma_N/\lambda$, where $\lambda$ is the localization length of the crosslinks, is the relevant parameter that determines the dynamics, in agreement with theoretical predictions \cite{dell2014theory,cai2015hopping}. Three dynamic regimes can be identified: Weak confinement ($C\lesssim 1$), strong confinement ($1 \lesssim C \lesssim 2)$, and extreme confinement ($C \gtrsim 3$).
In the weak confinement regime, the NPs can freely diffuse through the mesh. In the strong confinement regime, the MSD of the NP displays a subdiffusive transient, $\langle r_N^2(t) \rangle \propto t^\beta$ with $\beta<1$, before eventually recovering diffusive behavior. In the extreme confinement regime, a marked subdiffusive regime appears in the MSD on intermediate time scales during which the NP undergo a very heterogeneous dynamics. This heterogeneity is directly related to the frozen in disorder of the network and is an ingredient which, to the best of our knowledge, has so far not been properly been taken into account in the theoretical approaches to describe these systems.

In the strong and extreme confinement regimes, NP motion can basically only proceed through the mechanism of activated hopping \cite{dell2014theory,cai2015hopping}, \textit{i.e.}, by waiting for a thermal fluctuation of the mesh that allows a NP to  jump to a nearby cage. We find that the exponent $\beta(t)$ and the long-time NP diffusion coefficient, $D_N$, are controlled by $C$, but depend on the type of NP. The diffusion coefficient decreases dramatically with increasing $C$, displaying in one case a drop of five orders of magnitude upon an increase of $C$ by a factor $6$. The $C-$dependence of $D_N$ can be described very well with the theoretical predictions by Cai \textit{et al.} \cite{cai2015hopping} but also with the ones by Dell and Schweizer \cite{dell2014theory} and hence we conclude that with the present set of data it is not possible to decide which theory is more reliable. More quantitative calculations using these two approaches should therefore be done in the future, including, if possible, the disordered nature of the network.

Finally we study the details of the hopping dynamics by analyzing the van Hove function of the NPs and the non-Gaussian parameter. The time and space dependence of the self part of the van Hove function confirms that, in the strongly confined regime, the NPs move indeed through activated hopping in that the function shows several peaks from which one can infer that the jump length is close to the mean mesh size, in agreement with the predictions of Cai \textit{et al.} \cite{cai2015hopping} and with a recent simulation study \cite{cho2020tracer}. At the same time the motion is strongly non-Gaussian with values of the non-Gaussian parameter comparable to those found in deeply supercooled liquids \cite{kob1995testing}. The prevalence of hopping motion is confirmed by the analysis of the distinct part of the van Hove function

In conclusion, we have studied for the first time in simulations the diffusion of NPs in disordered and polydisperse networks, probing dynamic regimes up to extreme confinements of the NPs by the mesh. This work represents the natural extension of recent efforts in simulations of crosslinked nanocomposites \cite{kumar2019transport,chen2020nanoparticle,cho2020tracer}. Understanding the dynamics of NPs in crosslinked networks still poses a formidable challenge both for theoretical approaches and for simulations, in the latter case due to the enormous relaxation times of these systems. Although at present it is not possible to fully equilibrate such systems in the extreme confinement regime, many questions remain whose answer is within reach, and which will help to understand better real-life systems. Examples are the interplay between crosslinks and entanglements in polydisperse systems, the role of chemical \textit{versus} physical crosslinks, the effect of NP shape etc., questions that remain to be clarified.
Some of these questions can probably be addressed by equilibrating the systems using clever Monte Carlo algorithms, such as parallel tempering. 
Moreover, a more detailed analysis is required in order to compare qualitatively different theoretical predictions \cite{dell2014theory,cai2015hopping}. We are therefore convinced that the study of such systems will remain a fruitful and challenging topic of research in the future.

%%%%%%%%%%%%%%%%%%%%%%%%%%%%%%%%%%%%%%%%%%%
\begin{acknowledgement}

We thank K. Schweizer and M. Lenz for useful discussions. The analysis of the pore size distribution has been performed with the open-soursce \textit{baggianalysis} software by L. Rovigatti (DOI: 10.5281/zenodo.4588503). This work has been supported by LabEx NUMEV
(ANR-10-LABX-20) funded by the ‘‘Investissements d'Avenir’’
French Government program, managed by the French National Research Agency
(ANR). 

\end{acknowledgement}

%%%%%%%%%%%%%%%%%%%%%%%%%%%%%%%%%%%%%%%%%%%
%%%%%%%%%%%%%%%%%%%%%%%%%%%%%%%%%%%%%%%%%%%
\newpage
\section*{SUPPLEMENTARY MATERIAL}

In this Supplementary Material we give
more details on the properties of the systems, such as the swelling behavior, the structure factor, pore size distribution, and the mean squared displacement.

\setcounter{equation}{0}
\setcounter{figure}{0}
\setcounter{table}{0}
\setcounter{section}{0}

\renewcommand{\theequation}{S\arabic{equation}}
\renewcommand{\thefigure}{S\arabic{figure}}
\renewcommand{\thetable}{S\arabic{table}}
\renewcommand{\thesection}{S\arabic{section}}
%%%%%%%%%%%%%%%%%%%%%%%%%%%%%%%%%%%%%%%%%%%
%%%%%%%%%%%%%%%%%%%%%%%%%%%%%%%%%%%%%%%%%%%

%%%%%%%%%%%%%%%%%%%%%%%%%%%%%%%%%%%%%%%%%%%
\section{Properties of the simulated systems} \label{sec:tables}
%%%%%%%%%%%%%%%%%%%%%%%%%%%%%%%%%%%%%%%%%%%
Here we list some properties of the systems we simulated. Tables~\ref{tab:rnp} and \ref{tab:anp} are for the repulsive and attractive NP, respectively.

\begin{table}
\centering

\caption{Properties of the simulated systems (\textbf{RNPs}). $\sigma_N$: NP diameter. $N_N$: Number of NPs. $\rho_{m0}=M/V_0$: monomer number density in the neat system (independent of $\sigma_N$).  $\phi_m=\pi \rho_N \sigma^3 /6$: monomer volume fraction. $\phi_N=\pi \rho_N \sigma_N^3 /6$: NP volume fraction. $C\equiv \sigma_N/\lambda$: Confinement parameter ($\lambda$ = crosslink localization length). $D_N$: NP diffusion coefficient. $\tau_D \equiv \sigma_N^2 / 6 D_N$: Diffusion time. Values marked with $*$ are upper and lower bounds, for $D_N$ and $\tau_D$ respectively, for systems which have not equilibrated.}

$\begin{array}{c@{\hspace{1.25 em}}c@{\hspace{1.25 em}}c@{\hspace{1.25 em}}c@{\hspace{1.25 em}}c@{\hspace{1.25 em}}c@{\hspace{1.25 em}}c@{\hspace{1.25 em}}c@{\hspace{1.25 em}}}
\toprule
\sigma_N & N_N &  \rho_{m0} & \phi_m & \phi_N &C & D_N & \tau_D\\
\midrule
2.0 & 9755 & 0.1902 & 0.0823 & 0.0165 & 0.641 & 3.758\times 10^{-1} & 1.774\times 10^{0}\\ 
3.0 & 2890 & 0.1902 & 0.0900 & 0.0181 & 0.962 & 1.331\times 10^{-1} & 1.127\times 10^{1}\\ 
4.0 & 1219 & 0.1902 & 0.0928 & 0.0186 & 1.282 & 4.862\times 10^{-2} & 5.484\times 10^{1}\\ 
5.0 & 624 & 0.1902 & 0.0943 & 0.0189 & 1.603 & 1.634\times 10^{-2} & 2.550\times 10^{2}\\ 
6.0 & 361 & 0.1902 & 0.0951 & 0.0191 & 1.923 & 4.230\times 10^{-3} & 1.419\times 10^{3}\\ 
7.0 & 228 & 0.1902 & 0.0956 & 0.0192 & 2.244 & 9.627\times 10^{-4} & 8.483\times 10^{3}\\ 
8.0 & 152 & 0.1902 & 0.0958 & 0.0192 & 2.564 & *2.480\times 10^{-4} & *4.301\times 10^{4}\\ 
\midrule
2.0 & 6538 & 0.2898 & 0.1376 & 0.0181 & 0.997 & 1.444\times 10^{-1} & 4.615\times 10^{0}\\ 
3.0 & 1937 & 0.2898 & 0.1428 & 0.0188 & 1.496 & 3.222\times 10^{-2} & 4.655\times 10^{1}\\ 
3.5 & 1220 & 0.2898 & 0.1442 & 0.0190 & 1.745 & 1.343\times 10^{-2} & 1.520\times 10^{2}\\ 
4.0 & 817 & 0.2898 & 0.1448 & 0.0191 & 1.994 & 4.995\times 10^{-3} & 5.339\times 10^{2}\\ 
4.5 & 574 & 0.2898 & 0.1456 & 0.0192 & 2.244 & 1.596\times 10^{-3} & 2.115\times 10^{3}\\ 
5.0 & 418 & 0.2898 & 0.1459 & 0.0192 & 2.493 & *5.169\times 10^{-4} & *8.062\times 10^{3}\\ 
6.0 & 242 & 0.2898 & 0.1468 & 0.0193 & 2.991 & *3.887\times 10^{-5} & *1.544\times 10^{5}\\ 
\midrule
2.0 & 5028 & 0.3751 & 0.1833 & 0.0187 & 1.327 & 5.989\times 10^{-2} & 1.113\times 10^{1}\\ 
3.0 & 1490 & 0.3751 & 0.1878 & 0.0191 & 1.991 & 5.832\times 10^{-3} & 2.572\times 10^{2}\\ 
3.5 & 938 & 0.3751 & 0.1885 & 0.0192 & 2.323 & 1.287\times 10^{-3} & 1.586\times 10^{3}\\ 
4.0 & 629 & 0.3751 & 0.1892 & 0.0193 & 2.655 & *3.148\times 10^{-4} & *8.471\times 10^{3}\\ 
4.5 & 441 & 0.3751 & 0.1895 & 0.0193 & 2.987 & *4.429\times 10^{-5} & *7.620\times 10^{4}\\ 
5.0 & 322 & 0.3751 & 0.1897 & 0.0193 & 3.319 & *1.004\times 10^{-5} & *4.151\times 10^{5}\\ 
\bottomrule
\end{array}$
\label{tab:rnp}
\end{table}

\begin{table}
\centering

\caption{Properties of the simulated systems (\textbf{ANPs}). $\sigma_N$: NP diameter. $N_N$: Number of NPs. $\rho_{m0}=M/V_0$: monomer number density in the neat system (independent of $\sigma_N$).  $\phi_m=\pi \rho_N \sigma^3 /6$: monomer volume fraction. $\phi_N=\pi \rho_N \sigma_N^3 /6$: NP volume fraction. $C\equiv \sigma_N/\lambda$: Confinement parameter ($\lambda$ = crosslink localization length). $D_N$: NP diffusion coefficient. $\tau_D \equiv \sigma_N^2 / 6 D_N$: Diffusion time. Values marked with $*$ are upper and lower bounds, for $D_N$ and $\tau_D$ respectively, for systems which have not equilibrated.}

$\begin{array}{c@{\hspace{1.25 em}}c@{\hspace{1.25 em}}c@{\hspace{1.25 em}}c@{\hspace{1.25 em}}c@{\hspace{1.25 em}}c@{\hspace{1.25 em}}c@{\hspace{1.25 em}}c@{\hspace{1.25 em}}}
\toprule
\sigma_N & N_N &  \rho_{m0} & \phi_m & \phi_N &C & D_N & \tau_D\\
\midrule
1.0 & 78038 & 0.1902 & 0.1054 & 0.0212 & 0.321 & 5.086\times 10^{-1} & 3.277\times 10^{-1}\\ 
2.0 & 9755 & 0.1902 & 0.1027 & 0.0206 & 0.641 & 1.810\times 10^{-1} & 3.682\times 10^{0}\\ 
3.0 & 2890 & 0.1902 & 0.1005 & 0.0202 & 0.962 & 9.593\times 10^{-2} & 1.564\times 10^{1}\\ 
4.0 & 1219 & 0.1902 & 0.0995 & 0.0200 & 1.282 & 5.340\times 10^{-2} & 4.994\times 10^{1}\\ 
5.0 & 624 & 0.1902 & 0.0991 & 0.0199 & 1.603 & 2.832\times 10^{-2} & 1.471\times 10^{2}\\ 
6.0 & 361 & 0.1902 & 0.0988 & 0.0199 & 1.923 & 1.339\times 10^{-2} & 4.482\times 10^{2}\\ 
7.0 & 228 & 0.1902 & 0.0987 & 0.0198 & 2.244 & 5.149\times 10^{-3} & 1.586\times 10^{3}\\ 
8.0 & 152 & 0.1902 & 0.0985 & 0.0198 & 2.564 & 1.520\times 10^{-3} & 7.017\times 10^{3}\\ 
9.0 & 107 & 0.1902 & 0.0983 & 0.0197 & 2.885 & *3.423\times 10^{-4} & *3.943\times 10^{4}\\ 
10.0 & 78 & 0.1902 & 0.0981 & 0.0197 & 3.205 & *1.112\times 10^{-4} & *1.499\times 10^{5}\\ 
\midrule
1.0 & 52300 & 0.2898 & 0.1635 & 0.0216 & 0.499 & 3.060\times 10^{-1} & 5.446\times 10^{-1}\\ 
2.0 & 6538 & 0.2898 & 0.1525 & 0.0201 & 0.997 & 9.891\times 10^{-2} & 6.740\times 10^{0}\\ 
2.5 & 3347 & 0.2898 & 0.1515 & 0.0200 & 1.246 & 6.042\times 10^{-2} & 1.724\times 10^{1}\\ 
3.0 & 1937 & 0.2898 & 0.1507 & 0.0199 & 1.496 & 3.703\times 10^{-2} & 4.051\times 10^{1}\\ 
3.5 & 1220 & 0.2898 & 0.1505 & 0.0198 & 1.745 & 2.100\times 10^{-2} & 9.721\times 10^{1}\\ 
4.0 & 817 & 0.2898 & 0.1502 & 0.0198 & 1.994 & 1.096\times 10^{-2} & 2.433\times 10^{2}\\ 
4.5 & 574 & 0.2898 & 0.1499 & 0.0198 & 2.244 & 4.880\times 10^{-3} & 6.916\times 10^{2}\\ 
5.0 & 418 & 0.2898 & 0.1498 & 0.0197 & 2.493 & 1.836\times 10^{-3} & 2.269\times 10^{3}\\ 
\midrule
1.0 & 40225 & 0.3751 & 0.2040 & 0.0208 & 0.664 & 2.234\times 10^{-1} & 7.462\times 10^{-1}\\ 
2.0 & 5028 & 0.3751 & 0.1953 & 0.0199 & 1.327 & 5.412\times 10^{-2} & 1.232\times 10^{1}\\ 
2.5 & 2574 & 0.3751 & 0.1940 & 0.0198 & 1.659 & 2.537\times 10^{-2} & 4.105\times 10^{1}\\ 
3.0 & 1490 & 0.3751 & 0.1940 & 0.0198 & 1.991 & 1.059\times 10^{-2} & 1.417\times 10^{2}\\ 
3.5 & 938 & 0.3751 & 0.1936 & 0.0197 & 2.323 & 3.132\times 10^{-3} & 6.519\times 10^{2}\\ 
4.0 & 629 & 0.3751 & 0.1935 & 0.0197 & 2.655 & *8.280\times 10^{-4} & *3.221\times 10^{3}\\ 
5.0 & 322 & 0.3751 & 0.1930 & 0.0196 & 3.319 & *2.680\times 10^{-5} & *1.555\times 10^{5}\\ 
6.0 & 186 & 0.3751 & 0.1924 & 0.0196 & 3.982 & *2.954\times 10^{-6} & *2.031\times 10^{6}\\ 
\bottomrule
\end{array}$
\label{tab:anp}
\end{table}

%%%%%%%%%%%%%%%%%%%%%%%%%%%%%%%%%%%%%%%%%%%
%%%%%%%%%%%%%%%%%%%%%%%%%%%%%%%%%%%%%%%%%%%
\section{Structure}
%%%%%%%%%%%%%%%%%%%%%%%%%%%%%%%%%%%%%%%%%%%
%%%%%%%%%%%%%%%%%%%%%%%%%%%%%%%%%%%%%%%%%%%

%%%%%%%%%%%%%%%%%%%%%%%%%%%%%%%%%%%%%%%%%%%
\subsection{Chain length distribution}
%%%%%%%%%%%%%%%%%%%%%%%%%%%%%%%%%%%%%%%%%%%

\begin{figure}
\centering
\includegraphics[width=0.48 \textwidth]{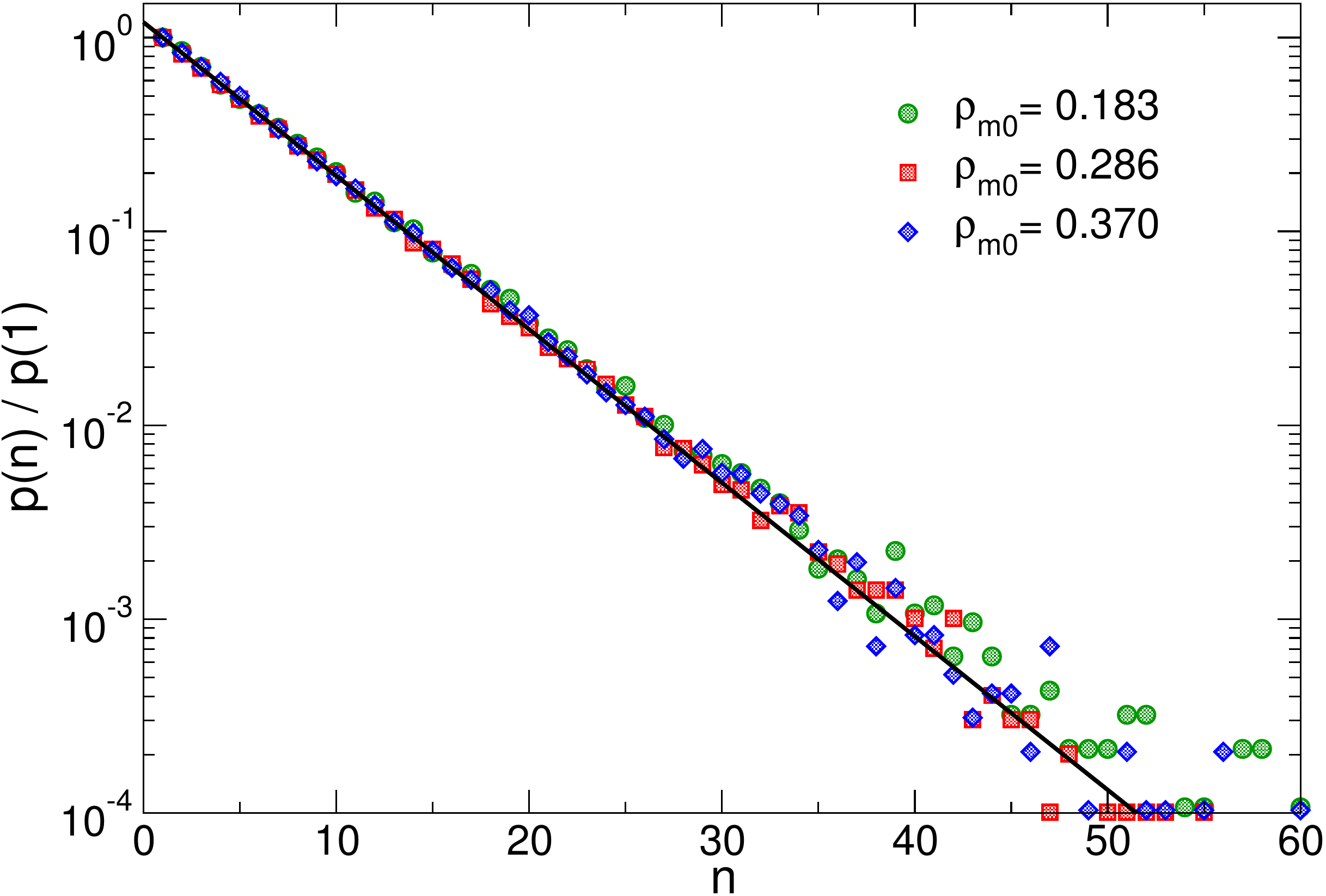}
\caption{Rescaled chain length distribution $p(n)/p(1)$ for the simulated networks. Solid line: Theoretical prediction given by Eq.~\eqref{eq:pn_fs}.}
\label{fig:chain_length}
\end{figure}

In the main text we have mentioned that the distribution of the chain length, $p(n)$, is given by an exponential. In Fig.~\ref{fig:chain_length} we show $p(n)$, divided by $p(1)=(N_m-N_c)/\langle n\rangle^2$, for the three simulated networks. We observe that $p(n)$ is independent of $\rho_{m0}$ and decreases indeed exponentially \cite{rovigatti2017internal}, \textit{i.e.} the functional form given by the Flory-Stockmayer formula \cite{flory1953principles,stockmayer1943theory}, Eq.~(\ref{eq:pn_fs}) of the main text.

The mean chain length can be thus be calculated as \cite{rovigatti2017internal} $ N_x \equiv \langle n\rangle = {2 (c^{-1}-1)/3}$ and for $c=0.1$ one obtains $N_x=6$.

%%%%%%%%%%%%%%%%%%%%%%%%%%%%%%%%%%%%%%%%%%%
\subsection{Swelling and shrinking of the network upon the addition of NPs}
%%%%%%%%%%%%%%%%%%%%%%%%%%%%%%%%%%%%%%%%%%%

\begin{figure}
\centering
\includegraphics[width=0.48 \textwidth]{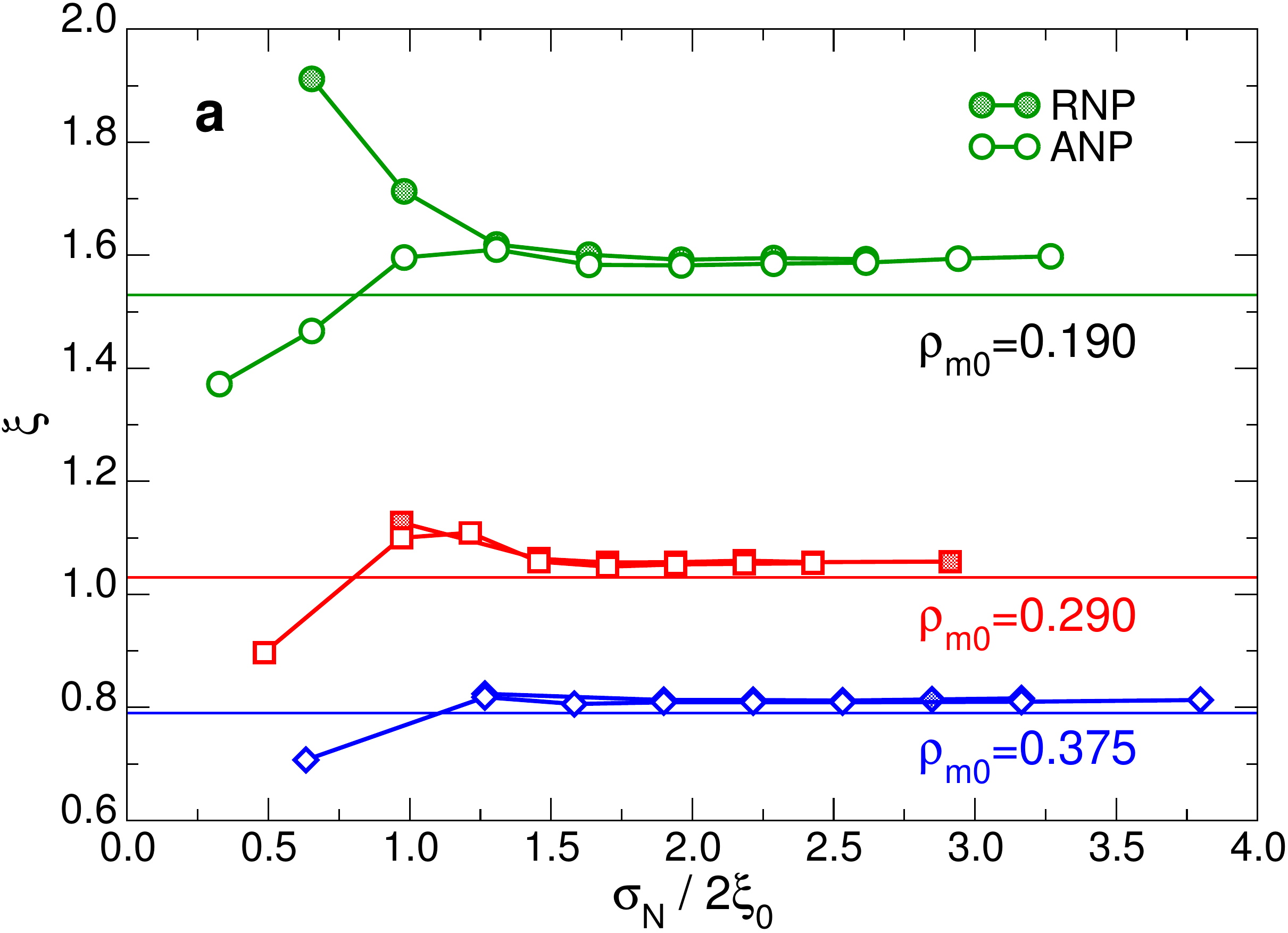}
\includegraphics[width=0.48 \textwidth]{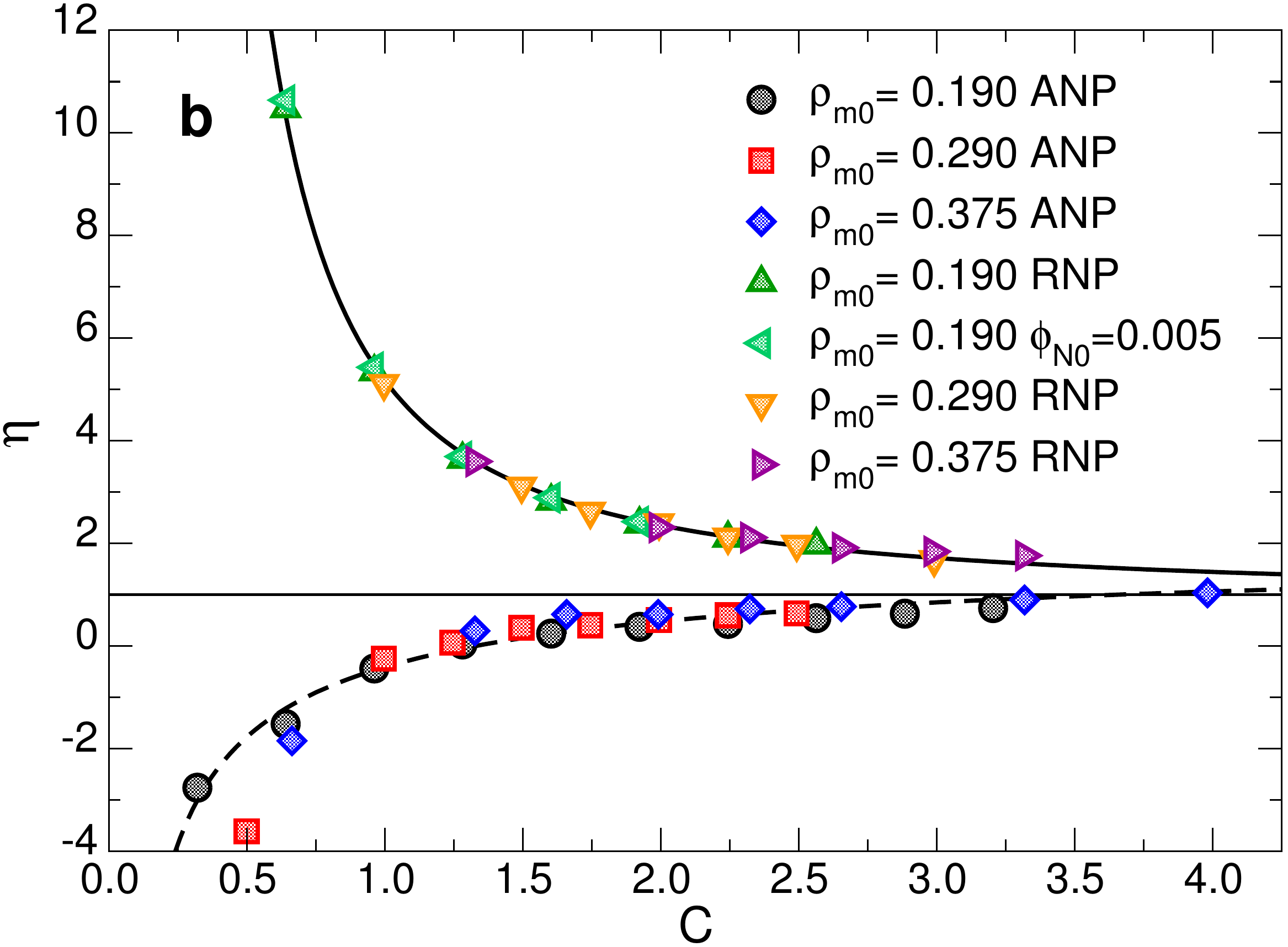}
\includegraphics[width=0.48 \textwidth]{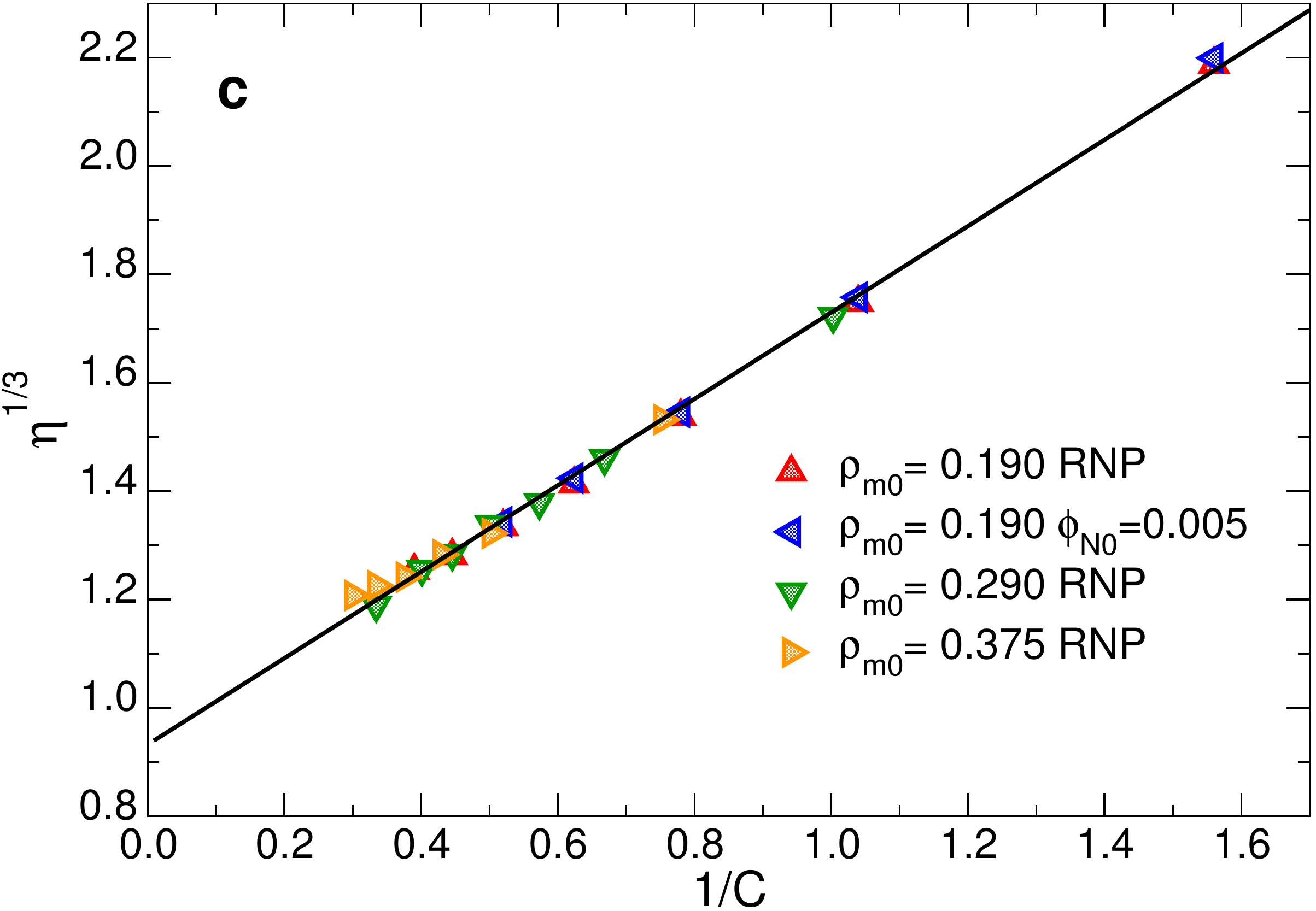}
\caption{(\textbf{a}): Mean mesh size (most probable pore radius) $\xi \equiv r_\text{max}$ as a function of $\sigma_N/2 \xi_0$, where $\xi_0$ is the mean mesh size of the neat system. The value of $\xi_0$ is $1.53$, $1.03$ and $0.79$ for $\rho_{m0}=0.190$, $0.290$ and $0.375$, respectively. Filled symbols: RNP. Open symbols: ANP. Solid lines: $\xi_0$. (\textbf{b}): Ratio $\eta \equiv (V-V_0)/V_N$ as a function of the confinement parameter $C$. Dashed line: $\eta=\alpha+\beta C^\gamma$, with $\alpha=2.16$, $\beta=-2.57$, and $\gamma=-0.61$. Solid line: same as in (c). (\textbf{c}): Representation of Eq.~\eqref{eq:eta_onethird}. Solid line: Linear fit, $\eta^{1/3} = 0.93+0.80 \times C^{-1}$.}
\label{fig:swell}
\end{figure}

The addition of the NP to the neat system makes that the density of the latter changes. In Fig.~\ref{fig:swell}a we show for all simulated systems with $\phi_{N0}=0.02$ the mean mesh size $\xi \equiv r_\text{max}$ as a function of the NP diameter $\sigma_N$ divided by $2\xi_0$, with $\xi_0$ the mesh size of the neat system (represented by the horizontal solid lines). Naively one could expect that if a volume fraction $\phi_{N0}=0.02$ of NPs is added to the network, the mesh size would increase by a factor of  $(1+\phi_{N0})^{1/3} = 1.0067 \simeq 1$. For $\sigma_N/2\xi_0>1$, we observe indeed an increase of $\xi$ for both the ANPs and the RNPs, although somewhat larger than this value. However, for values of $\sigma_N/2\xi_0<1$, a reduction of $\xi$ is observed for the ANPs, \textit{i.e.} the network is shrinking, whereas for the RNP a strong increase of $\xi$ is found (network swelling). Furthermore we find that for the ANPs, the maximum swelling is reached for $\sigma_N \simeq 2\xi_0$. These observations can be understood intuitively as follows: Small NPs are sufficiently small so that their excluded volume does not perturb the local mesh. However, they have, for a fixed NP volume fraction ($\phi_{N0}=0.02$), a much larger total surface than large NPs, $S_N =  \pi \sigma_N^2 N_N = 6 \phi_{N0} / \sigma_N V_0$, and therefore give rise to a significant change in the total energy of the system (here $V_0$ is the volume of neat system). For the case of the ANP, the NP-monomer interaction gives rise to an effective attraction between the network strands, and thus to a shrinkage of the system, while for the RNP the effective interaction is repulsive, leading to an expansion of the network~\cite{sorichetti2018structure}. 

Large NPs also perturb the local mesh, as one can see from the secondary peak in Fig.~\ref{fig:psd_rho02}, but have a limited effect on the network as a whole, since their total surface is much smaller at fixed volume fraction, as one can conclude from the fact that the main peak in Figs.~\ref{fig:psd_rho02}a and b remains basically unperturbed, \textit{i.e.} $\xi \simeq \xi_0$. 

In Fig.~\ref{fig:swell}b we report the quantity 

\begin{equation}
\eta \equiv (V-V_0)/V_N
\end{equation}

\noindent
as a function of the confinement parameter $C$, where $V$ is the volume of the filled system, and $V_N\equiv \pi \sigma_N^3 N_N /6$ is the total NP volume. The choice of this quantity is motivated by the fact that, if one approximates the network as an incompressible medium that expands uniformly due to the excluded volume of the NPs, one would have $\eta \simeq 1$
(we recall that the NP volume fraction with respect to $V_0$, $\phi_{N0}\equiv V_N/V_0$, is constant, $\phi_{N0}=0.02$). The graph demonstrates that for both types of NP, $\eta$ depends only on $C$, and it is independent of the network's density. Hence one can conclude that the confinement parameter controls the swelling of the network. This result is highly non-trivial in that it hints that $C$ is not only the relevant parameter for the dynamics, but also for the static properties of the system. If this conclusion is indeed true, it justifies the use of theoretical approaches that use the static properties of the system in order to predict its dynamics. 

Fig.~\ref{fig:swell}b also shows that the swelling/shrinking induced by the NPs is, in absolute value, significantly different from the simple estimate $V_N$ (horizontal line $\eta=1$ in Fig.~\ref{fig:swell}b) for intermediate and small values of $C$. For the RNPs, the behavior of $\eta(C)$ can be interpreted as follows: We have seen how the RNPs have a larger effective diameter than the ANPs. This is also seen in the behavior of the pore size distribution, since the presence of the NPs creates pores of diameter $\simeq \sigma_N+\delta$ (Fig.~4 in the main text and Sec.~\ref{sec:psd_si}). The expansion caused by the RNPs can therefore estimated to be

\begin{equation}
V = V_0 + V_N \left(1+\frac{\delta}{\sigma_N}\right)^3 \ \rightarrow \ \eta^{1/3} = \left(1+\frac{\delta}{\lambda} C^{-1}\right).
\label{eq:eta_onethird}
\end{equation}

In Fig.~\ref{fig:swell}c, we plot $\eta^{1/3}$ as a function of $C^{-1}$ for the RNPs: The data are well described by a straight line, as expected from Eq.~(\ref{eq:eta_onethird}); In particular, we find with good accuracy $\eta^{1/3} = 0.93+0.80 \times C^{-1}$, which implies $\delta = 0.80\times \lambda$, which is in qualitative agreement with the pore size distribution data, since one finds that $\delta$ decreases (like $\lambda$) with increasing network density (Sec.~\ref{sec:psd_si}). We note that the intercept is slightly smaller than unity, which is not surprising since the network is not incompressible.

For the ANPs, one can fit $\eta(C)$ by a (phenomenological) power-law, see solid line in Fig.~\ref{fig:swell}b, the origin of which is currently unknown (the values of the parameters are given in the caption of Fig.~\ref{fig:swell}). For values of $C$ corresponding to extreme confinement we find that $\eta$ is 1.4 for the RNP and very close to 1.0 for the ANP. We can however expect that these limits might depend on the details of the architecture of the gel and hence are not universal.  

%%%%%%%%%%%%%%%%%%%%%%%%%%%%%%%%%%%%%%%%%%%
\subsection{NP radial distribution function} \label{sec:rdf_si}
%%%%%%%%%%%%%%%%%%%%%%%%%%%%%%%%%%%%%%%%%%%

\begin{figure}
\centering
\includegraphics[width=0.48 \textwidth]{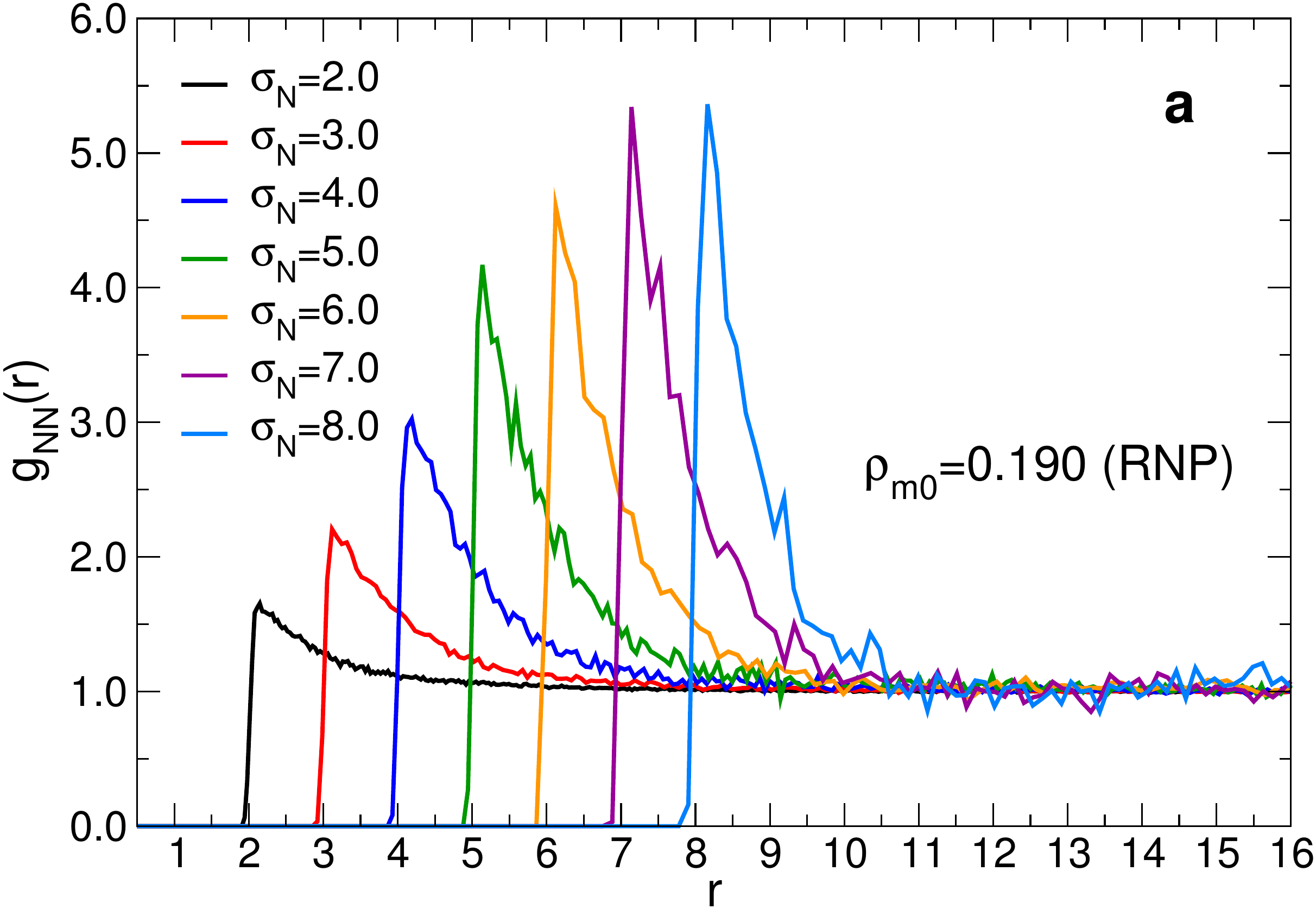}
\includegraphics[width=0.48 \textwidth]{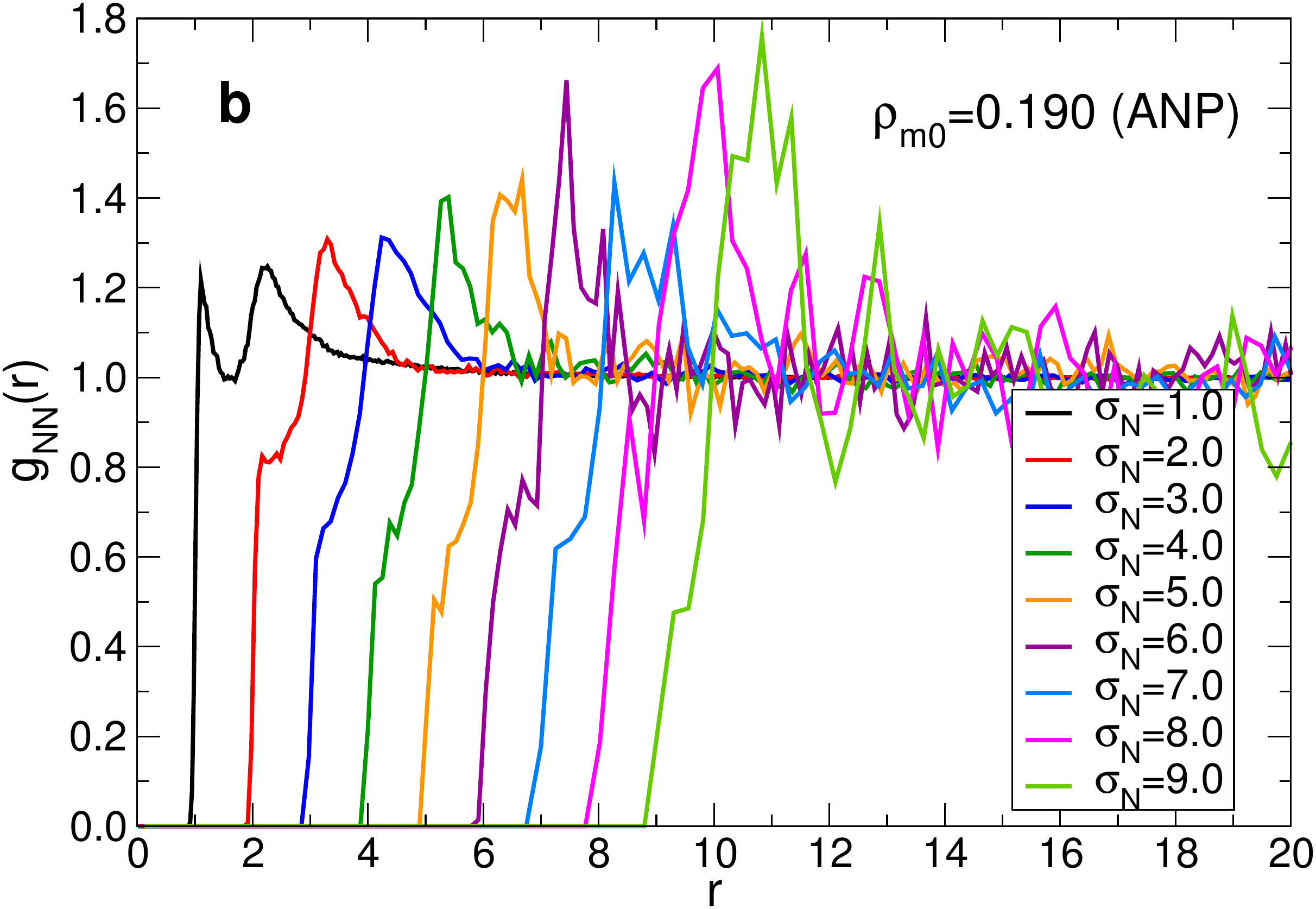}
\includegraphics[width=0.48 \textwidth]{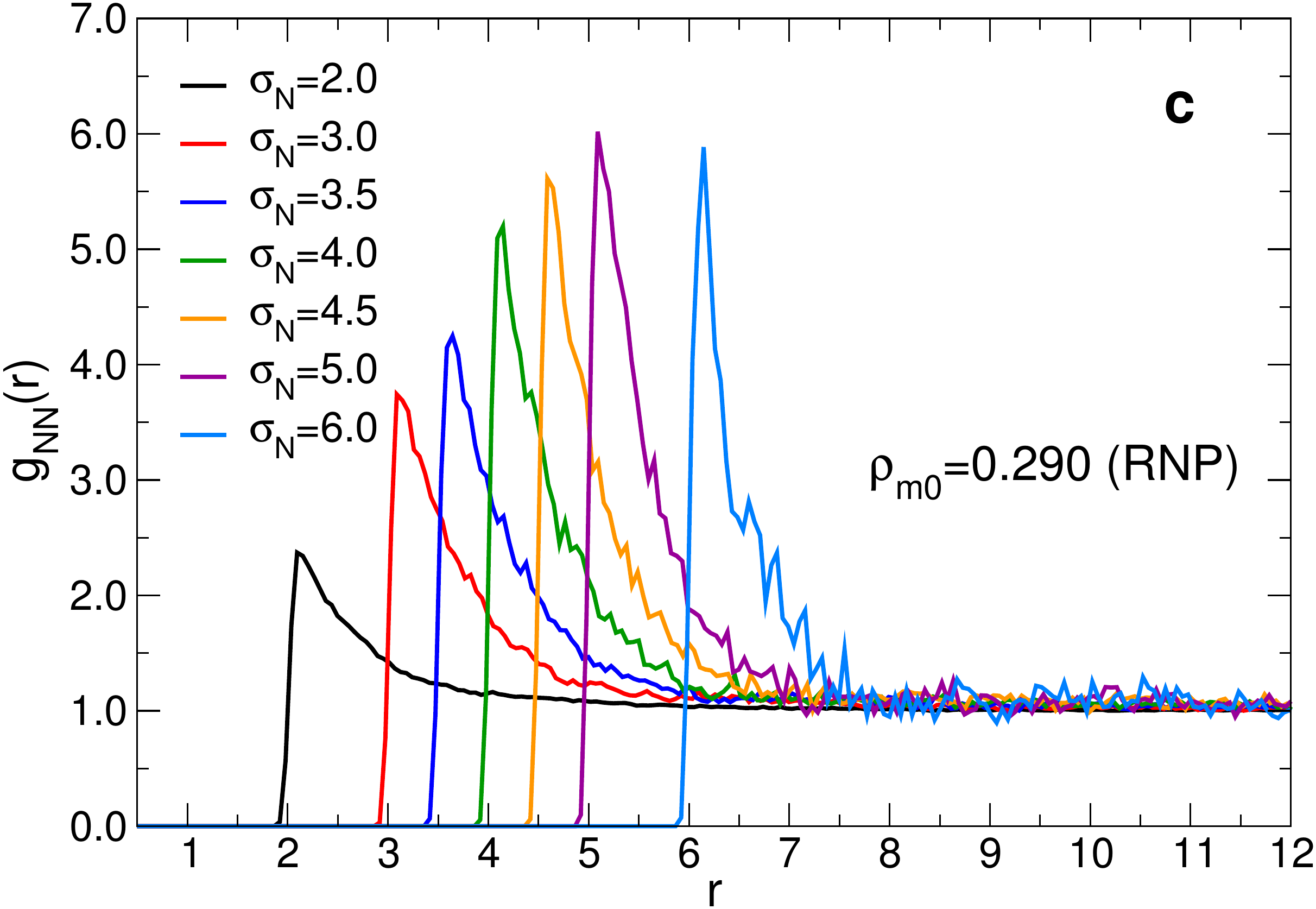}
\includegraphics[width=0.48 \textwidth]{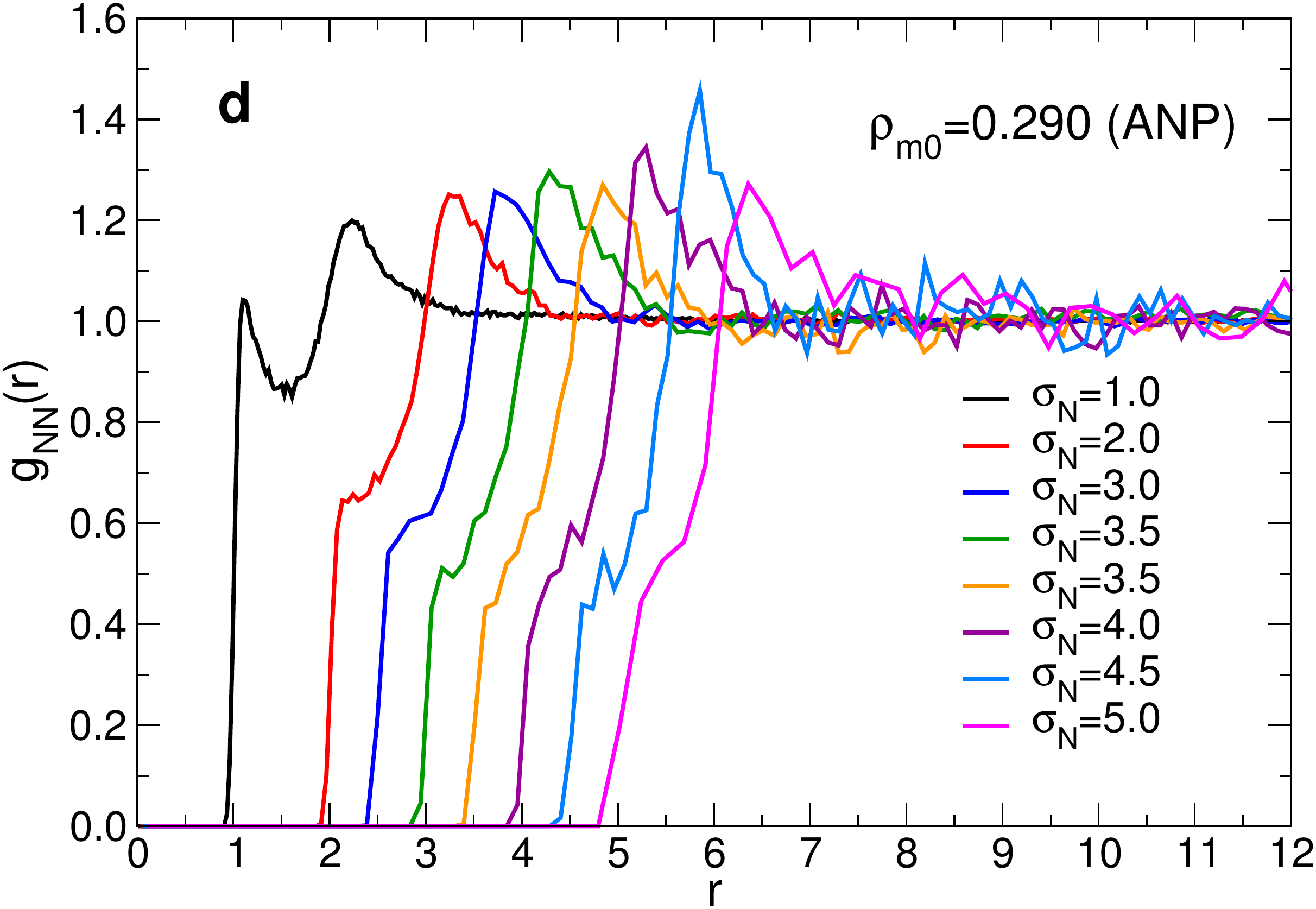}
\includegraphics[width=0.48 \textwidth]{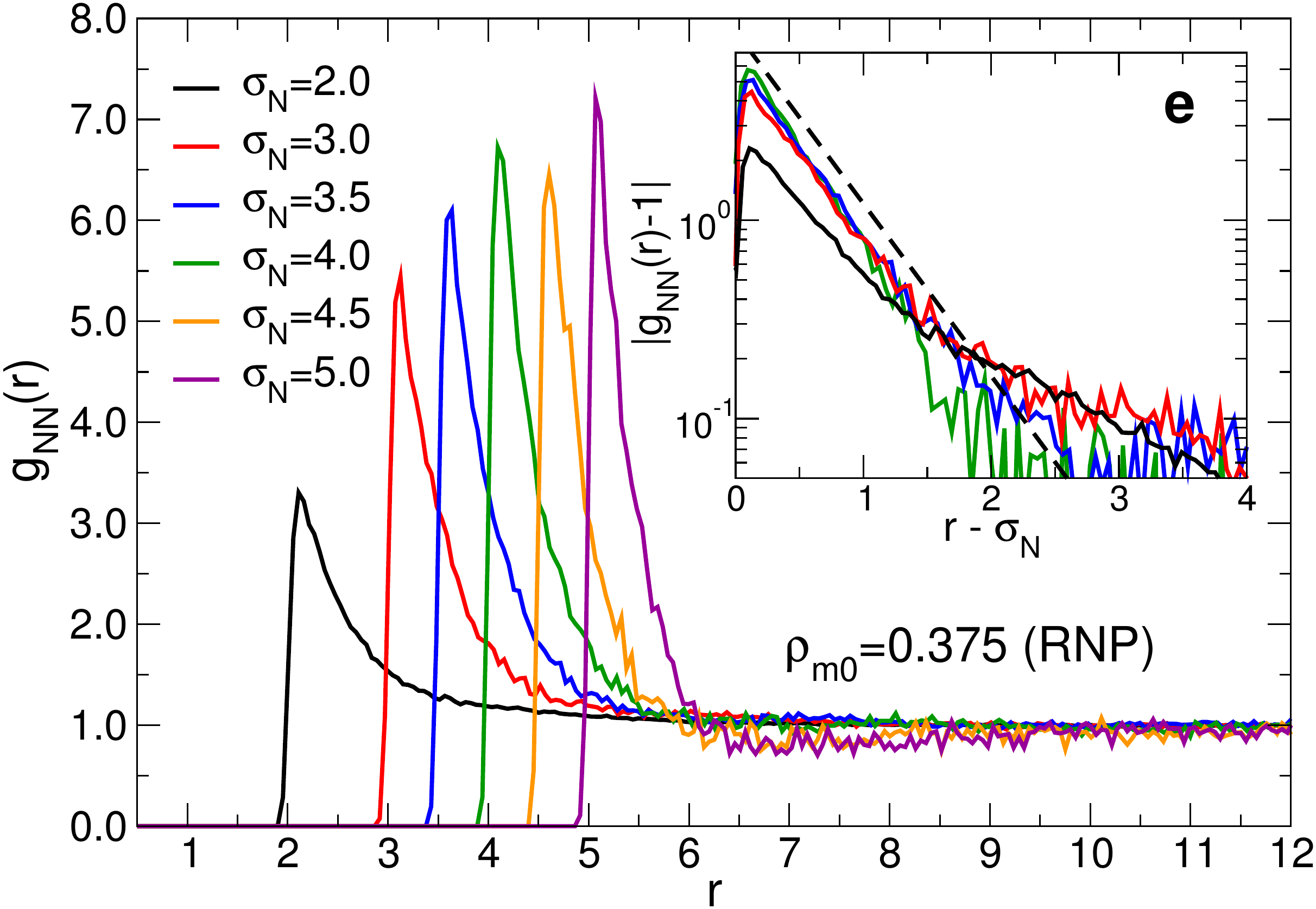}
\includegraphics[width=0.48 \textwidth]{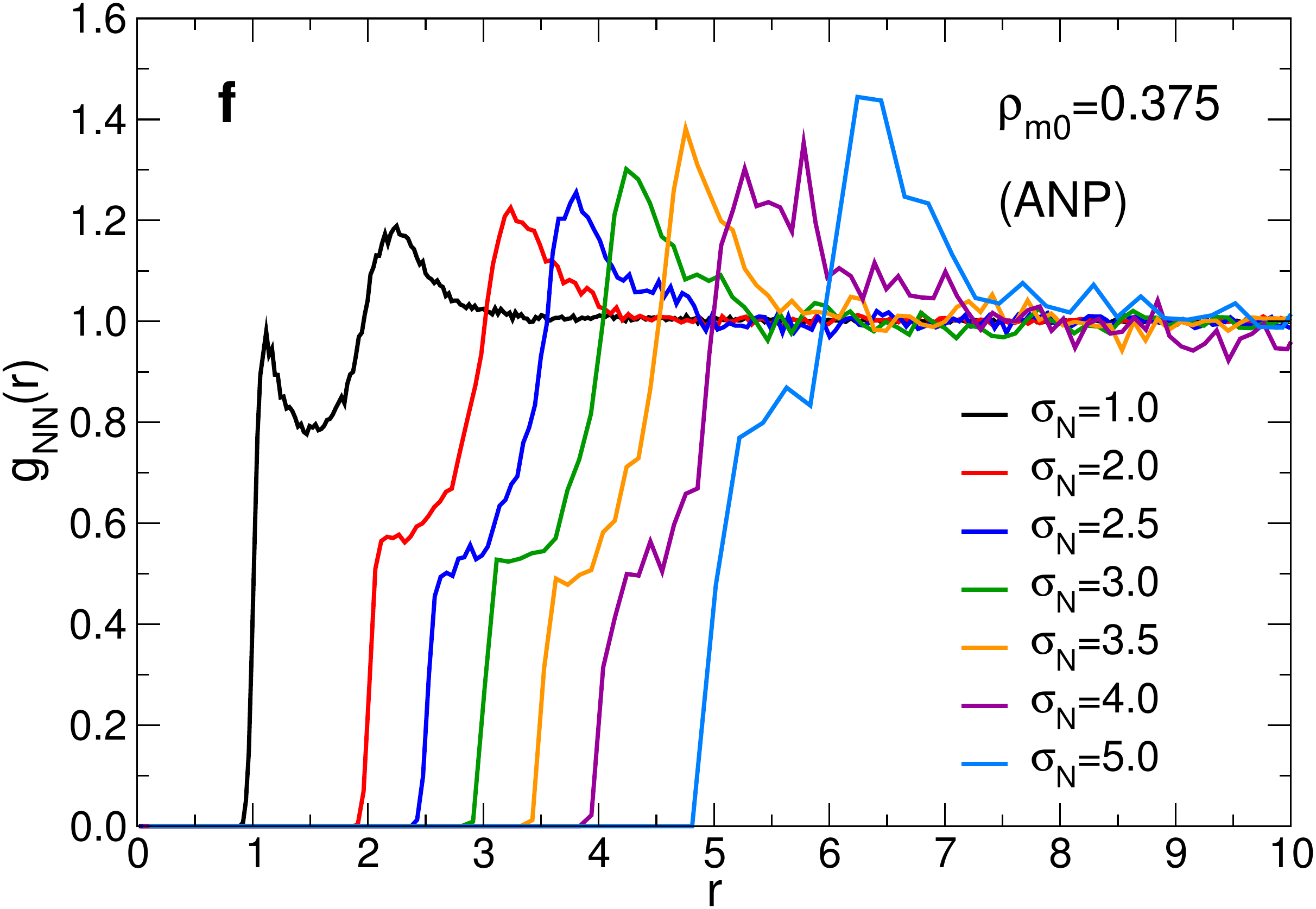}
\caption{NP radial distribution function for the various simulated systems (see labels). Inset of Fig.~{\bf (e)}: $|g_{NN}(r)-1|$ as a function of $r-\sigma_N$. Dashed line: Exponential with slope $-2$.
}
\label{fig:np_rdf_si}
\end{figure}

In the main text we have discussed the radial distribution function for different values of $\sigma_N$ at a given density. For the sake of completeness we show in Fig.~\ref{fig:np_rdf_si} $g_{NN}(r)$ for the other densities as well as further values of $\sigma_N$.

For the RNPs, the RDF displays a peak at the contact distance $r_c \equiv \sigma_N + (2^{1/6}-1) = \sigma_N + 0.122$. Since the height of this peak is rather large, one can deduce that the probability to find two NPs in contact with each other is significantly higher than in a NP fluid with the same density $\rho_N$. The decay of $g_{NN}$ on the right hand side of the peak is basically exponential, as shown in the inset of Fig.~\ref{fig:np_rdf_si}e, where we report $|g_{NN}(r)-1|$ as a function of $r-\sigma_N$ for $\rho_{m0}=0.375$ in a semi-logarithmic representation. We note that for $\sigma_N>2$ the shape of the peak, \textit{i.e.} the decay length, depends only very weakly on the NP diameter, as evidenced by the fact that data for different values of $\sigma_N$ fall almost on the same master curve. This result is thus in harmony with the discussion in the main text that RNP have the tendency to form small clusters and that the probability to find such a cluster with a given size decreases exponentially with its size.

\begin{figure}
\centering
\includegraphics[width=0.48 \textwidth]{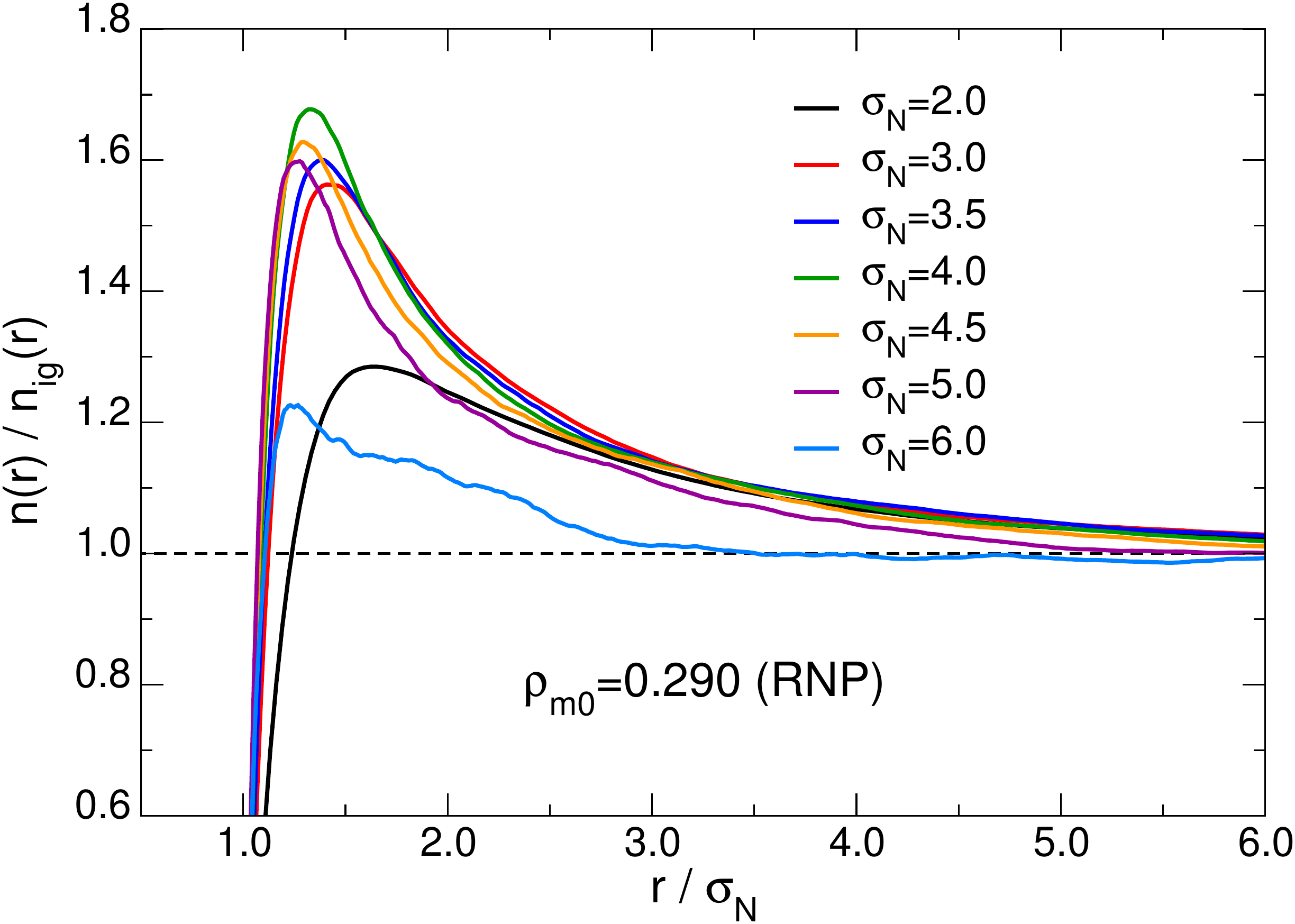}
\caption{Coordination number of the RNP (Eq.~\eqref{eq:coor}), divided by the ideal gas value (Eq.~\eqref{eq:coor_ig}) for $\rho_{m0}=0.290$ and different values of $\sigma_N$.}
\label{fig:np_coordination}
\end{figure}

These figures also demonstrate that the height of the peak of the RDF increases with increasing $\sigma_N$. This trend is simply due to the fact that $g(r)$ is divided by the monomer density $\rho_m$, and $\rho_m$ decreases with increasing $\sigma_N$ since $\phi_{N0}$ is kept fixed. In order to better compare the peak heights, it is therefore useful to consider the coordination number, defined as \cite{hansen1990theory}

\begin{equation}
n(r) \equiv 4 \pi \rho_m \int_0^r g(r') r'^2 dr'.
\label{eq:coor}
\end{equation}

\noindent
For the interpretation of the $r-$dependence of $n(r)$ it is useful to divide it by the same quantity calculated for an ideal gas:

\begin{equation}
n_\text{ig}(r) \equiv \frac 4 3 \pi \rho_m r^3 \quad .
\label{eq:coor_ig}
\end{equation}

\noindent
Thus the ratio $n(r)/n_\text{ig}(r)$ gives us the coordination number of the NPs normalized by the one for an ideal gas. In Fig.~\ref{fig:np_coordination}, we show this ratio for the case of the RNP at $\rho_{m0}=0.290$. One can see that for distances $1 < r/\sigma_N \lesssim 3$, this ratio is basically independent of $\sigma_N$ for $\sigma_N = 3.0, 3.5, 4.0$ and $5.0$, while it is smaller for $\sigma_N=2.0$ and $\sigma_N=6.0$. These two exceptions can be understood by arguing that small NP will be able to integrate into the matrix without much problem , \textit{i.e.} the particles are well dispersed. On the other hand very large NP will have difficulty to find holes in the mesh that are sufficiently large to host several NP, leading to a reduction of the coordination number. (We can, however, not exclude the possibility that the systems with very large NP are not fully equilibrated.) Overall, the absolute value of $n(r)/n_\text{ig}(r)$ remains modest, signaling that the NP clustering is not severe, \textit{i.e.}, only small transient clusters are formed.

For the ANPs, panels d-f, the main peak of the RDF shifts to $r= \sigma_N + (2^{7/6}-1) =\sigma_N+1.245$, which corresponds to a configuration in which two neighboring NPs are separated by a polymer strand \cite{sorichetti2018structure}. This is a clear signal of the fact that the NPs are well dispersed, since each NP is surrounded by a layer of polymers. Although a shoulder at the contact distance $r =\sigma_N + 0.122$ is still present, it reaches a height comparable with that of the main peak only for $\sigma_N=1$: This is a signal of the fact that for $\sigma_N=1$ even ANPs have a weak tendency to form clusters, which likely happens when several of them occupy the same hole in the mesh.

%%%%%%%%%%%%%%%%%%%%%%%%%%%%%%%%%%%%%%%%%%%
\subsection{Monomer structure factor}
%%%%%%%%%%%%%%%%%%%%%%%%%%%%%%%%%%%%%%%%%%%

\begin{figure}
\centering
\includegraphics[width=0.48 \textwidth]{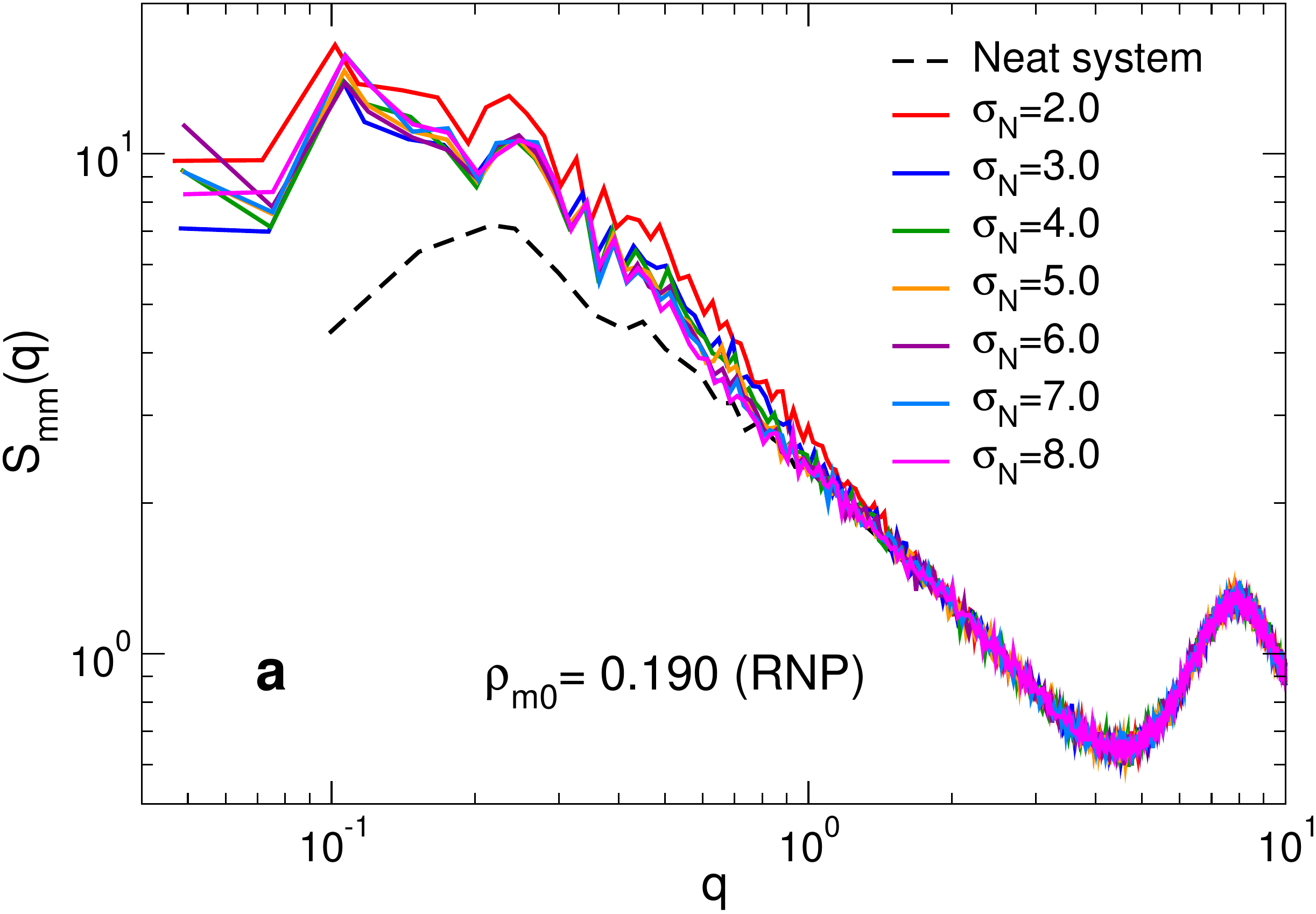}
\includegraphics[width=0.48 \textwidth]{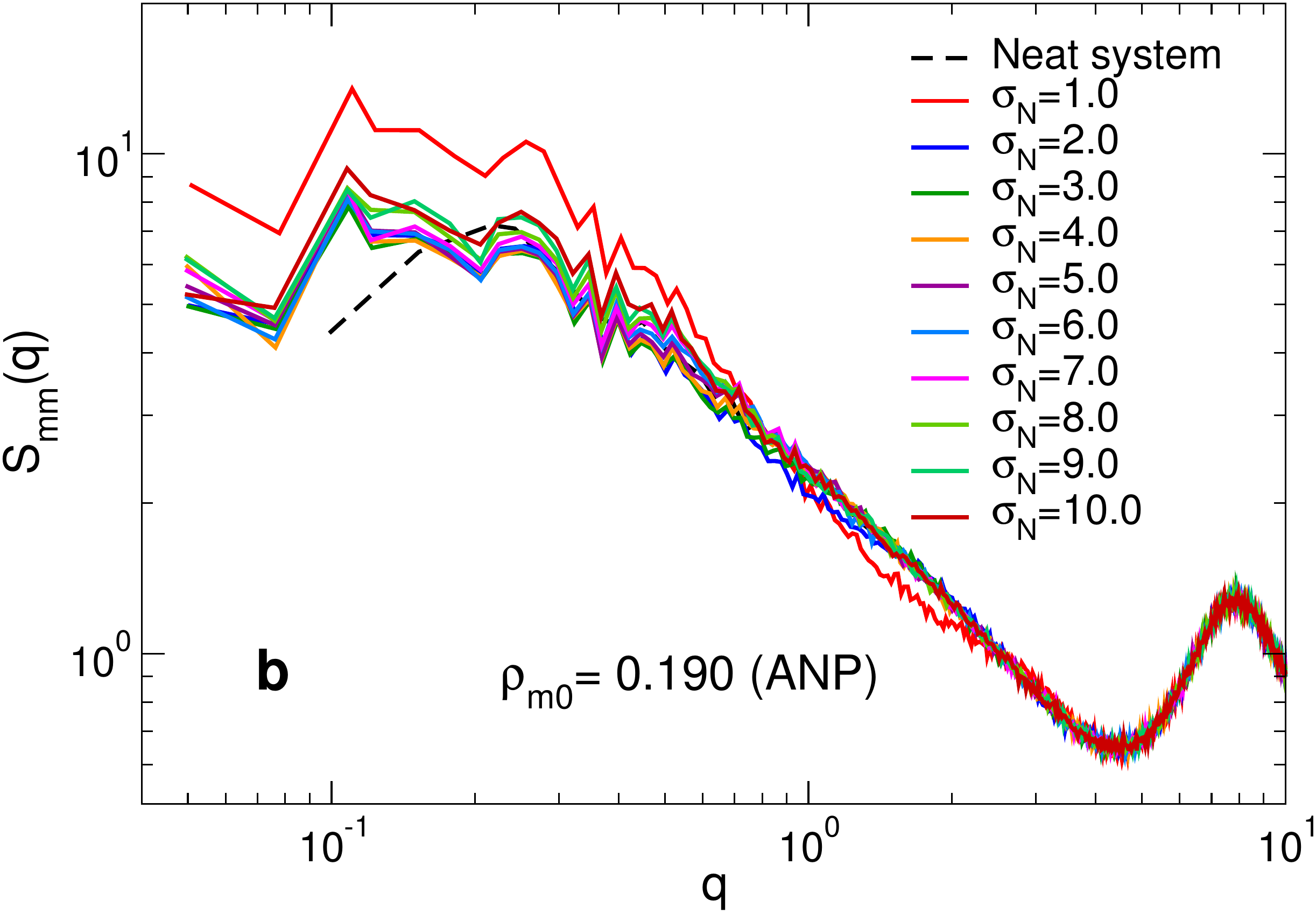}
\includegraphics[width=0.48 \textwidth]{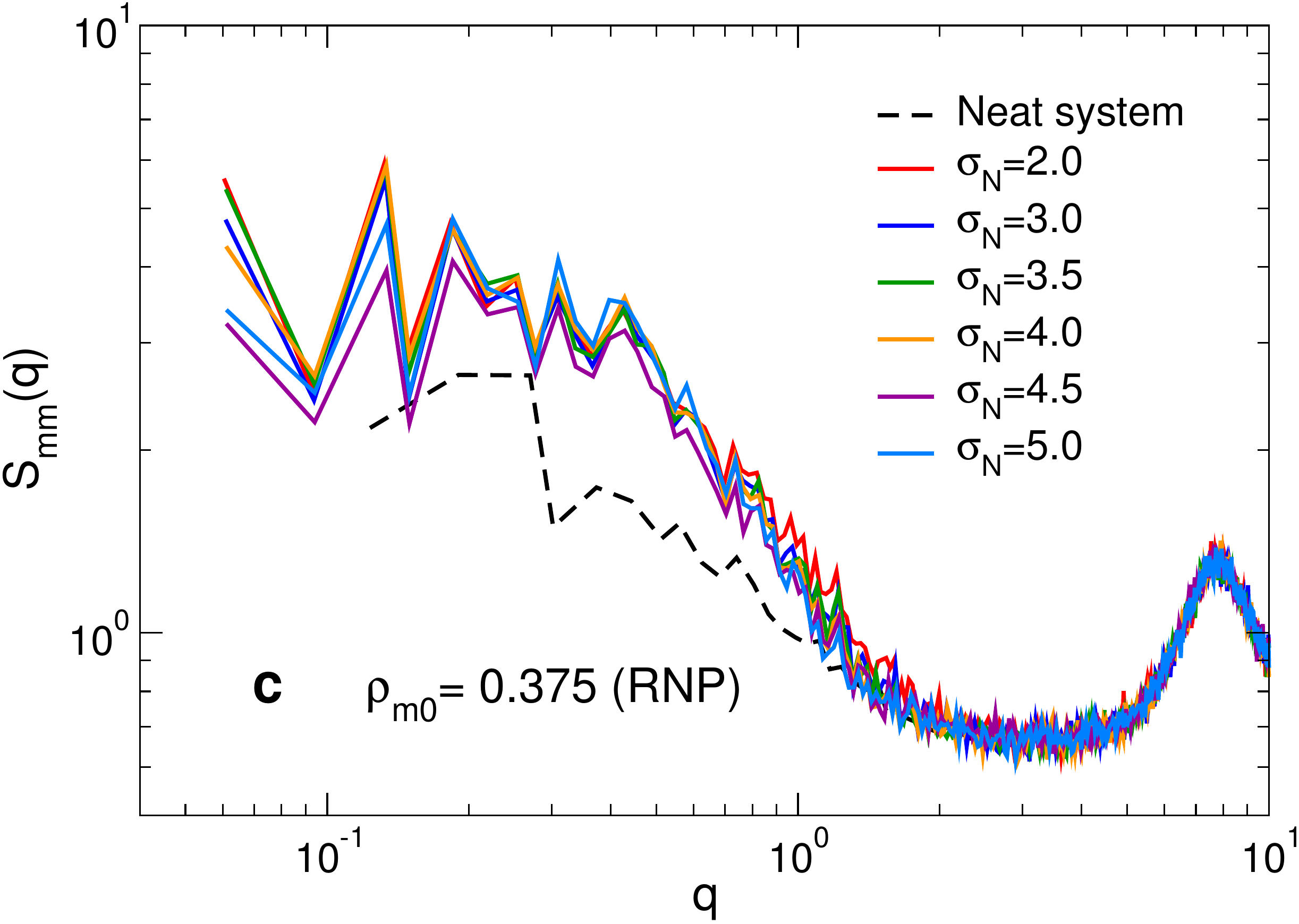}
\includegraphics[width=0.48 \textwidth]{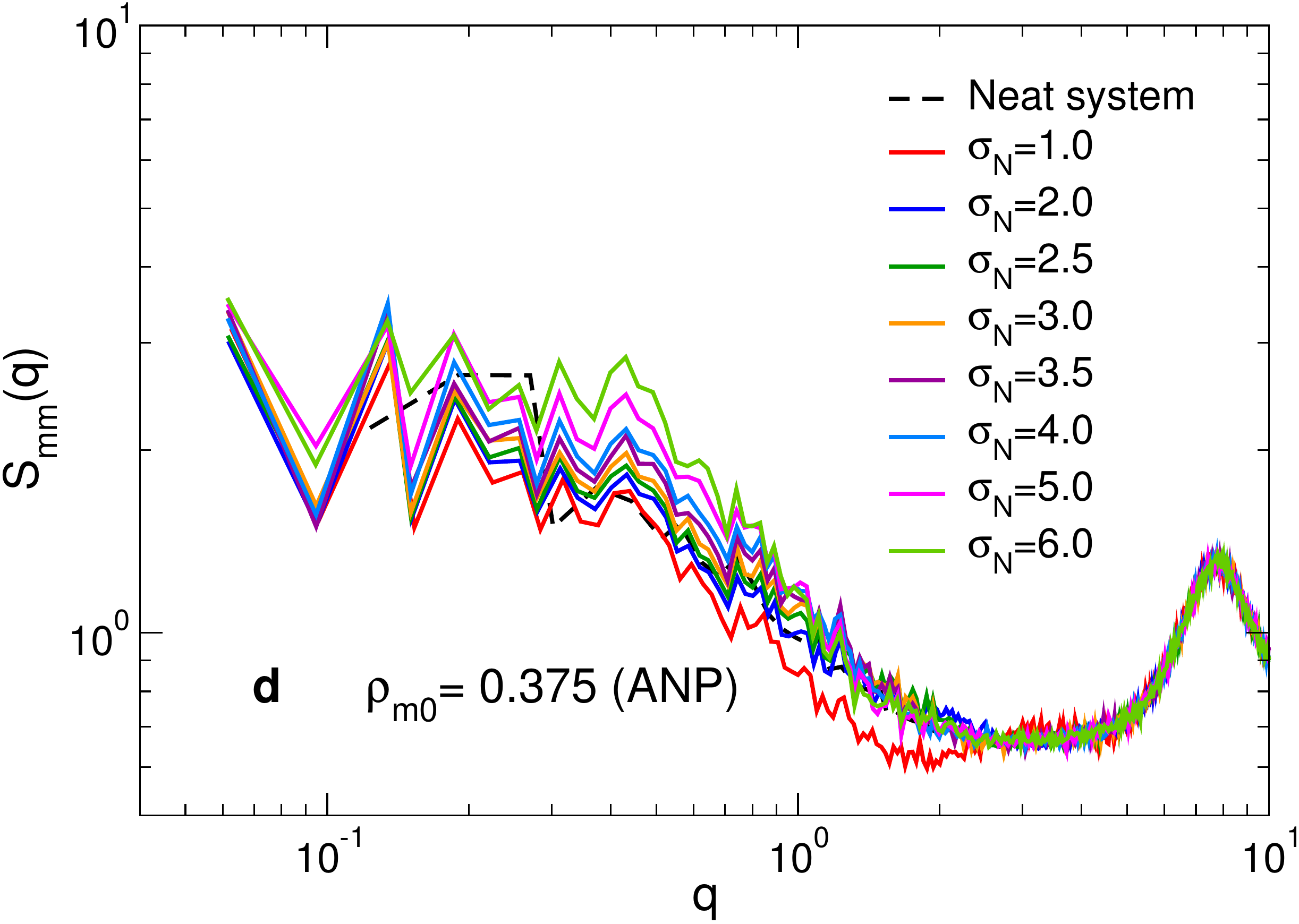}
\caption{Monomer structure factor. (a): $\rho_{m0}=0.190$, RNP (b): $\rho_{m0}=0.190$, ANP (c): $\rho_{m0}=0.375$, RNP (d): $\rho_{m0}=0.375$, ANP. Dashed line: Neat system ($N_m=5\times 10^4$).}
\label{fig:sq_mono}
\end{figure}

To study the structural properties of the system on large length scales it is useful to consider the static structure factor, as already defined in the main text. In Fig.~\ref{fig:sq_mono} we show the monomer structure factor $S_{mm}(q)$ for $\rho_{m0}=0.190$ (a-b) and $0.375$ (c-d), both for ANPs and RNPs. The data are compared with $S_{mm}(q)$ for a neat network with $N_m=5\times 10^4$ monomers. As also observed in the main text for $\rho_{m0}=0.290$, one can see that $S_{mm}(q)$ does not change much upon the addition of the ANPs, except if $\sigma_N$ is very small. This is consistent with the fact that ANPs are well dispersed, thus inducing only moderate deformation on the local mesh. The RNPs, on the other hand, induce a noticeable deformation of the mesh, causing an increase of $S_{mm}(q)$ at low $q$ which signals the appearance of larger holes and the swelling of the network, in agreement with the results shown in Fig.~\ref{fig:swell}b. This is due to the fact that RNPs partially aggregate to form small clusters, which deform the local mesh in a more substantial manner.

%%%%%%%%%%%%%%%%%%%%%%%%%%%%%%%%%%%%%%%%%%%
\subsection{NP structure factor} \label{sec:sq_si}
%%%%%%%%%%%%%%%%%%%%%%%%%%%%%%%%%%%%%%%%%%%

\begin{figure}
\centering
\includegraphics[width=0.48 \textwidth]{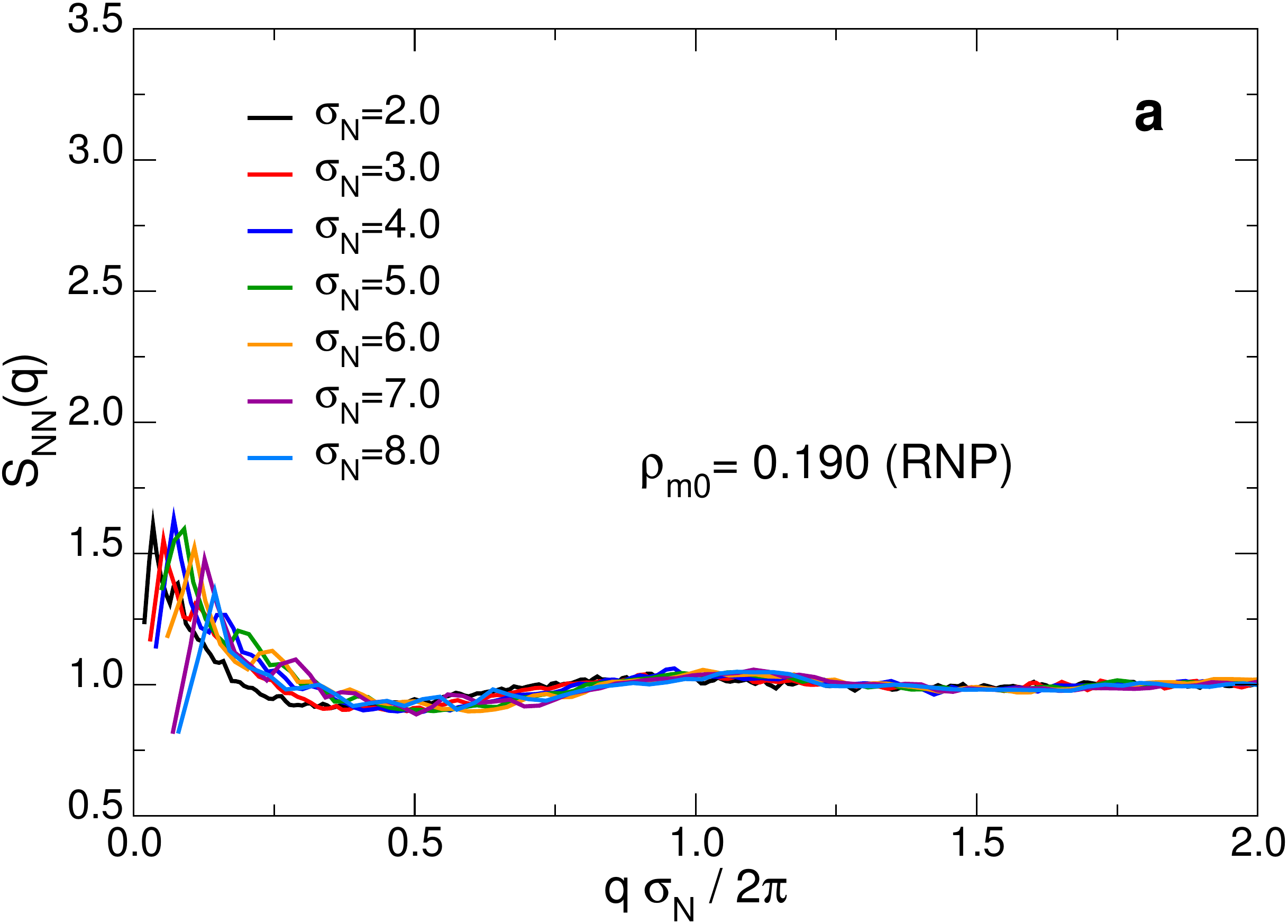}
\includegraphics[width=0.48 \textwidth]{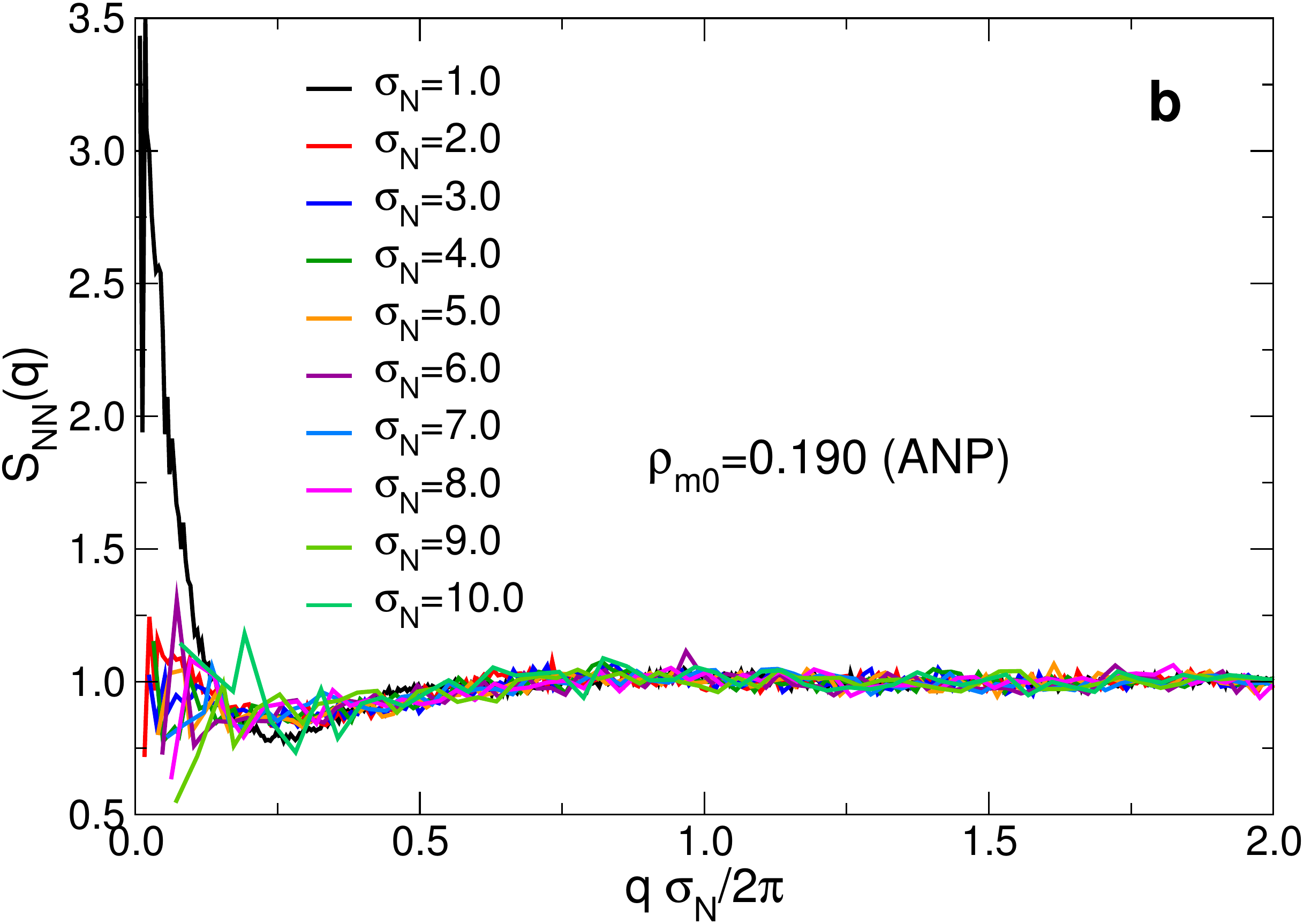}
\includegraphics[width=0.48 \textwidth]{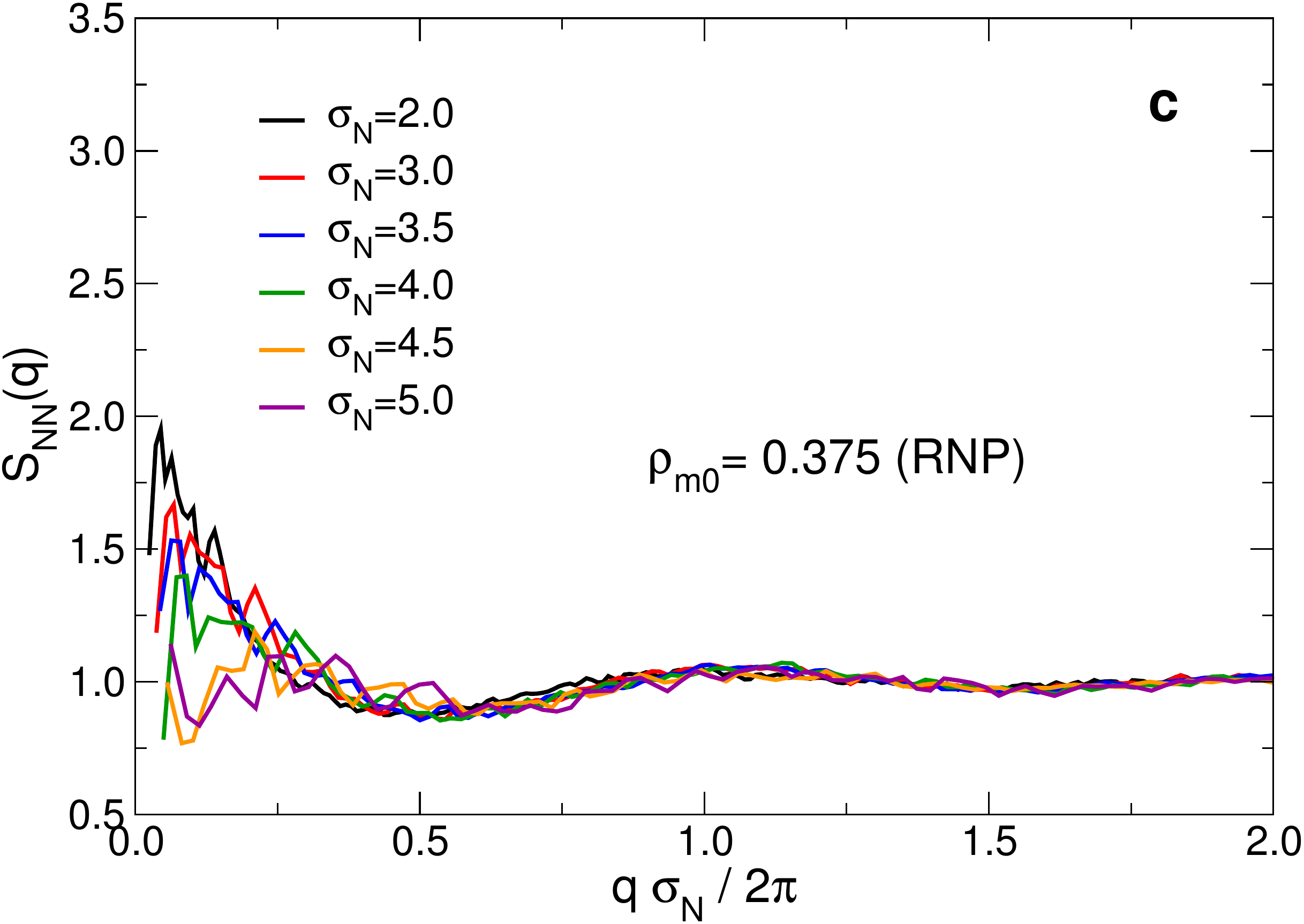}
\includegraphics[width=0.48 \textwidth]{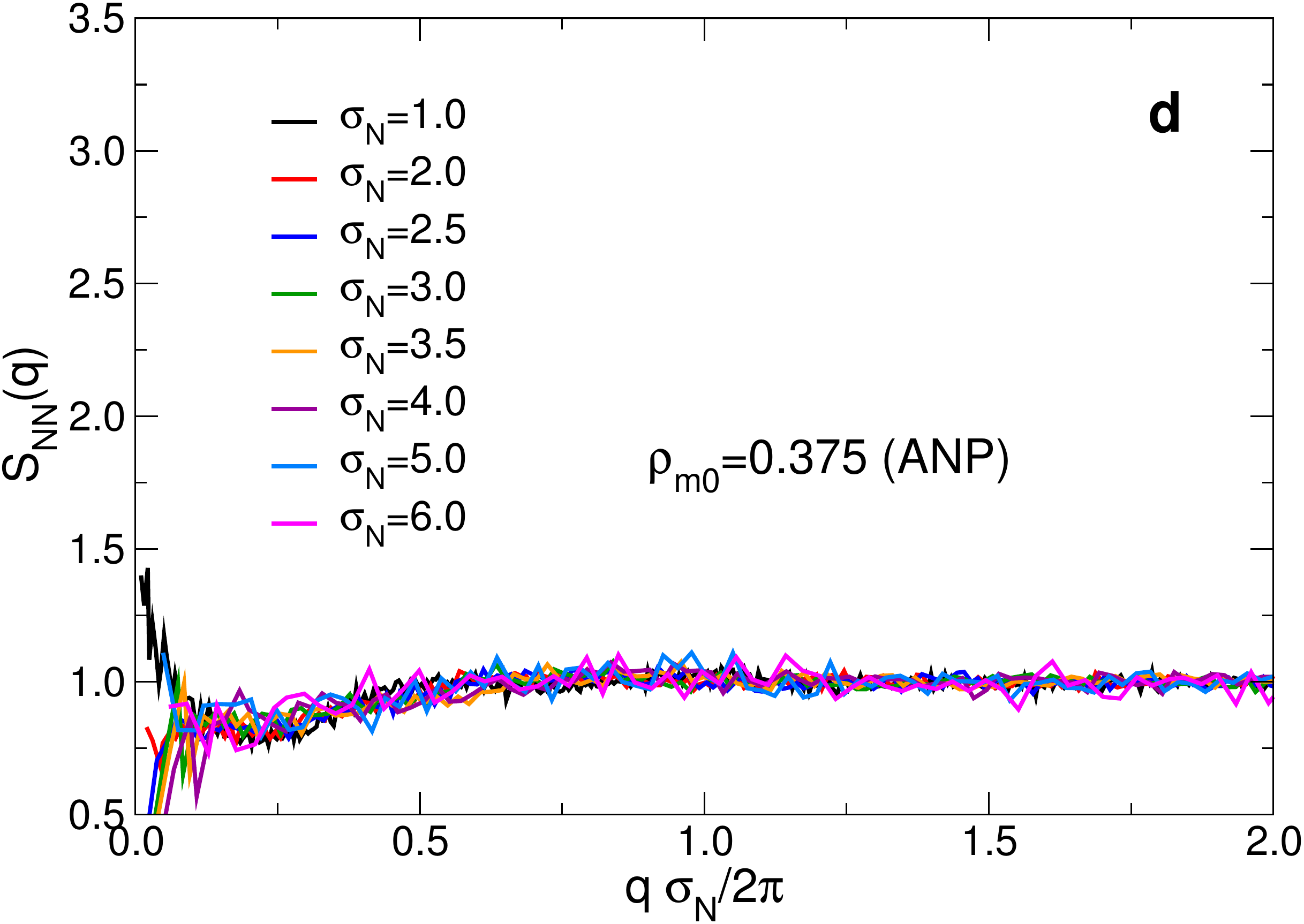}
\caption{NP-NP structure factor as a function of the rescaled wavevector $q \sigma_N /2 \pi$ for the RNPs (\textbf{a,c}) and the ANPs (\textbf{b,d}), for different values of $\sigma_N$ and $\rho_{m0}$.
}
\label{fig:np_sq_si}
\end{figure}

In Fig.~\ref{fig:np_sq_si} we present the NP structure factor $S_{NN}(q)$ as a function of the rescaled wavevector $q \sigma_N / 2 \pi$ for $\rho_{m0}=0.190$ and $0.375$, for both ANPs and RNPs. The $q-$dependence of $S_{NN}(q)$ is qualitatively the same as the one which was observed for $\rho_{m0}=0.290$ in the main text: For the RNPs, the data basically fall on a master curve, with a small peak at small $q$ which is due to partial NP clustering (as also observed from the RDF, see Sec.~\ref{sec:rdf_si}). In contrast to this one finds that for the higher density $\rho_{m0}=0.375$ the height of this peak depends significantly on $\sigma_N$, and the peak height increases with decreasing $\sigma_N$. This $\sigma_N-$dependence can be understood by realizing that with increasing density the typical size of the holes in the mesh decreases and hence the NP will have an increasing hard time to find holes that can host several NP. This has the consequence that the clustering of the NP is suppressed, \textit{i.e.} they become dispersed well in the gel, and hence the peak at small $q$ decreases.

Also for the ANPs the data fall on a master curve; in this case, the absence of a peak at $q \to 0$ confirms that the NPs are well dispersed. The only exception is the system with $\sigma_N=1$, for which clustering is observed, in agreement with what was observed from the RDF in Sec.~\ref{sec:rdf_si}.

%%%%%%%%%%%%%%%%%%%%%%%%%%%%%%%%%%%%%%%%%%%
\subsection{Pore size distribution} \label{sec:psd_si}
%%%%%%%%%%%%%%%%%%%%%%%%%%%%%%%%%%%%%%%%%%%

\begin{figure}
\centering
\includegraphics[width=0.48 \textwidth]{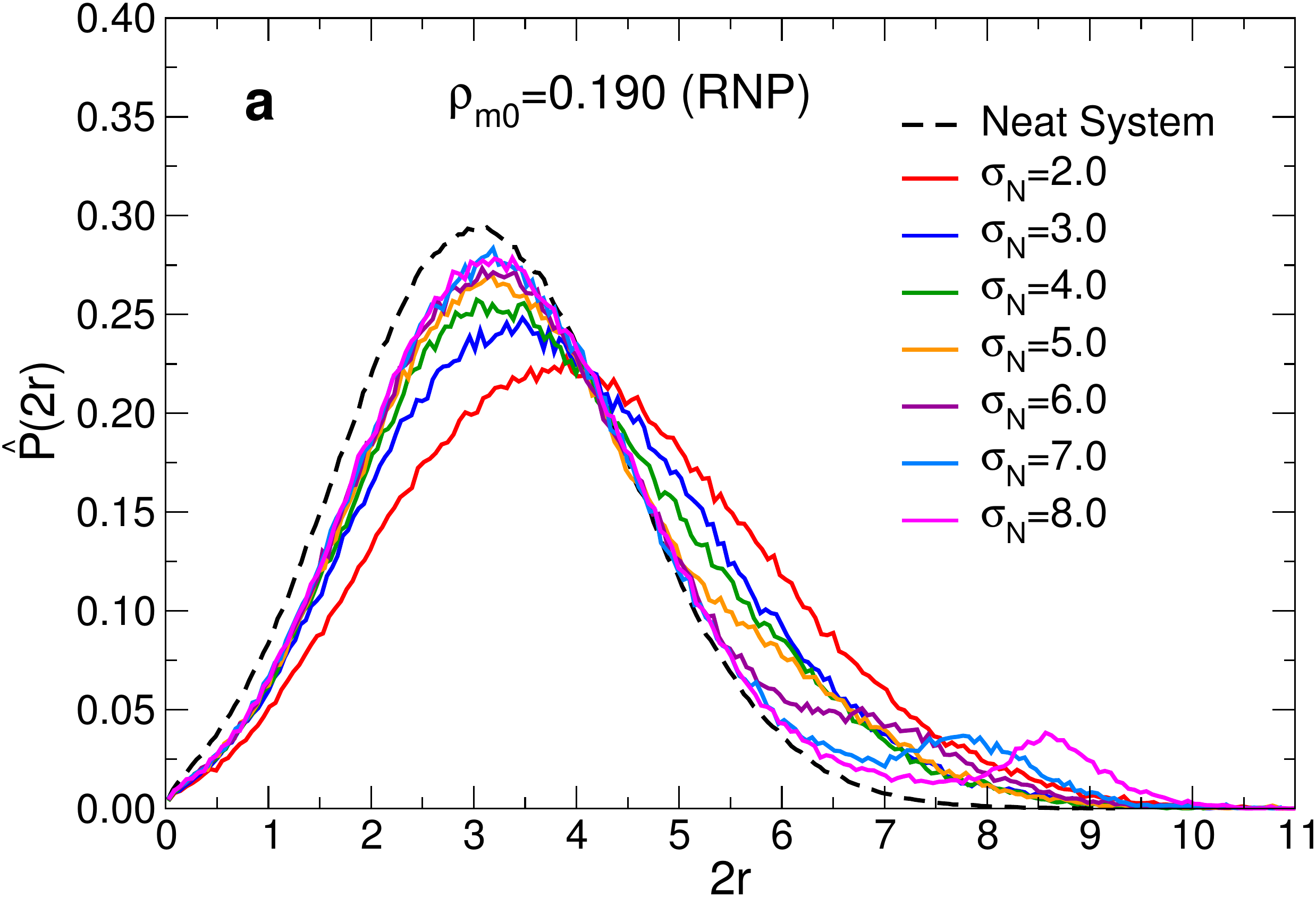}
\includegraphics[width=0.48 \textwidth]{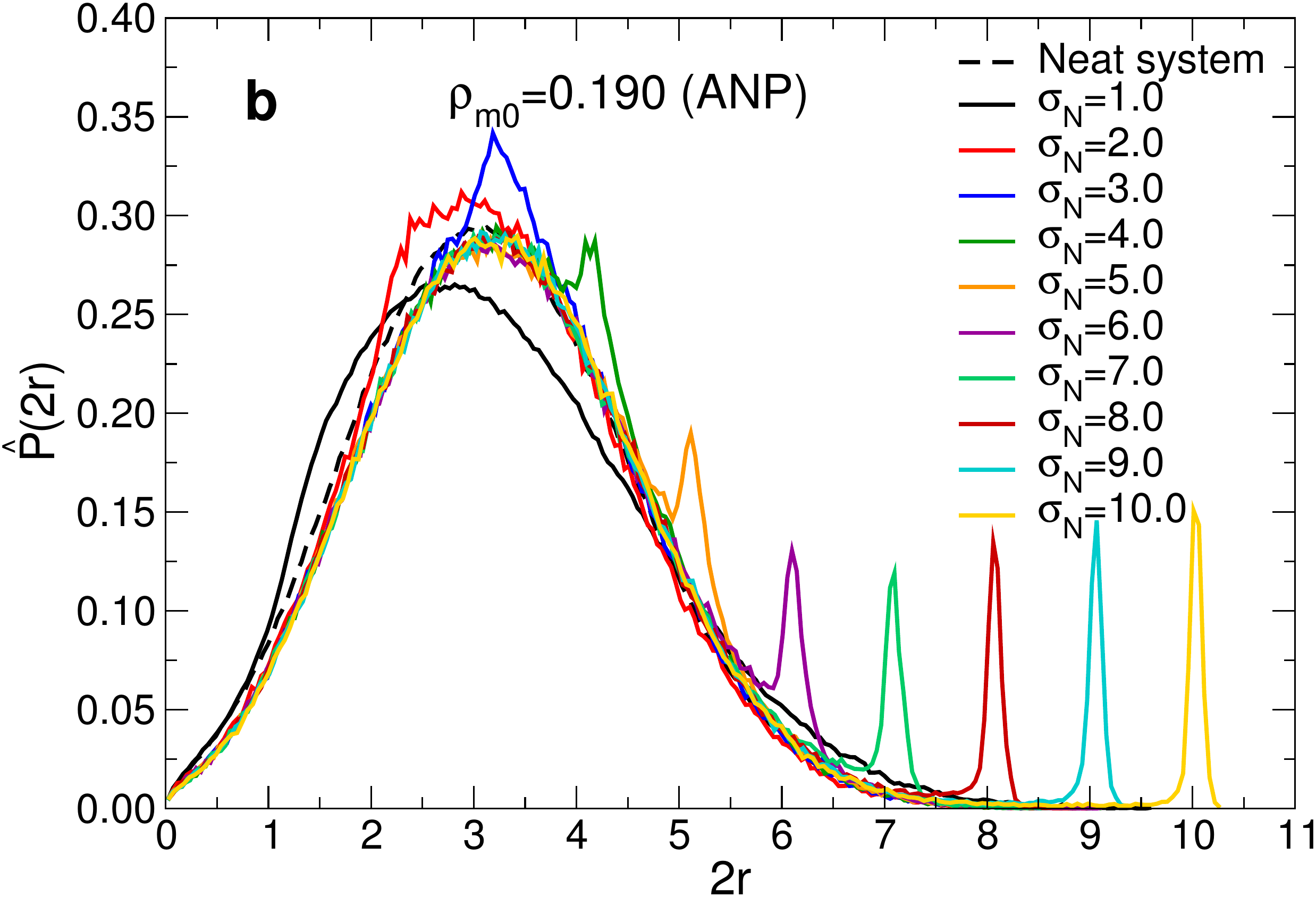}
\includegraphics[width=0.48 \textwidth]{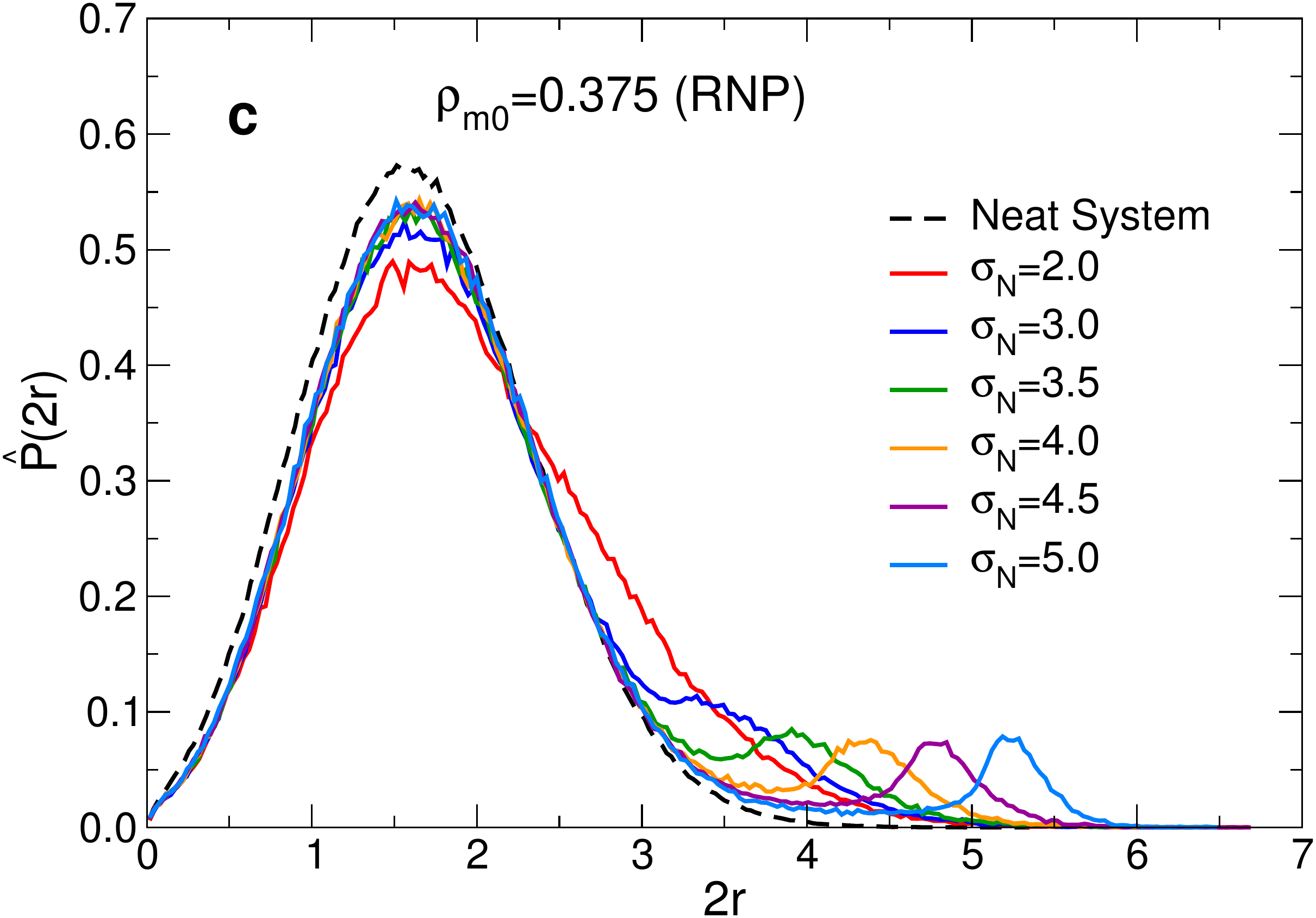}
\includegraphics[width=0.48 \textwidth]{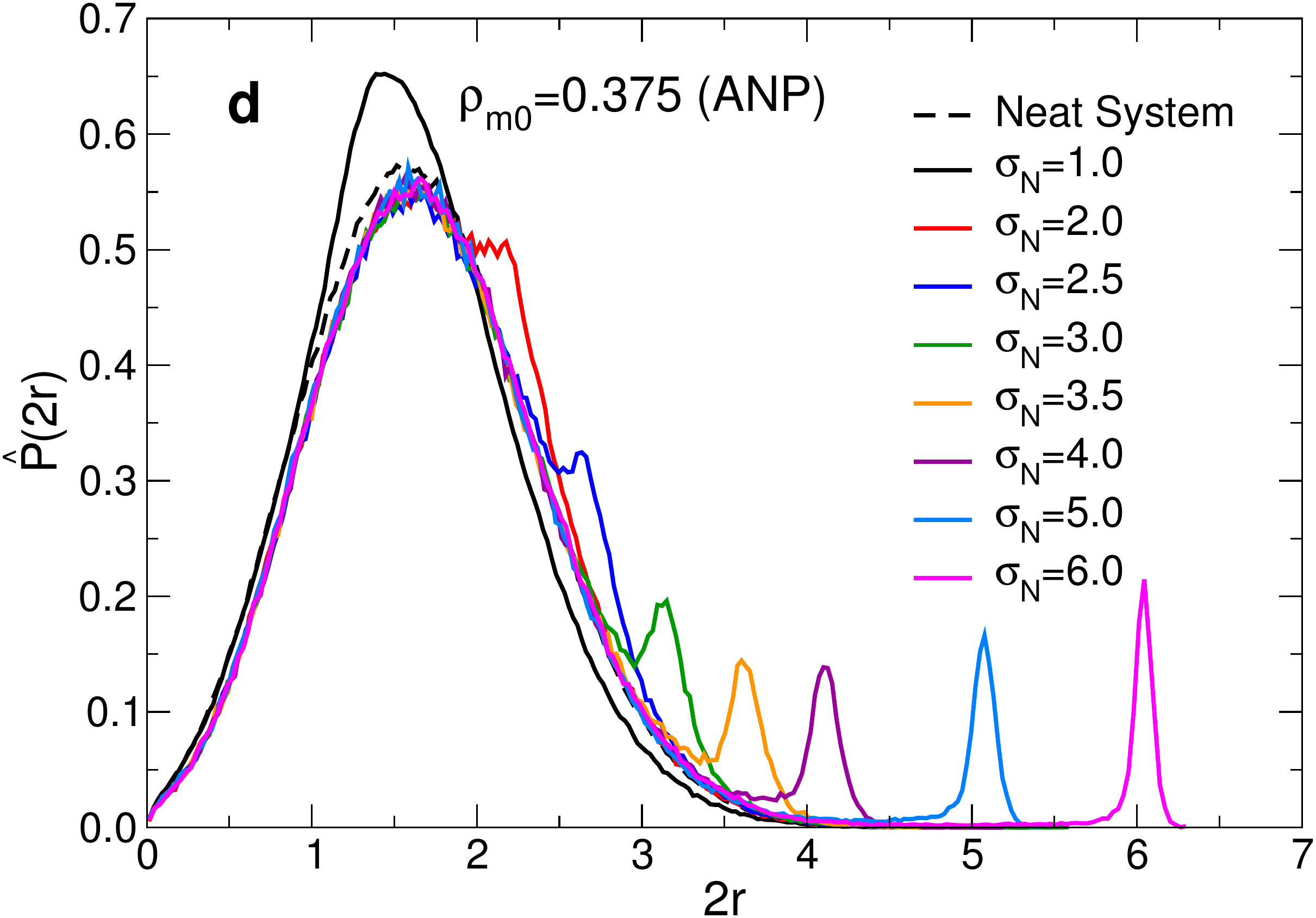}
\caption{Distribution of pore diameters, $\hat P(2r)$ for the RNPs (\textbf{a,c}) and the ANPs (\textbf{b,d}) for different values of $\sigma_N$ and $\rho_{m0}$.
}
\label{fig:psd_si}
\end{figure}

In Fig.~\ref{fig:psd_si} we report the pore size distribution distribution (PSD) of the pore diameters, $\hat P(2r)$, for the RNPs (\textbf{a,c}) and the ANPs (\textbf{b,d}) for different values of $\sigma_N$ and $\rho_{m0}=0.190$ (\textbf{a,b}) and $0.375$ (\textbf{c,d}). We observe a behavior which is qualitatively similar to the one shown in the main text for $\rho_{m0}=0.290$ (Fig.~\ref{fig:psd_rho02}): The NPs create  cavities in the network, resulting in peaks in the PSD located at $2r=\sigma_N$ for the ANPs and at $2r \simeq \sigma_N+\delta$ for the RNPs, where the value of $\delta$ decreases slightly with increasing network density. For the RNPs, the peak is significantly broader than for the ANPs with the same diameter, due to the fact that the RNPs repel the strands and deform locally the network. We refer to the main text for a more detailed discussion of these features. Note that the distribution becomes more narrow if the density is increased, a dependence that indicates that on average denser systems have smaller pores, in agreement with naive expectations.

%%%%%%%%%%%%%%%%%%%%%%%%%%%%%%%%%%%%%%%%%%%
%%%%%%%%%%%%%%%%%%%%%%%%%%%%%%%%%%%%%%%%%%%
\section{Dynamics}
%%%%%%%%%%%%%%%%%%%%%%%%%%%%%%%%%%%%%%%%%%%
%%%%%%%%%%%%%%%%%%%%%%%%%%%%%%%%%%%%%%%%%%%

%%%%%%%%%%%%%%%%%%%%%%%%%%%%%%%%%%%%%%%%%%%
\subsection{Subdiffusive regime: onset and apparent exponent}
%%%%%%%%%%%%%%%%%%%%%%%%%%%%%%%%%%%%%%%%%%%

\begin{figure}
\centering
\includegraphics[width=0.48 \textwidth]{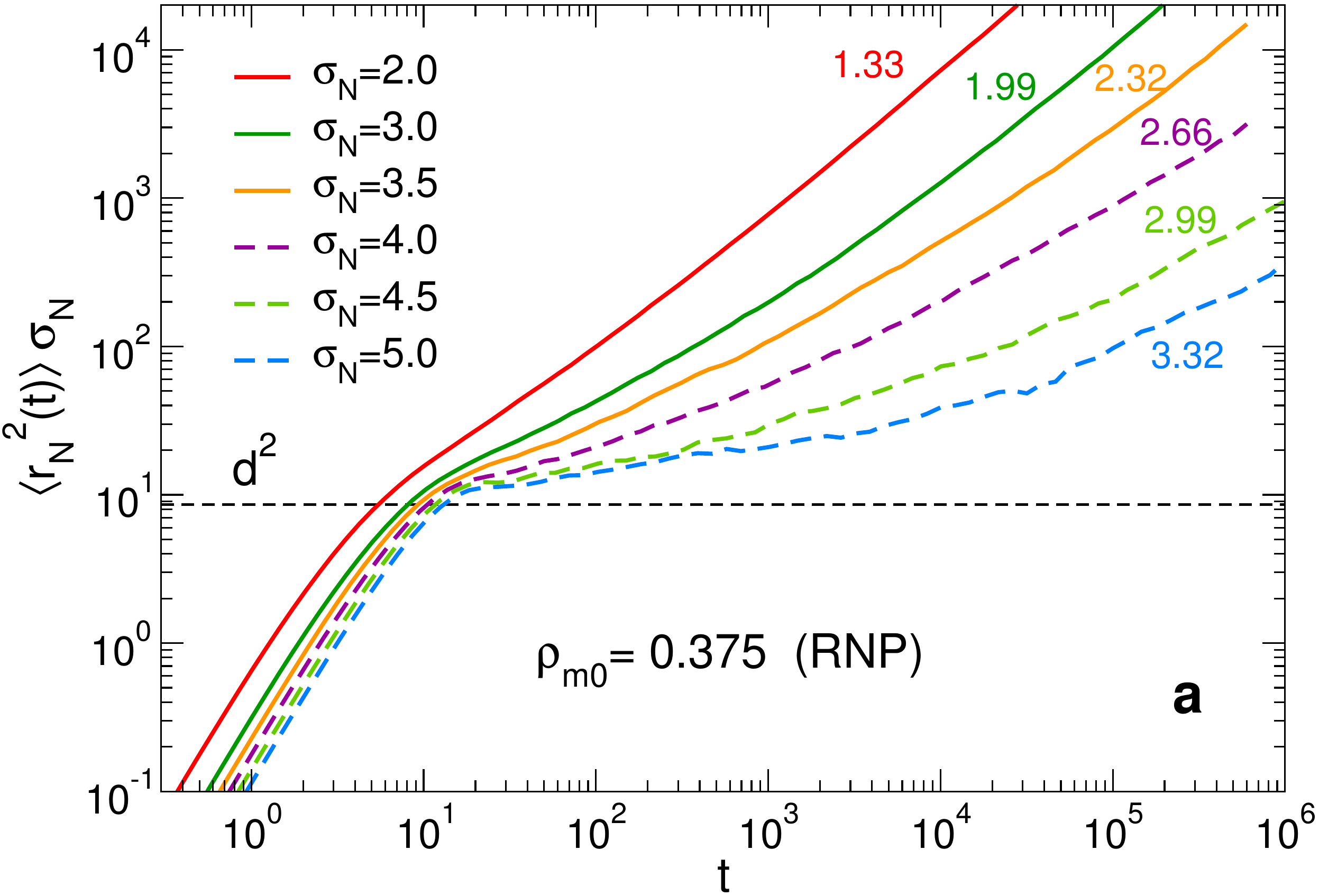}
\includegraphics[width=0.48 \textwidth]{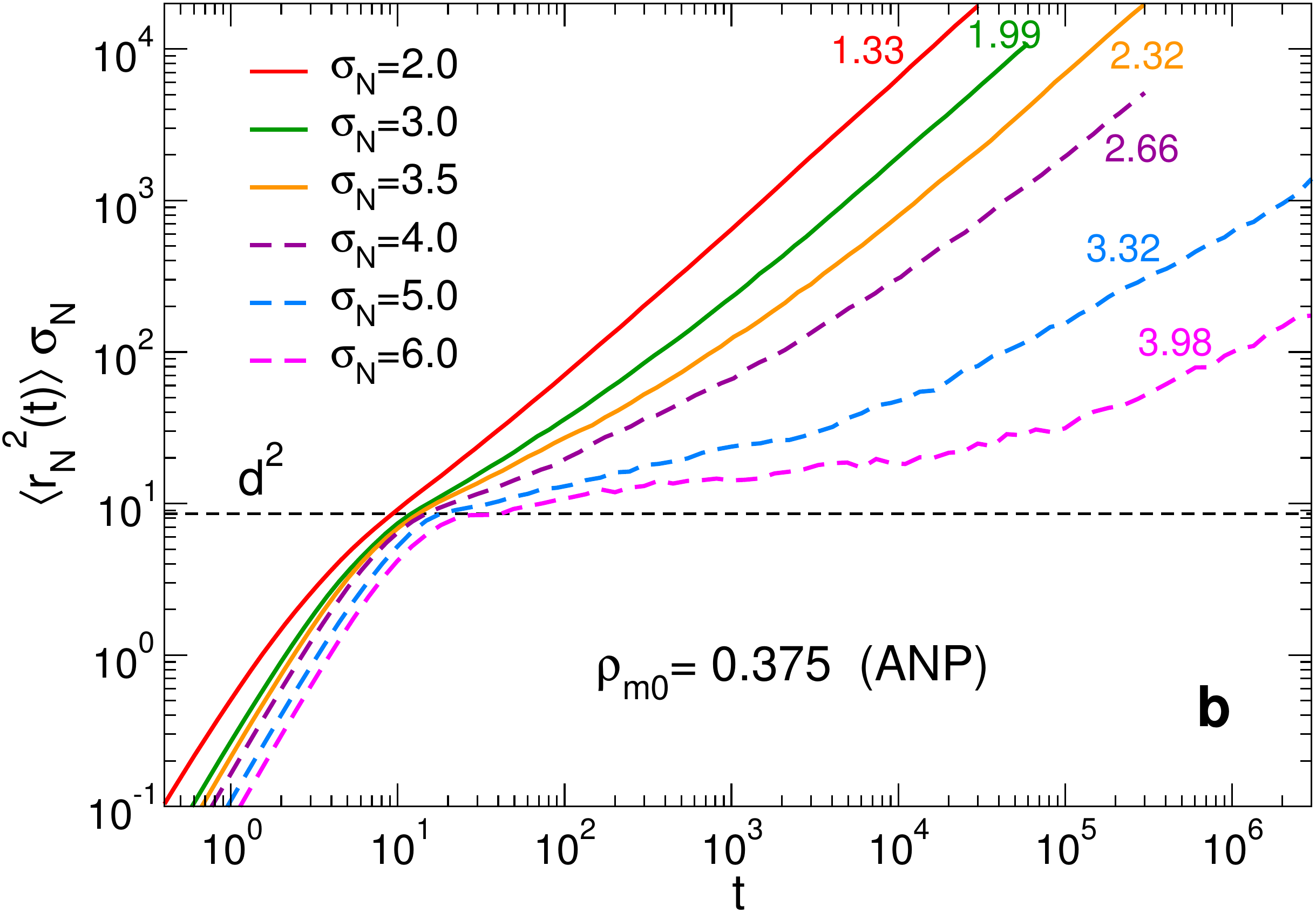}
\caption{MSD of the ANPs (\textbf{a}) and the RNPs (\textbf{b}), $\langle r_N^2(t) \rangle$, multiplied by the NP diameter $\sigma_N$, for $\rho_{m0}=0.375$ and for different $\sigma_N$ values. Dashed horizontal line: $d^2\sigma$, with $d$ the tube diameter (Tab.~\ref{tab:lambda_xi}). Dashed curves represent systems which have not reached the diffusive regime (see discussion in the text). Labels give the value of the confinement parameter $C$, Eq.~\eqref{eq:confinement}.}
\label{fig:onset_si}
\end{figure}

In Fig.~\ref{fig:onset_si}, we present the MSD of the NPs, $\langle r_N^2(t) \rangle$, multiplied by the NP diameter $\sigma_N$ for the RNPs (\textbf{a}) and the ANPs (\textbf{b}), for $\rho_{m0}=0.375$ and for different NP diameters. The theory of Cai \textit{et al.} \cite{cai2015hopping} predicts that $\langle r_N^2(t) \rangle$ reaches a plateau of height $d^2 \sigma / \sigma_N$, with $\sigma=1$ the monomer diameter and $d \simeq 1.94 \xi(\rho_{m0})$ the tube diameter, \textit{i.e.}~the product plotted in the graph should have a plateau that is independent of $\sigma_N$. For our systems, the plateau is not fully developed even for the largest values of $C$, but one notes that the onset of the subdiffusive regime (which develops into a plateau for large values of $C$) occurs at $\langle r_N^2(t) \rangle \sigma_N  \simeq d^2$ for all the simulated systems, in agreement with the theoretical prediction.

\begin{figure}
\centering
\includegraphics[width=0.48 \textwidth]{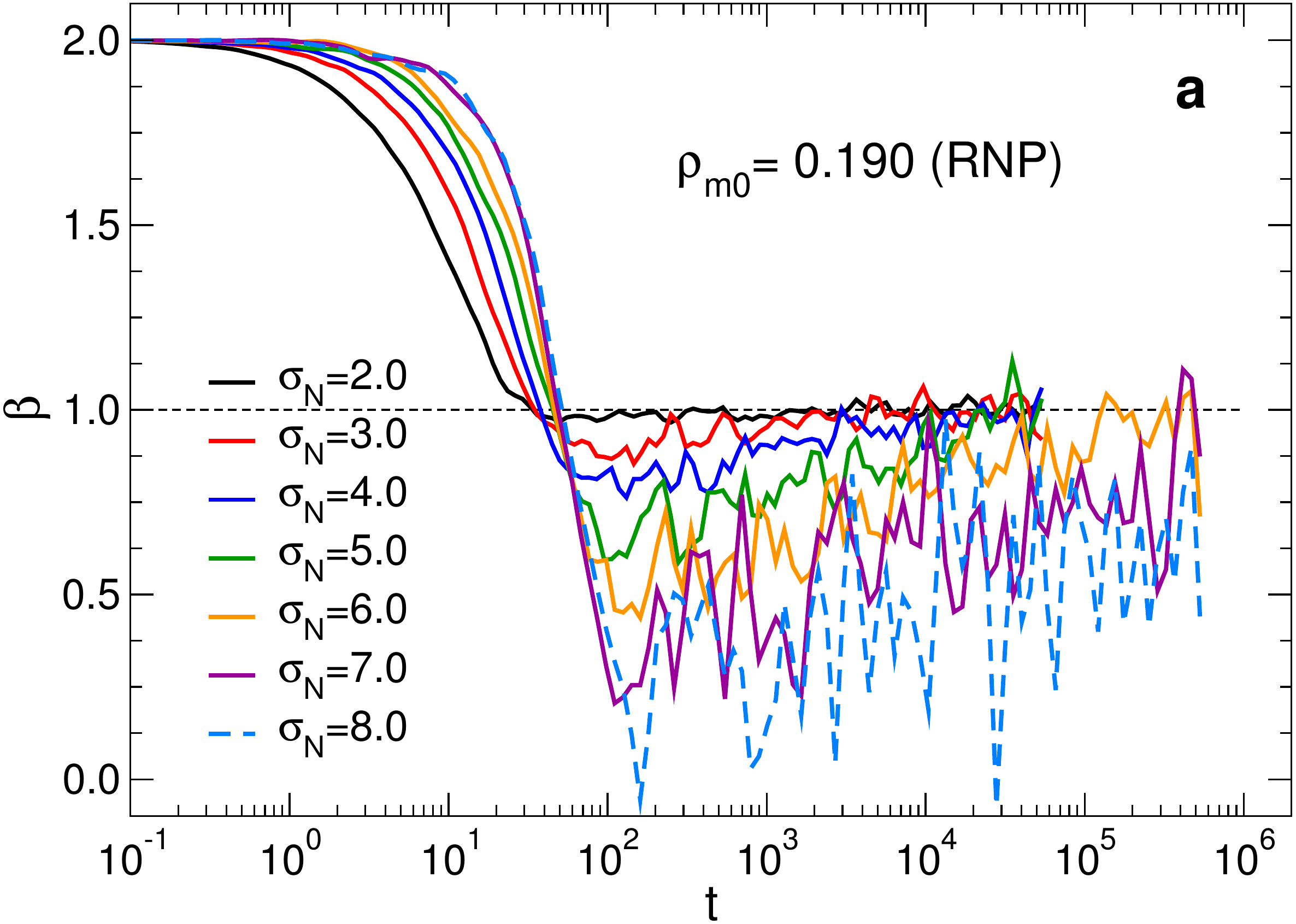}
\includegraphics[width=0.48 \textwidth]{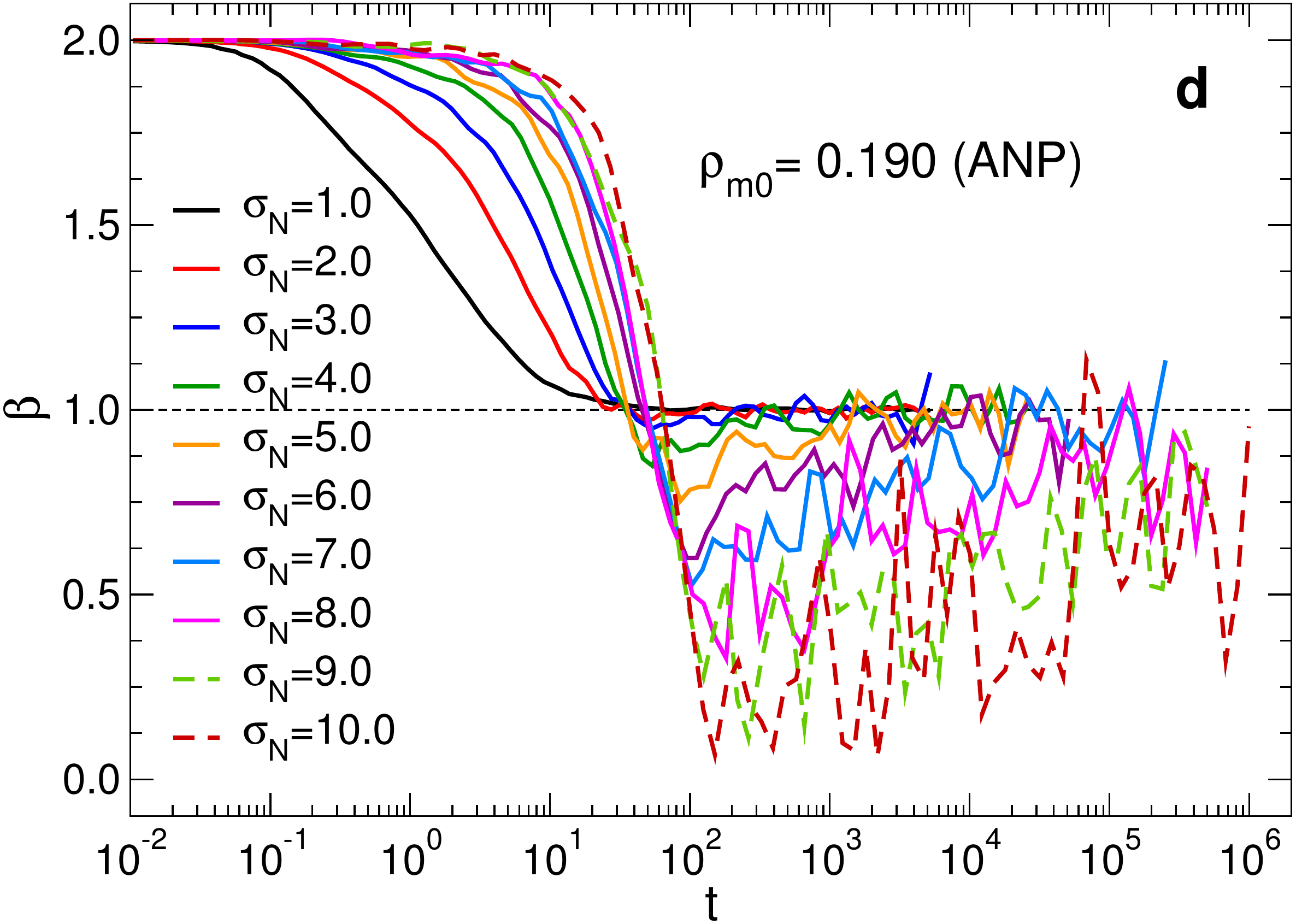}
\includegraphics[width=0.48 \textwidth]{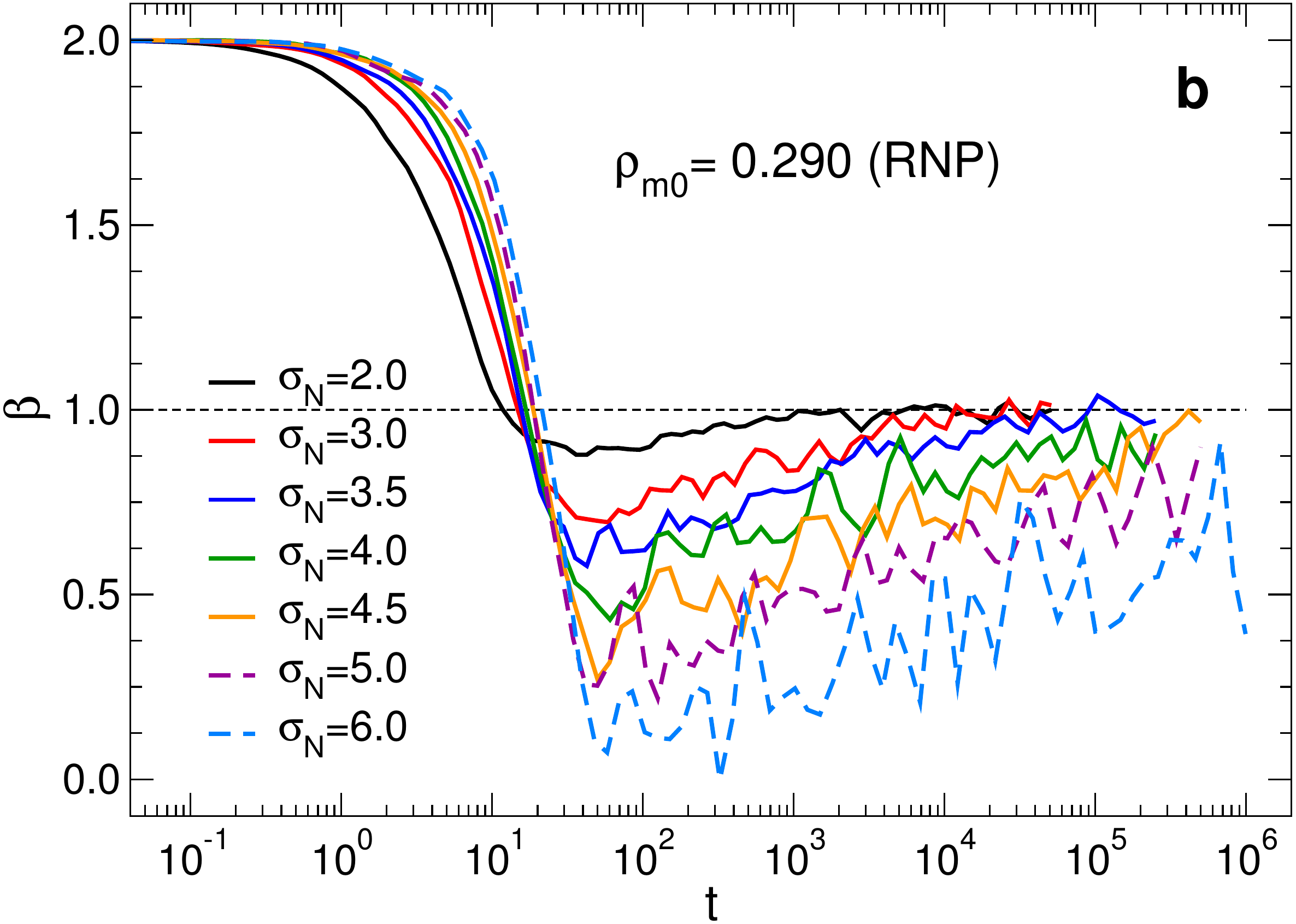}
\includegraphics[width=0.48 \textwidth]{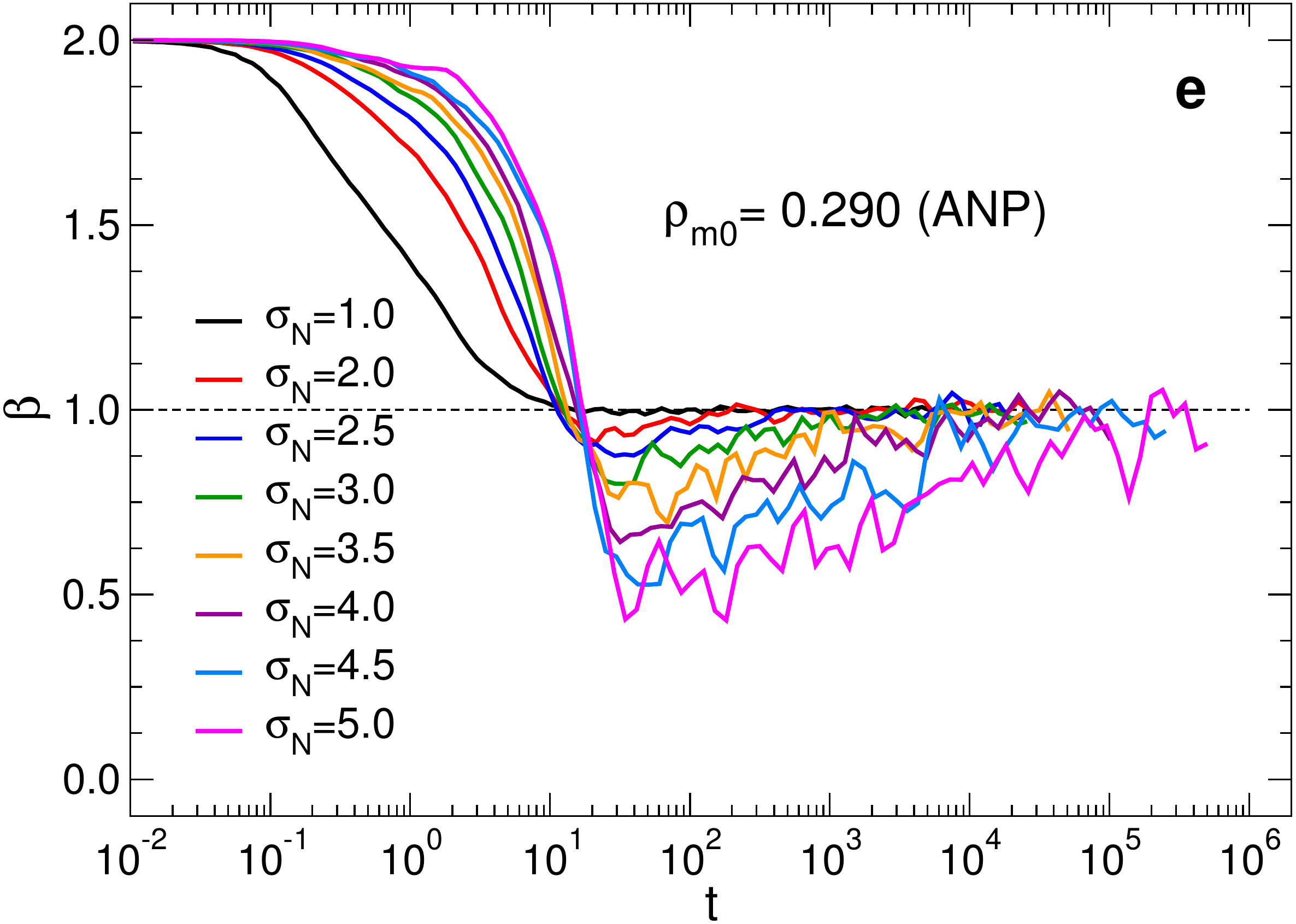}
\includegraphics[width=0.48 \textwidth]{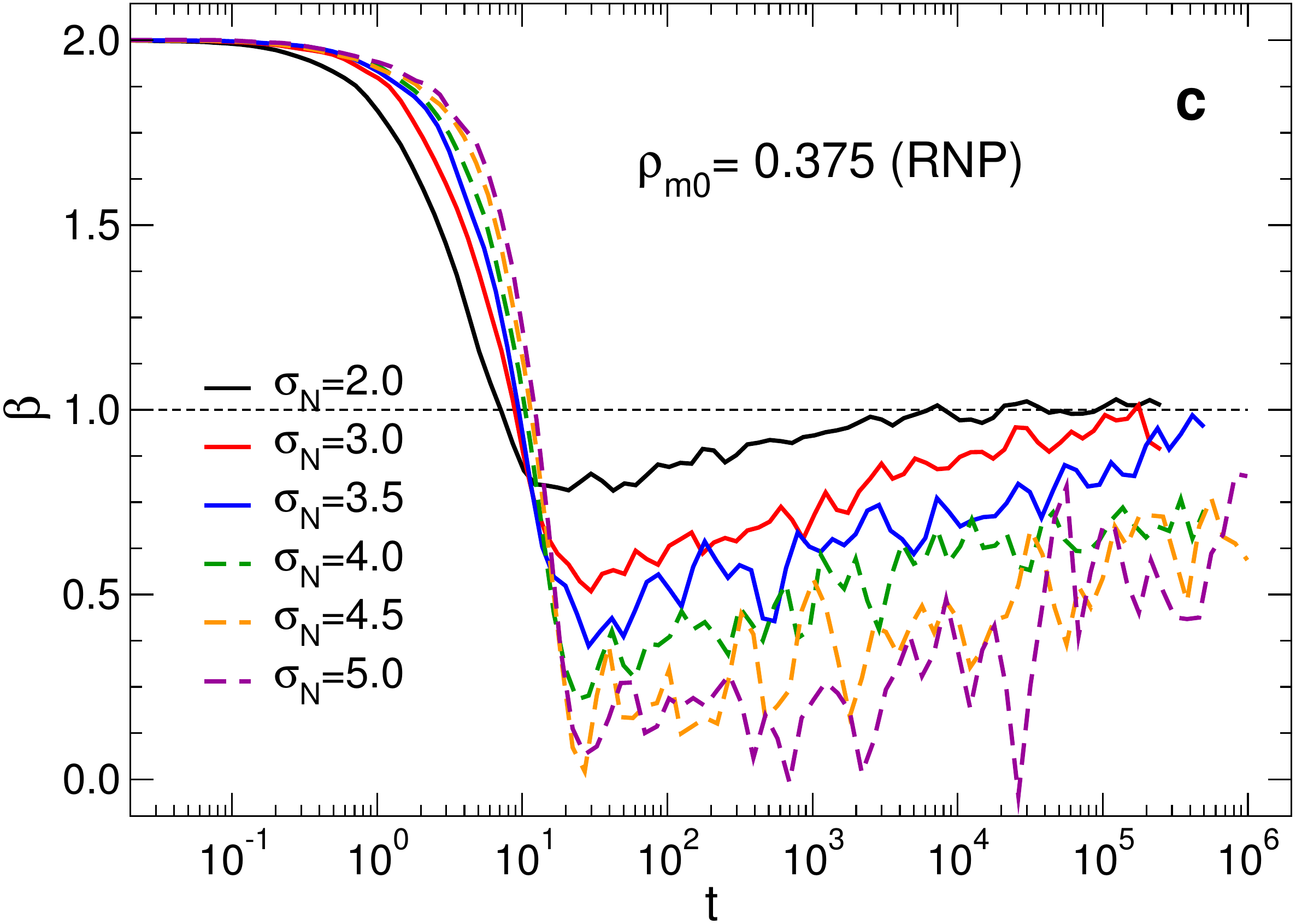}
\includegraphics[width=0.48 \textwidth]{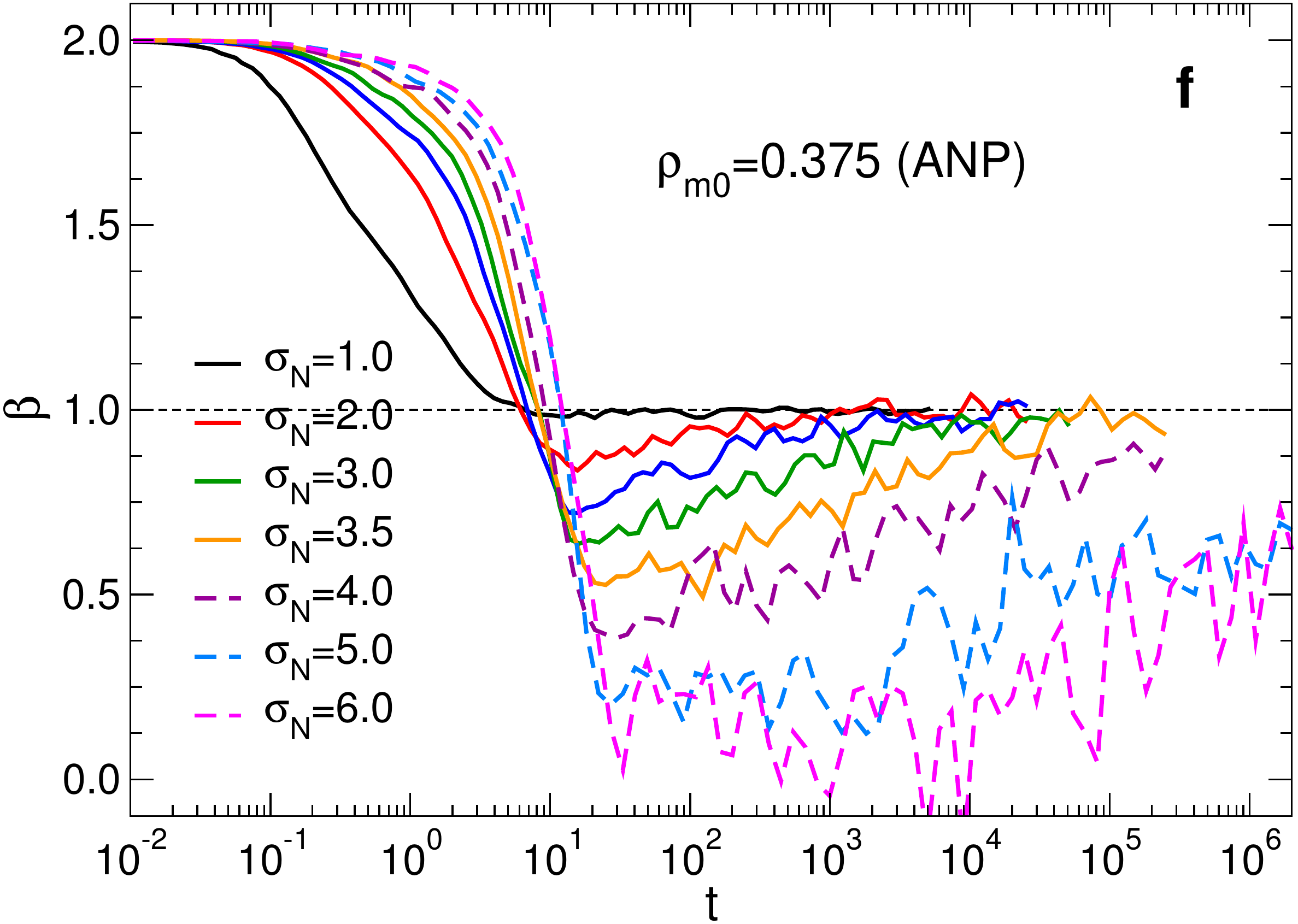}
\caption{Apparent subdiffusive exponent $\beta(t)$ of the NPs, Eq.~\eqref{eq:beta} in the main text, for the simulated systems (see labels). Dashed curves represent systems which have not reached the diffusive regime. ({\bf a})-({\bf c}): RNP, ({\bf d})-({\bf f}): ANP.}
\label{fig:beta_si}
\end{figure}

In Fig.~\ref{fig:beta_si} we show the apparent subdiffusive exponent $\beta(t)$ of the NPs, Eq.~\eqref{eq:beta} in the main text, for the simulated systems. The time dependence of these curves are qualitatively similar to the one shown in the main text, Fig.~\ref{fig:beta}: Initially, $\beta=2$ (ballistic regime), then there is a sharp decrease as the NP enters in the subdiffusive regime, followed by a very slow transition to the diffusive regime $\beta=1$.
So these results show that the MSD of the NPs does not really show a plateau at intermediate times, which would correspond to $\beta=0$, since the structural heterogeneity of the network makes that NPs start to leave their cage on time scales that are extremely broadly distributed.

%%%%%%%%%%%%%%%%%%%%%%%%%%%%%%%%%%%%%%%%%%%
\subsection{NP mean-squared displacement}
%%%%%%%%%%%%%%%%%%%%%%%%%%%%%%%%%%%%%%%%%%%
\label{sec:np_msd_si}

\begin{figure}
\centering
\includegraphics[width=0.48 \textwidth]{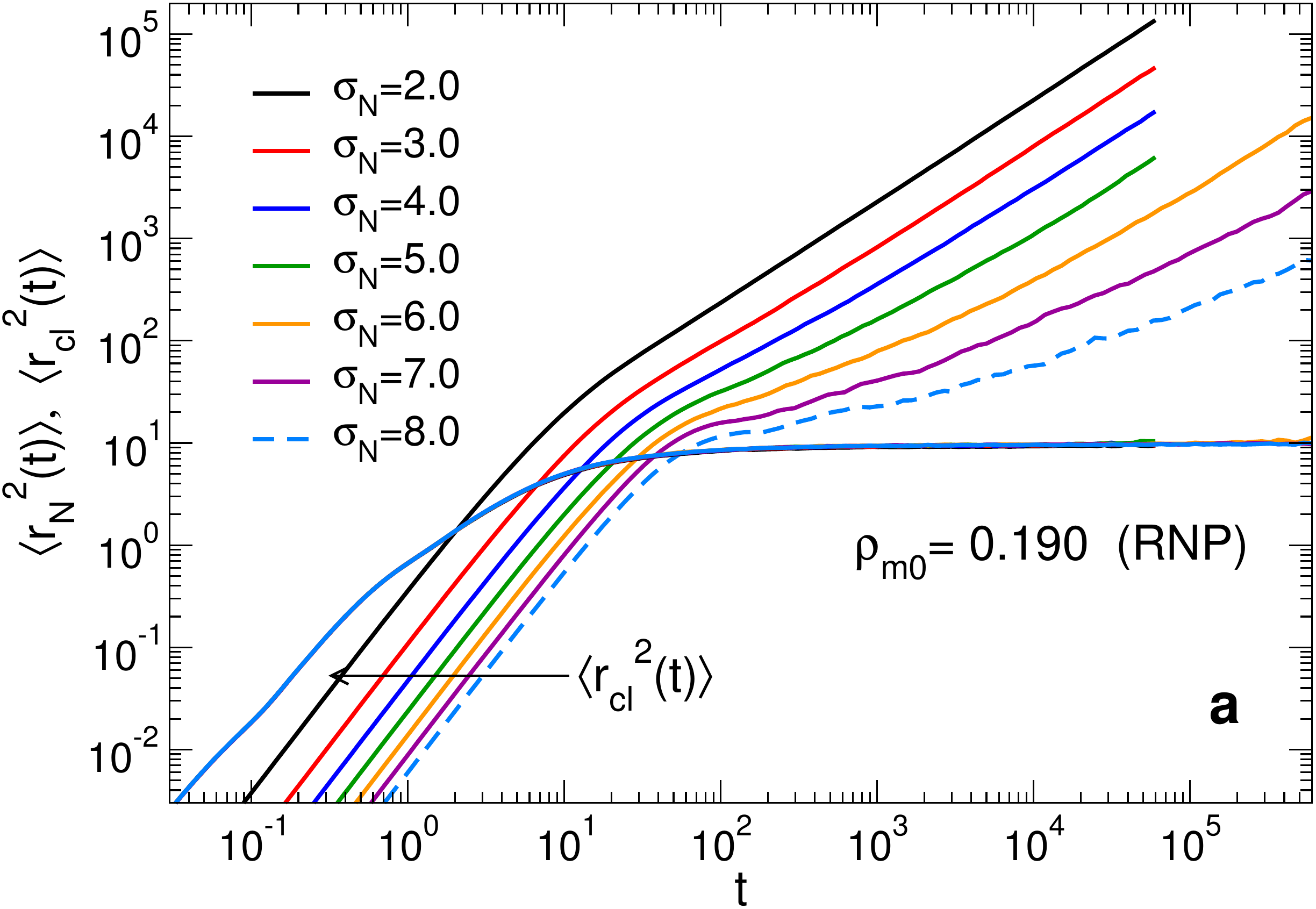}
\includegraphics[width=0.48 \textwidth]{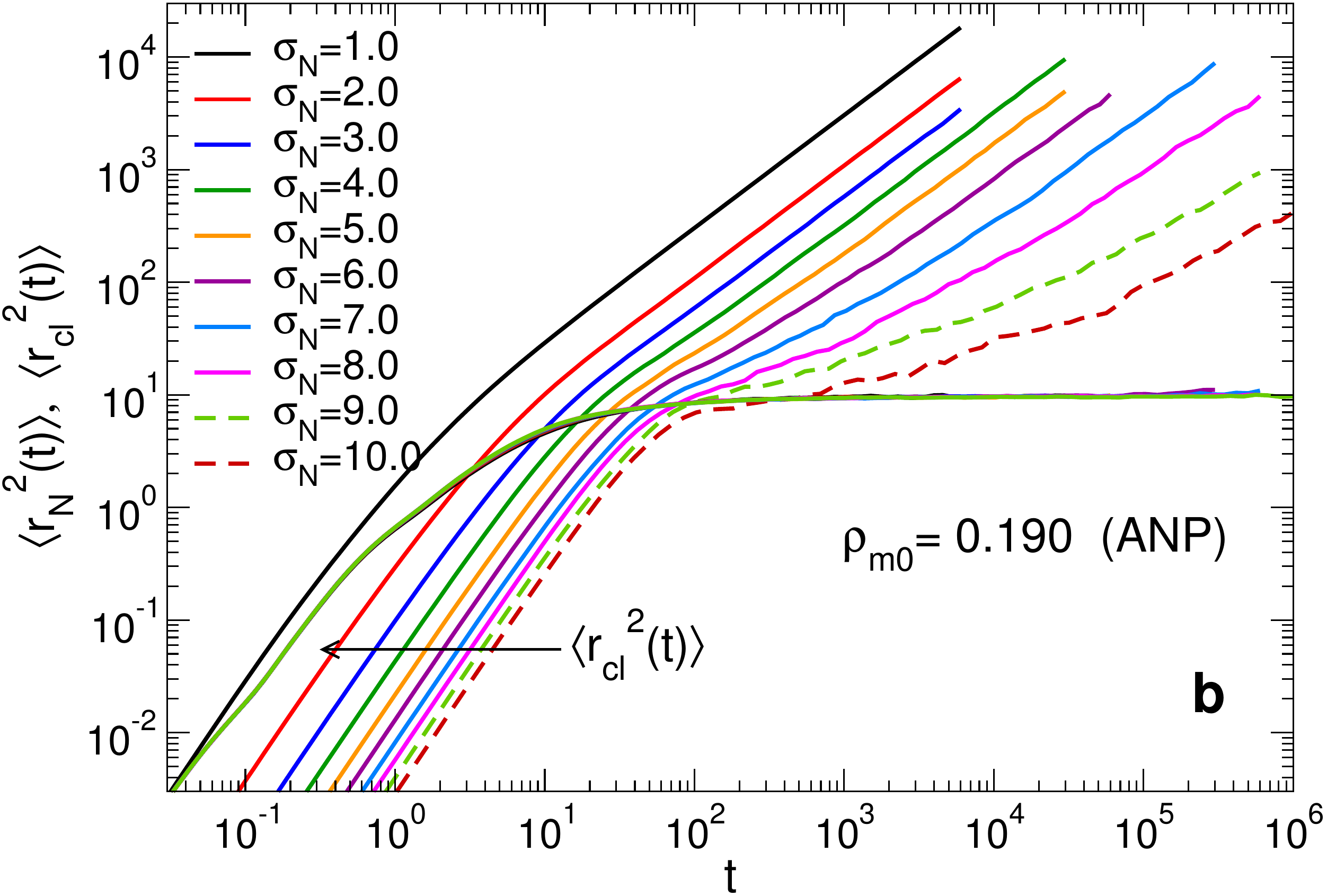}
\includegraphics[width=0.48 \textwidth]{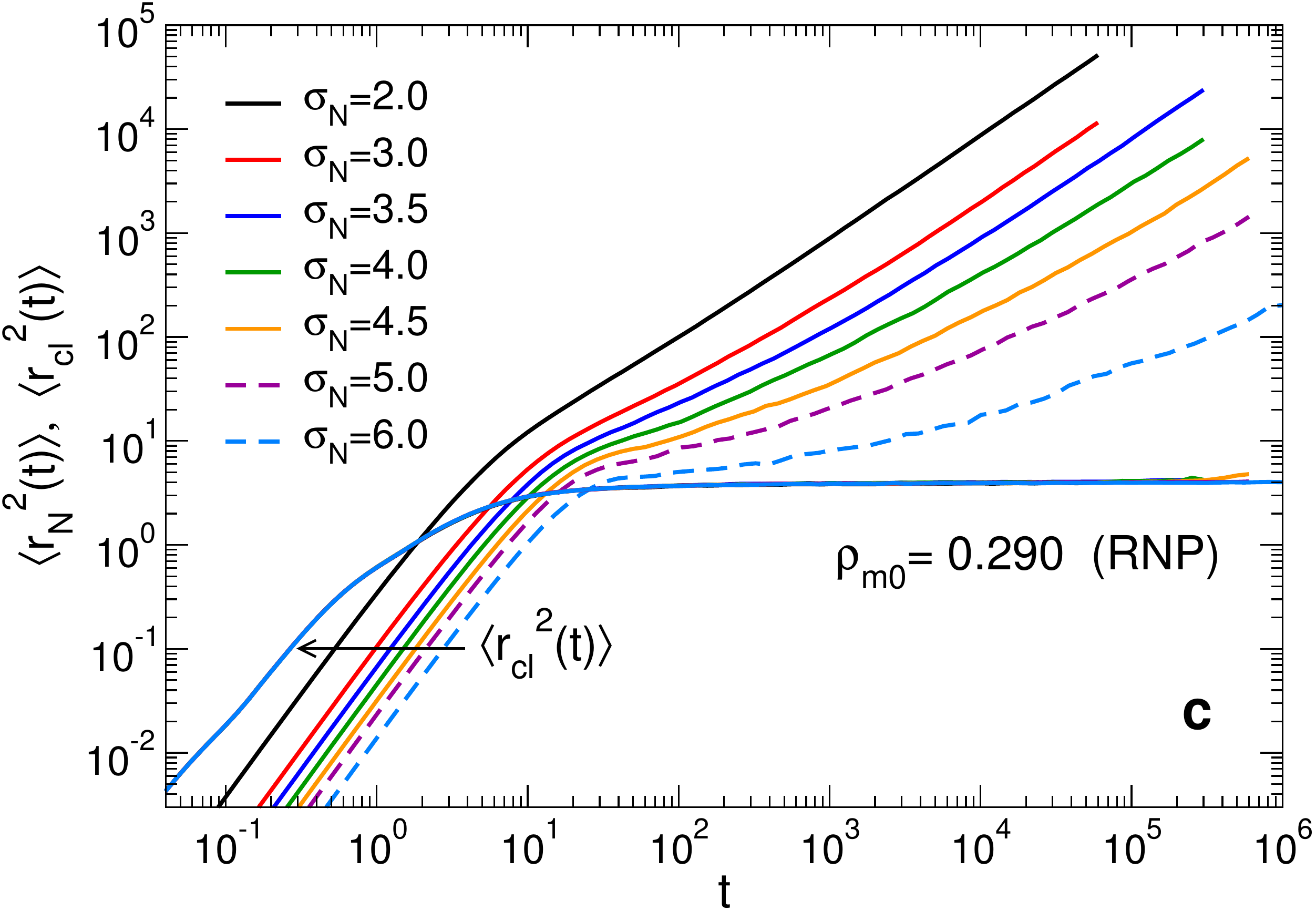}
\includegraphics[width=0.48 \textwidth]{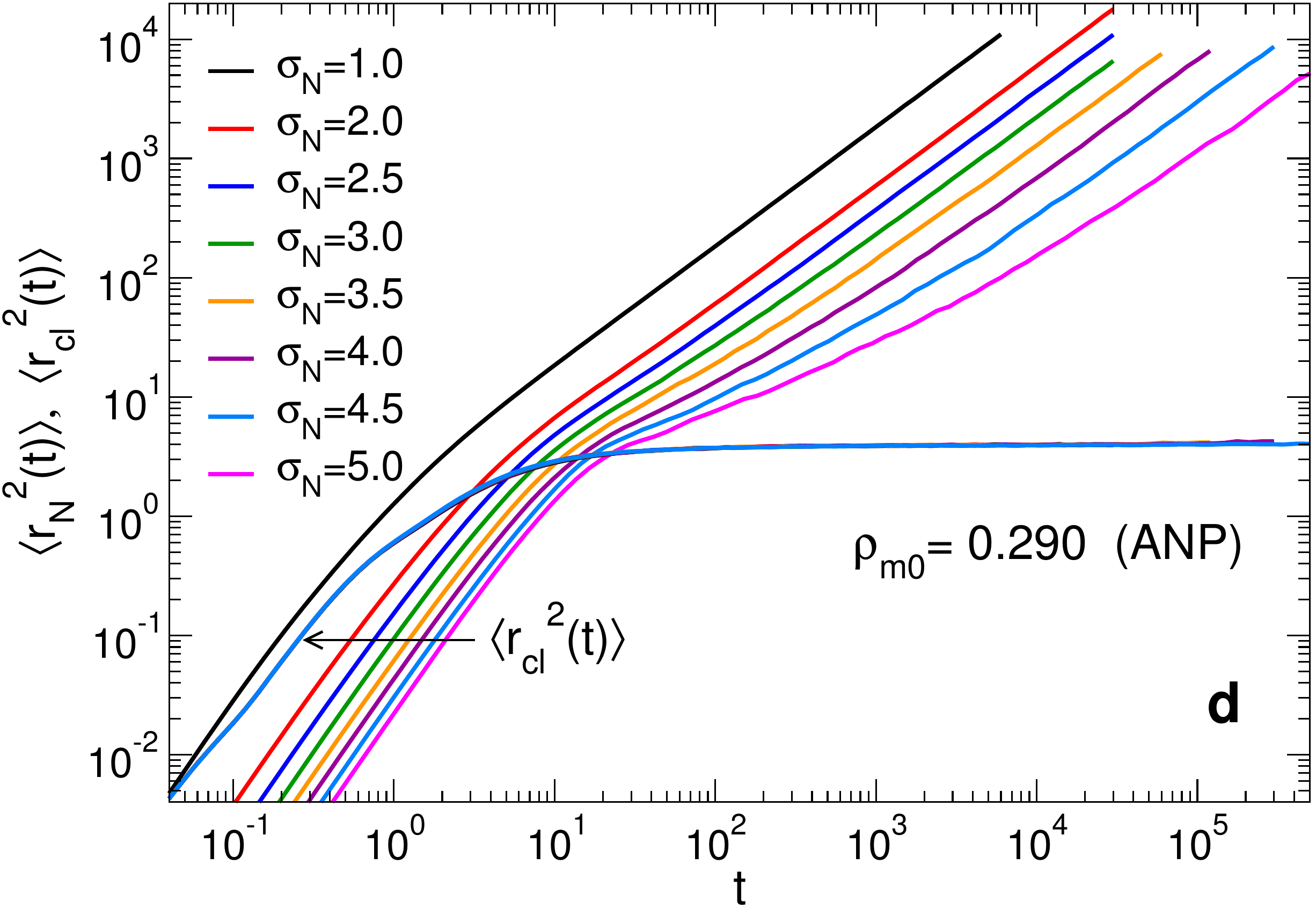}
\caption{MSD of the RNPs (\textbf{a,c}), and the ANPs (\textbf{b,d}), $\langle r_N^2(t) \rangle$, and of the crosslinks, $\langle r_\text{cl}^2(t) \rangle$, for different NP diameters and $\rho_{m0}$. Dashed curves represent systems which have not reached the diffusive regime (see discussion in text). Labels give the value of the confinement parameter $C$, Eq.~\eqref{eq:confinement}.
}
\label{fig:np_msd_si}
\end{figure}

In Fig.~\ref{fig:np_msd_si}, we display the MSD of the RNPs (\textbf{a,c}), and the ANPs (\textbf{b,d}), $\langle r_N^2(t) \rangle$, and of the crosslinks, $\langle r_\text{cl}^2(t) \rangle$, for different NP diameters and $\rho_{m0} = 0.190$ (\textbf{a,b}) and $0.375$ (\textbf{c,d}). These results are qualitatively the same as those reported in the main text for $\rho_{m0}=0.375$: Small NPs slip through the mesh, and go directly from the ballistic to the diffusive regime, whereas larger NPs are transiently trapped in the mesh, showing therefore a subdiffusive regime at intermediate times. At longer times, diffusive behavior is recovered as the NP is able to escape its local cage and diffuse \textit{via} hopping motion. We refer to the main text for a more detailed discussion.

%%%%%%%%%%%%%%%%%%%%%%%%%%%%%%%%%%%%%%%%%%%
\subsection{Non-Gaussian parameter}
%%%%%%%%%%%%%%%%%%%%%%%%%%%%%%%%%%%%%%%%%%%

\begin{figure}
\centering
\includegraphics[width=0.48 \textwidth]{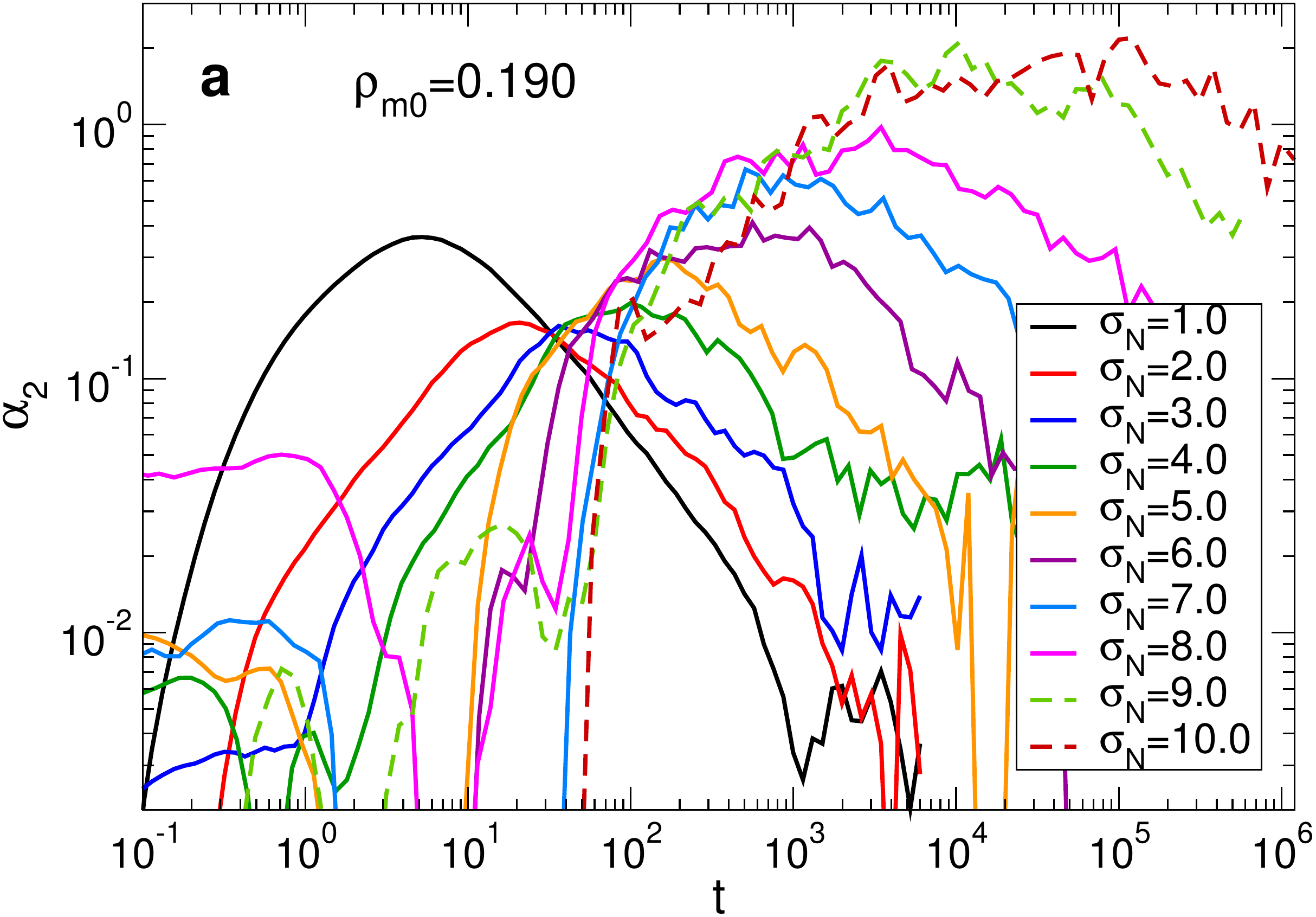}
\includegraphics[width=0.48 \textwidth]{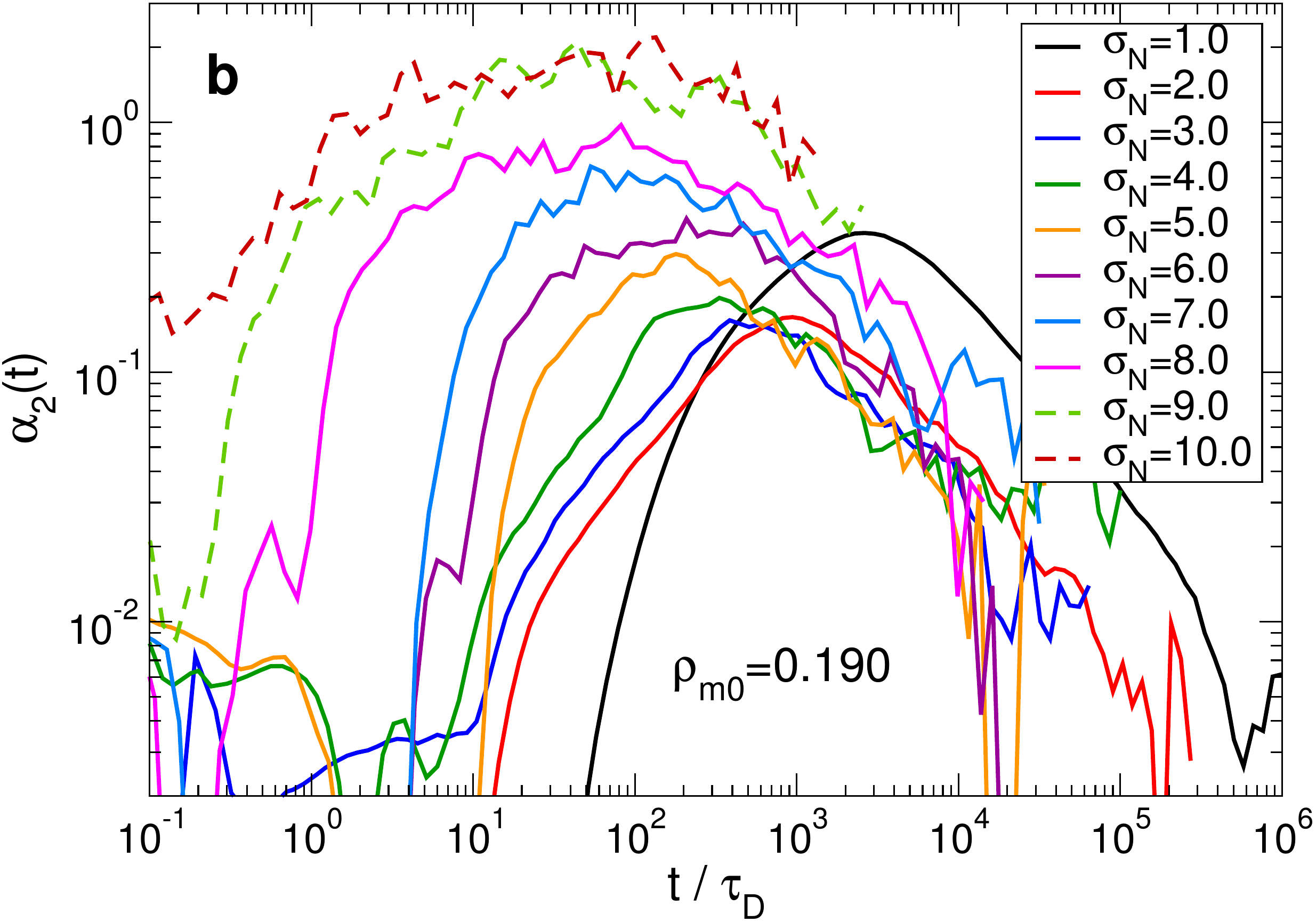}
\includegraphics[width=0.48 \textwidth]{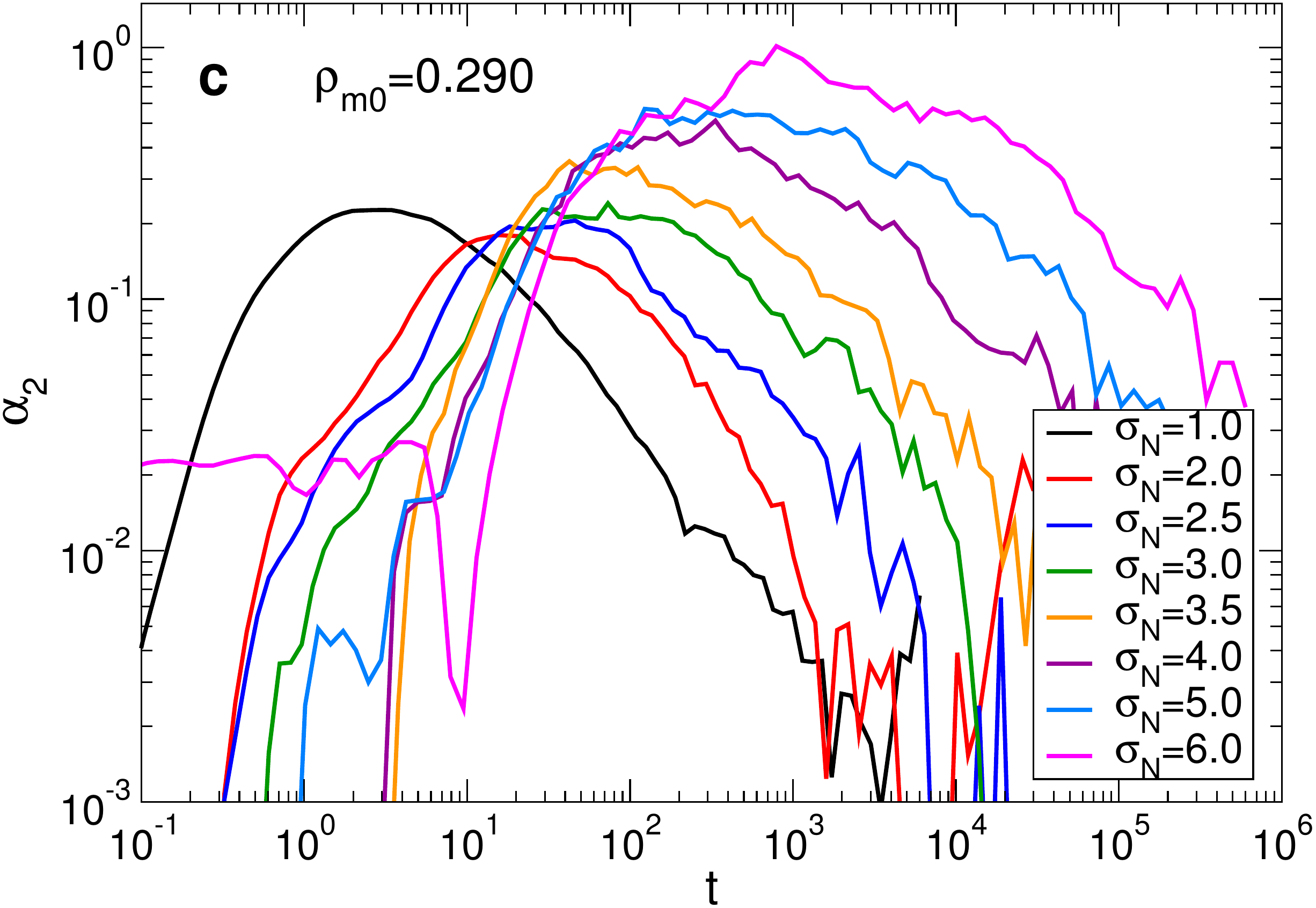}
\includegraphics[width=0.48 \textwidth]{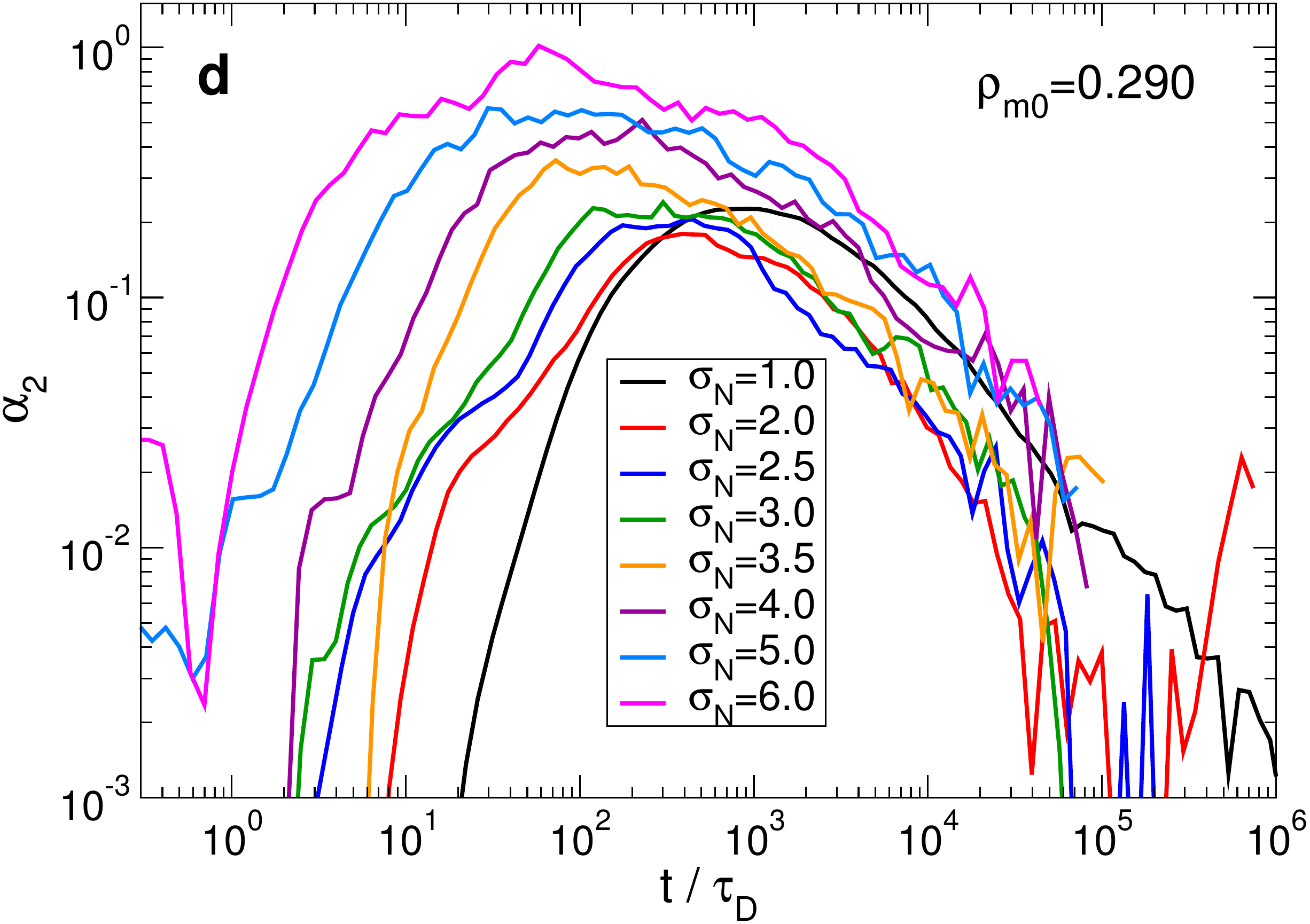}
\caption{Non-Gaussian parameter for ANPs with different values of $\sigma_N$ and $\rho_{m0}$ (see labels), shown as a function of time (\textbf{a,c}) and of the rescaled time $t/\tau_D$, with $\tau_D \equiv \sigma_N^2/6D_N$ (\textbf{b,d}). Dashed curves represent systems which have not reached the diffusive regime.}
\label{fig:nongaussian_attr_si}
\end{figure}

\begin{figure}
\centering
\includegraphics[width=0.48 \textwidth]{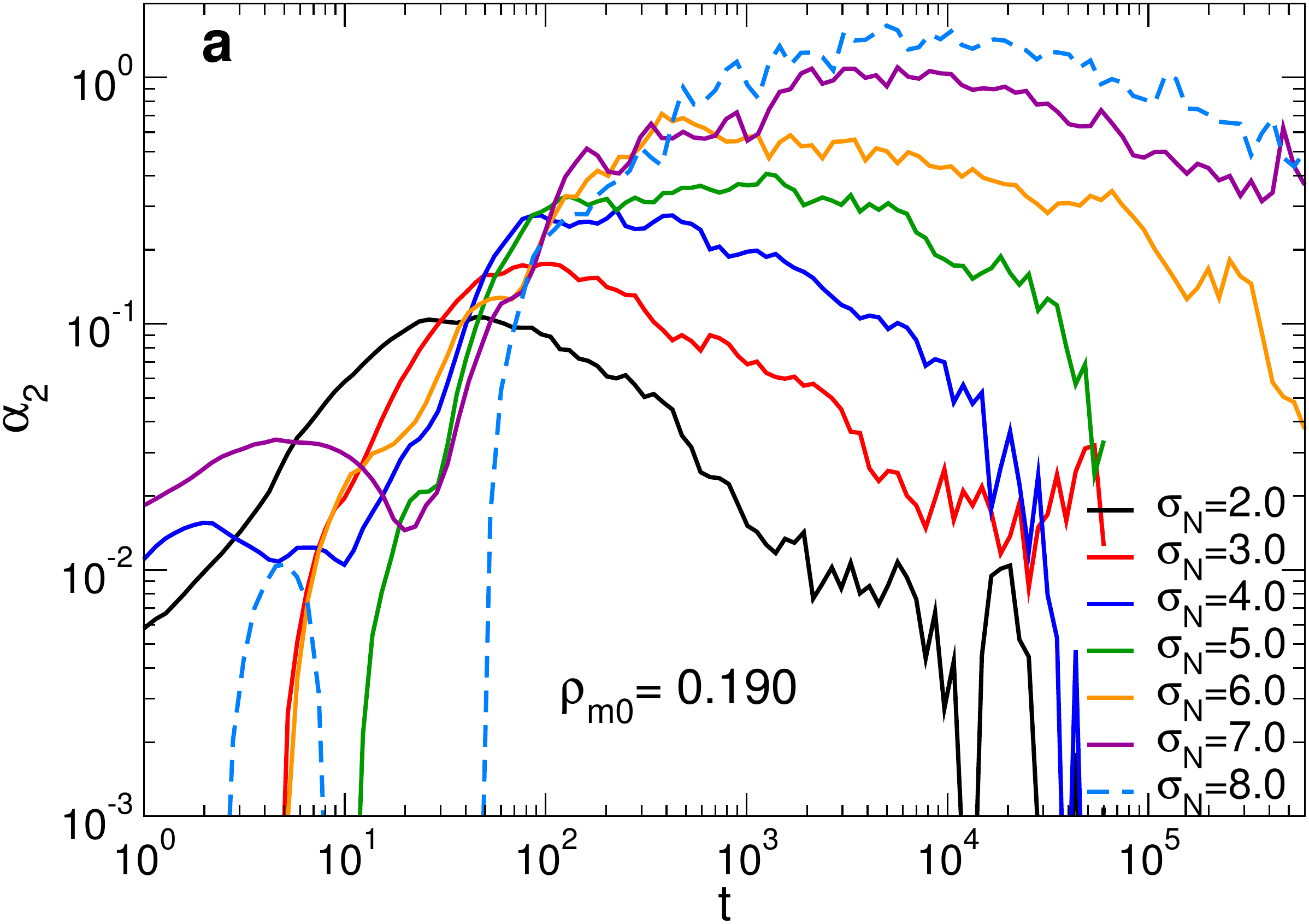}
\includegraphics[width=0.48 \textwidth]{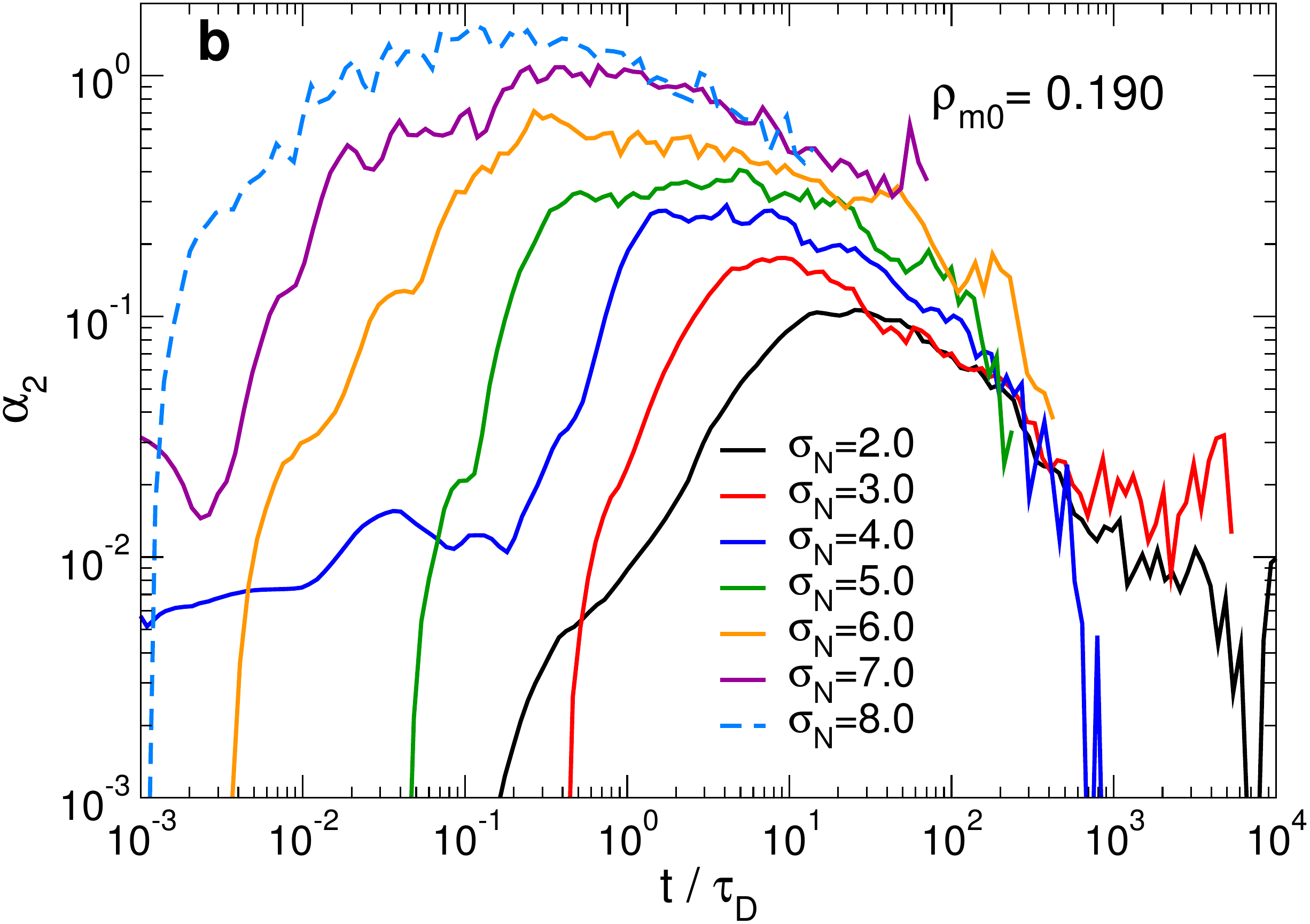}
\includegraphics[width=0.48 \textwidth]{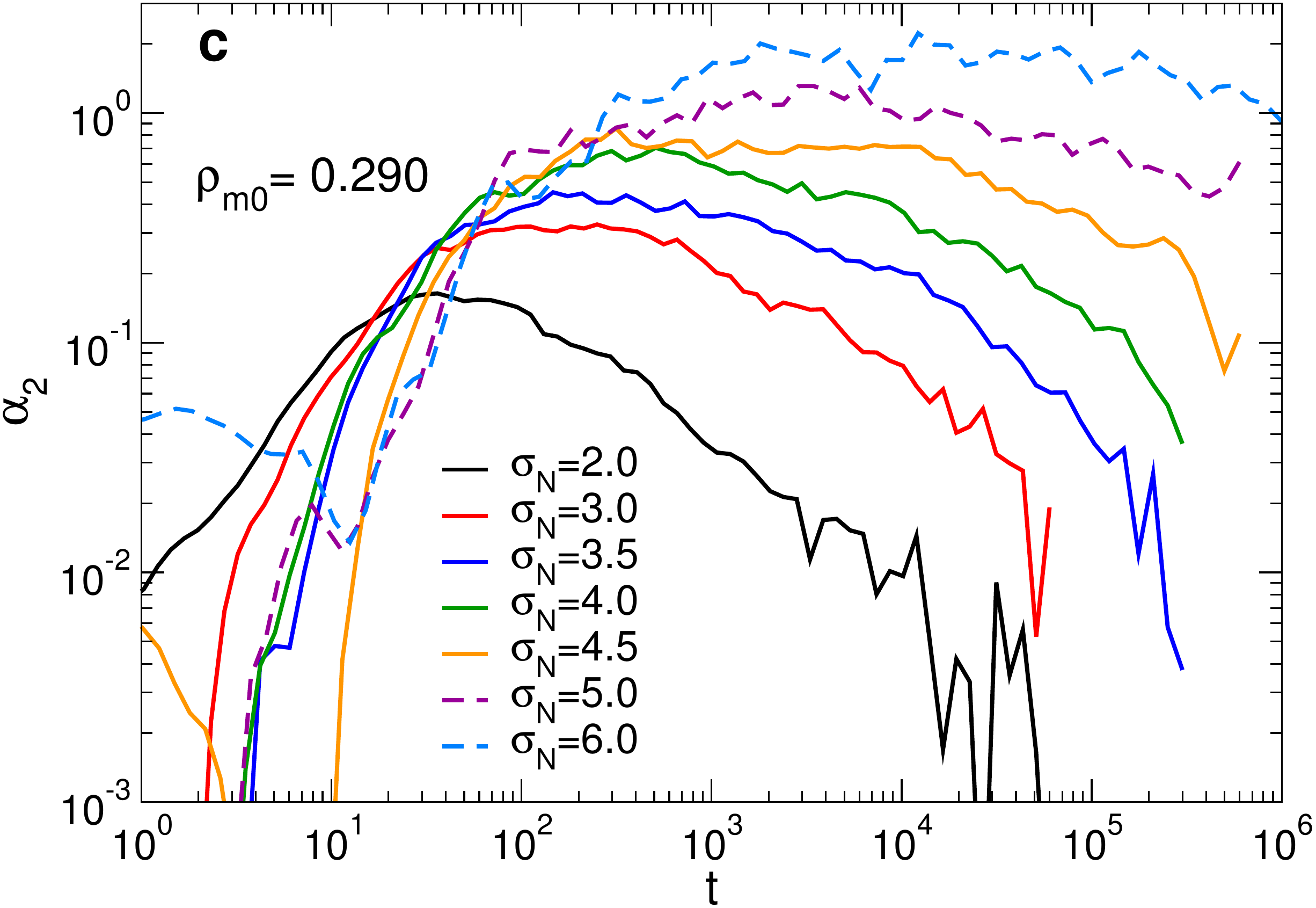}
\includegraphics[width=0.48 \textwidth]{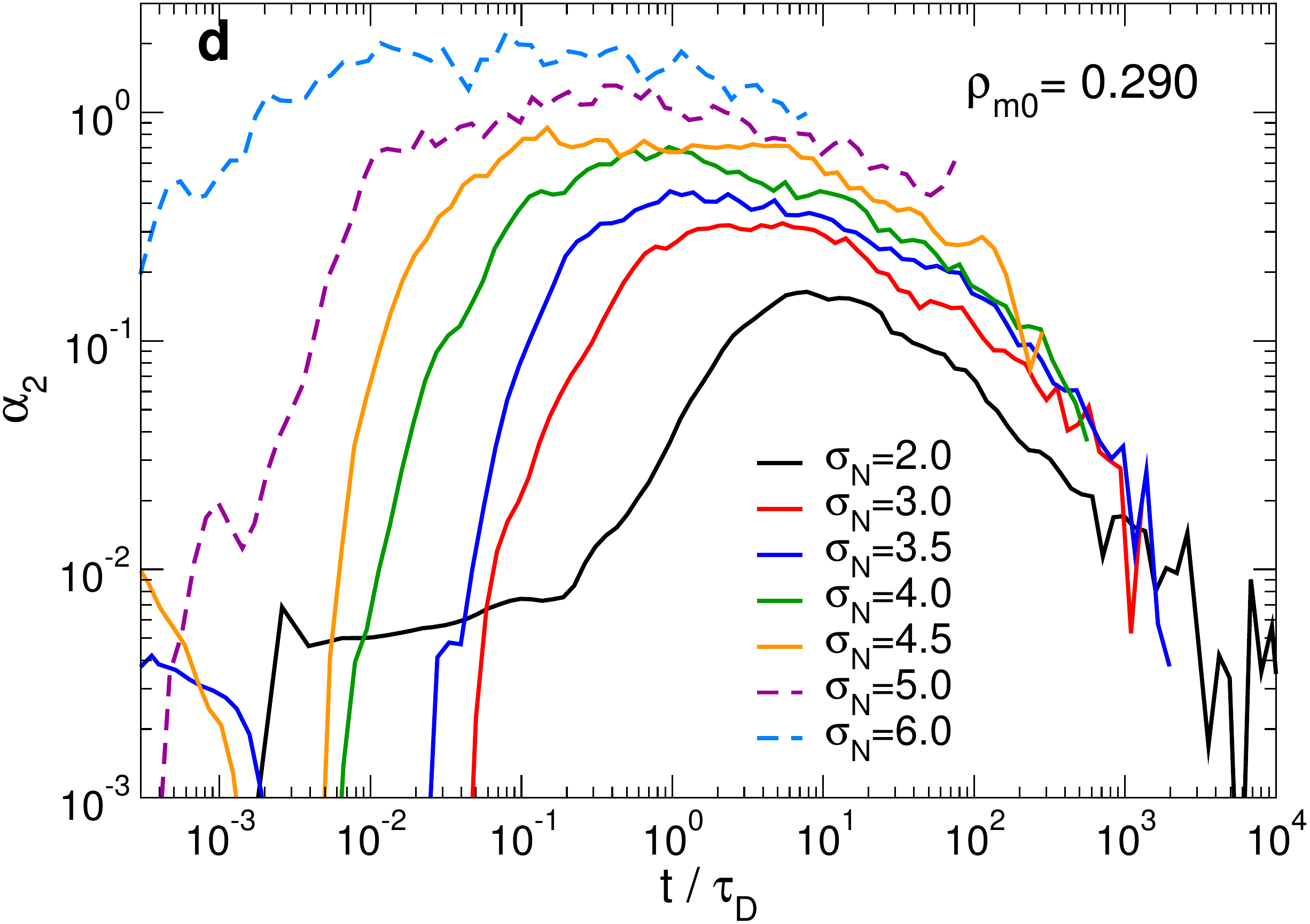}
\includegraphics[width=0.48 \textwidth]{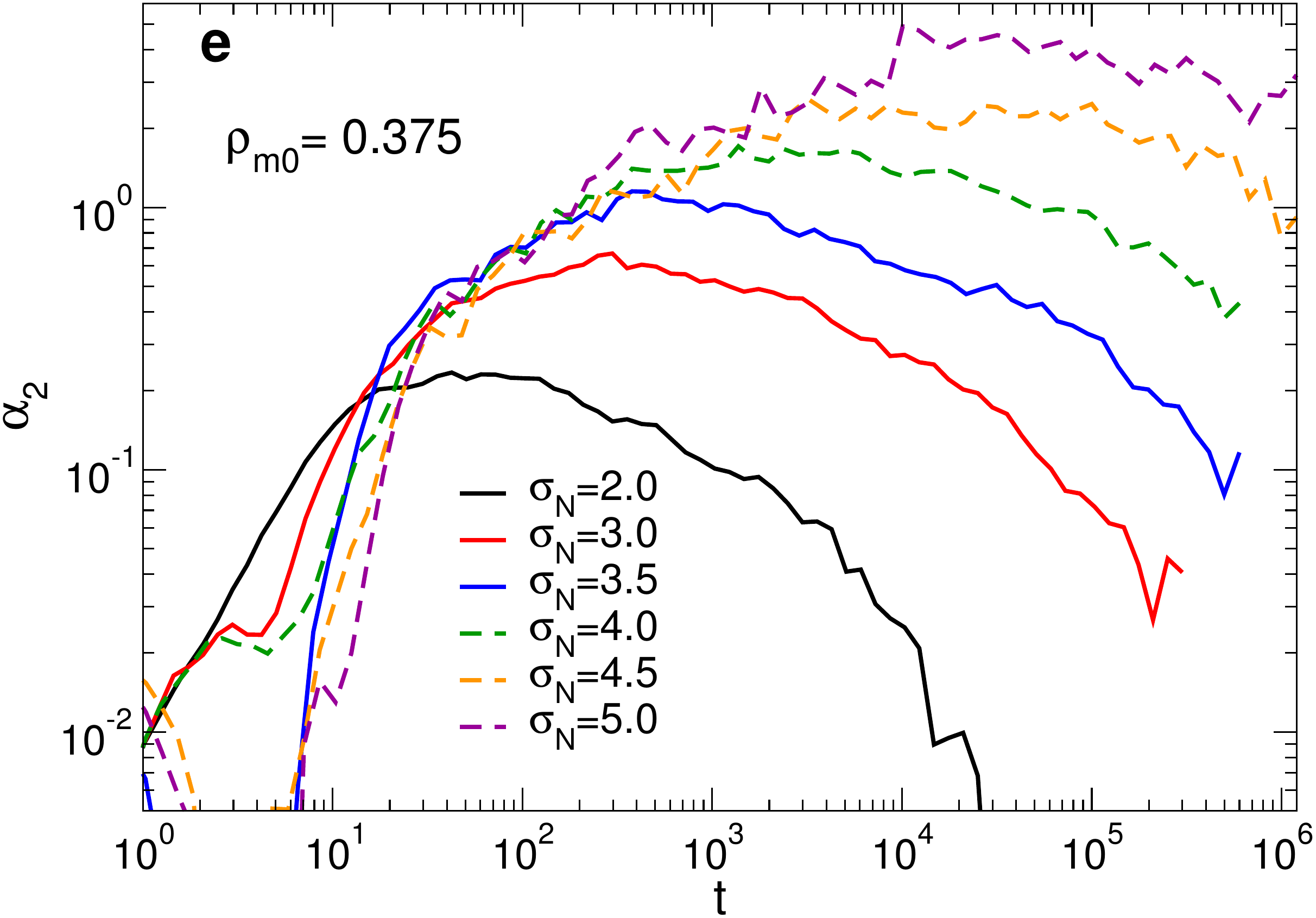}
\includegraphics[width=0.48 \textwidth]{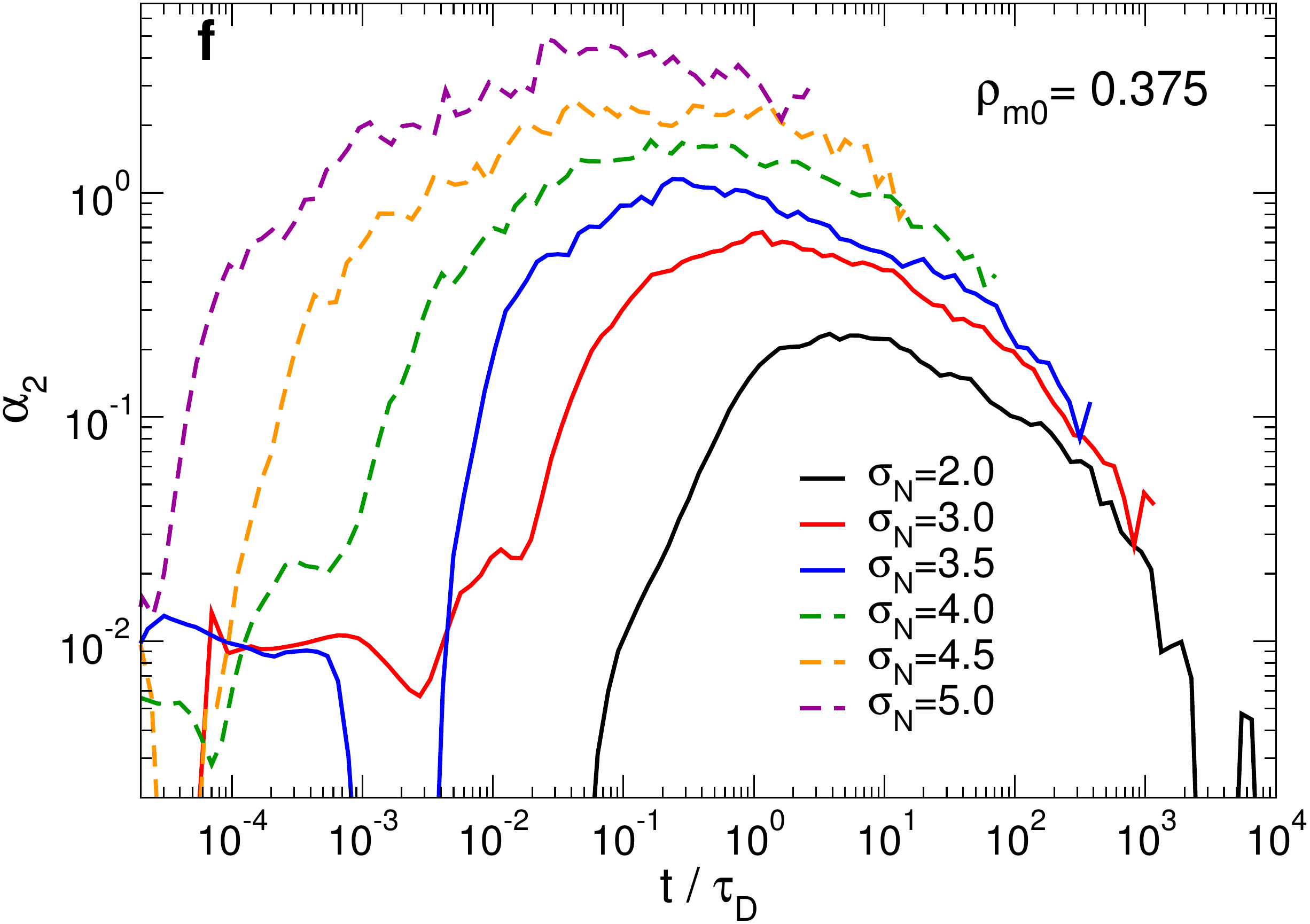}
\caption{Non-Gaussian parameter for RNPs with different values of $\sigma_N$ and $\rho_{m0}$ (see labels), shown as a function of time (\textbf{a,c,e}) and of the rescaled time $t/\tau_D$, with $\tau_D \equiv \sigma_N^2/6D_N$ (\textbf{b,d,f}). Dashed curves represent systems which have not reached the diffusive regime.}
\label{fig:nongaussian_rep_si}
\end{figure}

The time dependence of the non-Gaussian parameter $\alpha_2(t)$ is shown in Fig.~\ref{fig:nongaussian_attr_si} (ANP) and Fig.~\ref{fig:nongaussian_rep_si} (RNP).
This quantity is shown as a function of time (left side of the figures) and as a function of the rescaled time $t/\tau_D$, with $\tau_D \equiv \sigma_N^2/6D_N$. For the ANPs, the results are qualitatively the same as those reported in the main text for the ANPs for $\rho_{m0}=0.375$, and thus we refer to the main text for a detailed discussion. Also for the RNPs the behavior of $\alpha_2$ is qualitatively similar to that of the same quantity for ANPs, with the difference that the curves are somewhat broader. This is consistent with the observation that the time needed to reach the diffusive regime is larger for RNPs, as one can see from Fig.~\ref{fig:np_msd_si}. Furthermore one notices that the height of the maximum in $\alpha_2$ increases quickly with increasing density (keeping $C$ constant), which shows that the dynamical heterogeneities become more pronounced.

%%%%%%%%%%%%%%%%%%%%%%%%%%%%%%%%%%%%%%%%%%%%%%%%%
\bibliography{bibliography.bib}
\end{document}